\documentclass[fleqn,psfig,oneside,12pt]{book}

\usepackage{makeidx}
\usepackage{url}
\usepackage{graphics}
\usepackage{natbib}
\usepackage{amssymb}
\usepackage{comment}
\usepackage{hyperref}
\usepackage{tocloft}

\font\dunh=cmdunh10 scaled \magstep3
\def\ve#1{{\mbox{\boldmath$#1$}}}
\def\mat#1{{\hbox{\bf #1}}}

\def\newover#1{\mathop{\vtop{\ialign{##\crcr%
$\hfil\displaystyle{#1}\hfil$\crcr%
\noalign{\kern.1pt\nointerlineskip}%
\crcr\noalign{\kern.1pt}}}}\limits}

\def\newbot#1.#2{{\newover{#1}_{\scriptscriptstyle #2}}{}}
\def\nbot#1.#2{{\newover{#1}_{\scriptstyle #2}}{}}
\def\nbotx#1.#2{{\newover{#1}_{\displaystyle #2}}{}}

\def\uu{\hbox{\rm\char16}}
\def\mi{\buildrel\circ\over\uu}

\def\glos#1#2{{\bf #1}:\ \ #2\smallskip}
\def\grad{\hbox{\rm grad}}

\def\concept#1{{\sl #1}\index{#1}}

\renewcommand\Re{{\mathfrak{Re}}}
\renewcommand\Im{{\mathfrak{Im}}}

\makeindex

\begin{document}

\thispagestyle{empty}

\begin{center}

{

\Huge\bf

Lecture Notes on 
%Elementary Introduction in

\vskip 5mm
Basic 
Celestial Mechanics

\vskip 7cm

\dunh

Sergei A. Klioner

\vfill

%\today %2011

2011

}

\end{center}
\newpage

\tableofcontents

\newpage

%\listoftables

\listoffigures

\chapter{Introduction}

{\small {\bf Summary:}\
Research field of celestial mechanics. Historical overview: apparent
motion of planets, and solar and lunar eclipse as impetus for celestial
mechanics. Ancient celestial mechanics. Appolonius and the idea of
epicyclic motion. Ptolemy and the geocentric system.
Copernicus and the heliocentric system. Kepler and the three Kepler
laws. Galileo: satellites of Jupiter as a model for the Solar system,
the begin of mechanics. Newton: mathematical formulation of mechanics,
gravitational force. Einstein: the problem of perihelion advance of
Mercury and the general theory of relativity.

Three aspects of celestial mechanics: physics of motion, mathematics
of motion and (numerical) calculation of motion. The astronomical
objects and specific goals and problems of the modelling of their motion:
artificial satellites, the Moon, major planets, asteroids, comets,
Kuiper belt objects, satellites of the major planets, rings,
interplanetary dust, stars in binary and multiple systems, stars in
star clusters and galaxies.
}\bigskip

%%% BEGIN COMMENT
\begin{comment}

\chapter{Introduction into Maple}

{\small {\bf Summary:}\
Brief history of computer algebra systems. Maple as an integrated
computing environment for scientific calculations. The help system of
Maple. Input and output. Assignments. Maple as a pocket calculator:
numbers, operations, mathematical functions. Graphical capabilities:
2-dimensional plots, 3-dimensional plots, animation. Maple as a
symbolic calculator: elementary algebraic manipulations,
differentiation and integration, symbolic solution of equations,
limits, sums, products, solutions of differential equations,
substitutions. Maple as a library of numerical methods of computations.
Maple as a computer language: control structures, functions,
procedures.
}\bigskip

\end{comment}
%%% END COMMENT

\chapter{Two-body Problem}
\label{chapter-two-body-problem}

\section{Equations of motion}

{\small {\bf Summary:}\
Equations of motion of one test body around a motionless massive body.
Equations of general two-body problem. Center of mass. Relative
motion of two bodies. Motion relative to the center of mass.
}\bigskip

\begin{figure}
\begin{center}
\resizebox{!}{4.0cm}{\includegraphics{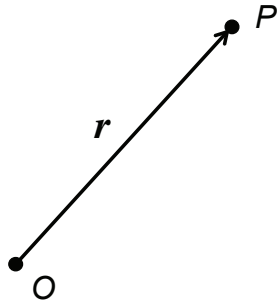}}
\end{center}
\caption[One body problem]{\small One body problem. Body $P$ with mass
$M$ is assumed to be at rest at the origin $O$ of the coordinate
system. The motion of a test particle, that is, of a body with
negligible mass $m\ll M$ is then considered assuming that $m$ is so
small that its influence on the body with mass $M$ can be totally
neglected. Position of that test particle is denoted by $\ve{r}$.
\label{Figure-one-body-problem}}
\end{figure}

Let us first consider the simplest case: the motion of a particle
having negligibly small mass $m$ in the gravitational field of a
body with mass $M$ ($m\ll M$).  Here we neglect the influence of the
smaller mass on the larger one and assume that the larger mass is at
rest at the origin of our coordinate system.  Let $\ve{r}$ be the
position of the mass $m$ (Fig. \ref{Figure-one-body-problem}). Then
according to the Newtonian law of gravity the force acting on the
smaller mass reads
\begin{displaymath}
\ve{F}=G\,{M\,m\over r^2}\,\cdot\,{\ve{r}\over r}=
-G {M\,m\over r^3}\,\ve{r}.
\end{displaymath}

\noindent 
Here and below the absolute value of a vector is designated
by the same symbol as the vector itself, but not in boldface (e.g.,
$r=|\ve{r}|$). In Newtonian mechanics force is equal to the product of
the mass and acceleration of the particle. Therefore, one has
\begin{displaymath}
\ve{F}=m\,{d^2\ve{r}\over dt^2}=m\,{\ddot \ve{r}}
\end{displaymath}
\noindent
(a dot over a symbol denote the time derivative of the corresponding
quantity and a double dot the second time derivative), and finally the
equations of motion of the mass $m$ read
\begin{equation}\label{1body-equation-of-motion}
{\ddot \ve{r}}+GM\,{\ve{r}\over r^3}=0.
\end{equation}

\begin{figure}
\begin{center}
\resizebox{!}{6.0cm}{\includegraphics{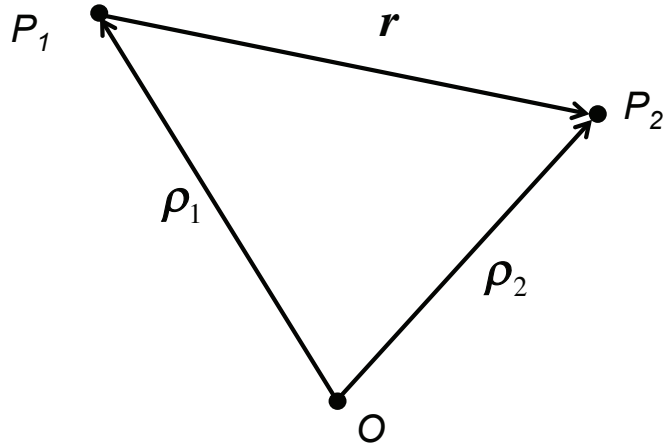}}
\end{center}
\caption[General two body problem]{\small General two body problem. Bodies $P_1$
and $P_2$ with masses $m_1$ and $m_2$ have positions $\ve{\rho}_1$ and
$\ve{\rho}_2$, respectively. The position of body $P_2$ relative to body $P_1$
is $\ve{r}=\ve{\rho}_2-\ve{\rho}_1$. Point $O$ is the origin of the
chosen coordinate system. \label{Figure-two-body}}
\end{figure}

Let us now consider the general case of two bodies experiencing
mutual gravitational attraction. Let vectors $\ve{\rho}_1$ and
$\ve{\rho}_2$ are the positions of bodies $P_1$ and $P_2$
with masses $m_1$ and $m_2$, respectively, in some
coordinate system, and $\ve{r}=\ve{\rho}_2-\ve{\rho}_1$ is the position of
body $P_2$ with respect to body $P_1$ (Fig. \ref{Figure-two-body}). Then, the equations
of motions of the two bodies read
\begin{eqnarray}\label{2body-equations-of-motion-2}
m_1\,{\ddot \ve{\rho}}_1&=&Gm_1m_2\,{\ve{r}\over r^3},
\nonumber\\
m_2\,{\ddot \ve{\rho}}_2&=-&Gm_1m_2\,{\ve{r}\over r^3}.
\end{eqnarray}
\noindent
This is a system of differential equations of order 12 (we have a
differential equation of order 2 for each of the 3 components of the
two vectors $\ve{\rho}_1$ and $\ve{\rho}_2$, the equations being
coupled to each other). Now, summing these two equations one gets that the
following linear combination of the position vectors remains zero at
any moment of time:
\begin{eqnarray}\label{2body-integral1}
&&m_1\,{\ddot \ve{\rho}}_1+m_2\,{\ddot \ve{\rho}}_2=0.
\end{eqnarray}
\noindent
Since the masses are considered to be constant in our consideration,
this equation can be integrated twice:
\begin{eqnarray}
\label{2body-cm-velocity}
&&m_1\,{\dot \ve{\rho}}_1+m_2\,{\dot \ve{\rho}}_2=\ve{A},
\\
\label{2body-cm-position}
&&m_1\,\ve{\rho}_1+m_2\,\ve{\rho}_2=\ve{A}\,t+\ve{B},
\end{eqnarray}
\noindent
where vectors $\ve{A}$ and $\ve{B}$ are some arbitrary integration constants.
Clearly these equations express that the barycenter (center of mass)
of the system of two bodies moves uniformly and rectilinear (that is,
with a constant velocity proportional to $\ve{A}$). The position of the barycenter is
\begin{displaymath}
\ve{R}={m_1\,\ve{\rho}_1+m_2\,\ve{\rho}_2\over m_1+m_2},
\end{displaymath}
\noindent
so that one has
\begin{displaymath}
{\dot \ve{R}}={\ve{A}\over m_1+m_2}={\rm const},
\end{displaymath}
\begin{displaymath}
\ve{R}={\ve{A}\over m_1+m_2}\ t+{\ve{B}\over m_1+m_2}.
\end{displaymath}

Quantities which remain constant during the motion are called
\concept{integrals of motion}. Here we have found 6 integrals 
of motion (3 components
of $\ve{A}$ and 3 components of $\ve{B}$). One often can use the
integrals of motion to reduce the order of the system of differential
equations describing the motion. Let us demonstrate that using
(\ref{2body-cm-position}) one can reduce the order of
(\ref{2body-equations-of-motion-2}) by 6. Two ways of thinking are
possible here. First, let us consider the motion of body $P_2$ relative
body $P_1$. In this case we need an equation for $\ve{r}$. Cancelling
factors $m_1$ and $m_2$ in the first and second equation of
(\ref{2body-equations-of-motion-2}), respectively,
\begin{eqnarray}\label{2body-equations-of-motion-2-prime}
{\ddot \ve{\rho}}_1&=&Gm_2\,{\ve{r}\over r^3},
\nonumber\\
{\ddot \ve{\rho}}_2&=-&Gm_1\,{\ve{r}\over r^3}
\nonumber
\end{eqnarray}
\noindent
and subtracting the resulting equations one gets
\begin{equation}\label{2body-equations-of-motion-relative}
{\ddot \ve{r}}+G(m_1+m_2)\,{\ve{r}\over r^3}=0.
\end{equation}
\noindent
This is a system of differential equations of order 6 (
components of $\ve{r}$ are defined by a system of 3 coupled equations of order 2).
Having a solution of this equation (that is, assuming that $\ve{r}$ as a
function of time is known) one has two linear equations for vectors
$\ve{\rho}_1$ and $\ve{\rho}_2$:
\begin{eqnarray}
m_1\,\ve{\rho}_1+m_2\,\ve{\rho}_2&=&\ve{A}\,t+\ve{B},
\nonumber\\
\ve{\rho}_2-\,\ve{\rho}_1&=&\ve{r}.
\nonumber
\end{eqnarray}
\noindent
The 6 constants $\ve{A}$ and $\ve{B}$
can be chosen arbitrarily (for example, computed from the initial
values for the positions $\ve{\rho}_1$ and $\ve{\rho}_2$, and
velocities ${\dot \ve{\rho}}_1$ and ${\dot \ve{\rho}}_2$).

Another possible way to use the integrals of motion
(\ref{2body-cm-position}) to reduce the order of
(\ref{2body-equations-of-motion-2}) is to consider the motion of each
body relative to the barycenter. This corresponds to choosing the
coordinate system with the origin at the barycenter and setting
$\ve{A}=0$ and $\ve{B}=0$. This is always possible due to the Galilean
relativity principle stating that coordinate systems moving with a
constant velocity relative to each other are equivalent and can be
equally used to describe the motion. From
\begin{eqnarray}
m_1\,\ve{\rho}_1+m_2\,\ve{\rho}_2&=&0,
\nonumber\\
\ve{\rho}_2-\,\ve{\rho}_1&=&\ve{r}
\nonumber
\end{eqnarray}
\noindent
one has
\begin{eqnarray}
\ve{\rho}_2&=&-\displaystyle{m_2\over m_1+m_2}\,\ve{r},
\nonumber
\\
\ve{\rho}_1&=&\displaystyle{m_1\over m_1+m_2}\,\ve{r}.
\nonumber
\end{eqnarray}
\noindent
Substituting these two equations into
(\ref{2body-equations-of-motion-relative}) one gets two uncoupled
equations for $\ve{\rho}_1$ and $\ve{\rho}_2$:
\begin{eqnarray}\label{2body-equations-of-motion-cm}
&&{\ddot \ve{\rho}}_1+G{m_2^3\over (m_1+m_2)^2}\,{\ve{\rho}_1\over \rho_1^3}=0,
\nonumber\\
&&{\ddot \ve{\rho}}_2+G{m_1^3\over (m_1+m_2)^2}\,{\ve{\rho}_2\over \rho_2^3}=0.
\end{eqnarray}
\noindent
Note that the second equation in (\ref{2body-equations-of-motion-cm})
can be derived from the first one by interchanging the indices $1\leftrightarrow 2$.

Now we notice that in all cases considered above the equations of motion
(\ref{1body-equation-of-motion}),
(\ref{2body-equations-of-motion-relative}) and (\ref{2body-equations-of-motion-cm})
have the form

\begin{equation}\label{2body-equations-of-motion}
{\ddot \ve{r}}+\kappa^2\,{\ve{r}\over r^3}=0,
\end{equation}

\noindent
where $\kappa>0$ is a constant depending on the masses of the bodies.
In the following we consider the equations of motion in their generic
form (\ref{2body-equations-of-motion}).

\bigskip
{\small
{\bf Remark.}
Note that the same equation (\ref{2body-equations-of-motion})
describes also the position of any body of a system of $N$ bodies
with $N>2$ when some special configuration (special set of positions
and velocities) of the $N$ bodies is considered. Such a configuration
must possess certain symmetry so that the sum of all gravitational
forces acting on each of the $N$ bodies is always directed toward
the center of mass of the $N$-body system. Such configurations
are called \concept{central configurations}.
}
\bigskip

{\small
{\bf Exercise.}
Find some examples of the central configuration. {\it Hint:} consider the
bodies at the vertices of equilateral polygons.
}

\section{Integrals of angular momentum and energy}

{\small {\bf Summary:}\
Integral of angular momentum (the law of areas).
The second Kepler's law. Integral of energy.
Integrals of angular momentum and energy in polar coordinates.
}\bigskip

The equations (\ref{2body-equations-of-motion}) can be further simplified by
using the so-called integrals of area (or angular momentum) and 
energy. Let us first compute the cross product of (\ref{2body-equations-of-motion})
by $\ve{r}$:
\begin{displaymath}
\ve{r}\times\left({\ddot \ve{r}}+\kappa^2\,{\ve{r}\over r^3}\right)=0
\end{displaymath}
\noindent
which implies
\begin{displaymath}
\ve{r}\times{\ddot \ve{r}}=0.
\end{displaymath}
\noindent
The latter equation can be integrated to give
\begin{equation}\label{2body-area-cartesian}
\ve{r}\times{\dot \ve{r}}=\ve{c},
\end{equation}
\noindent
where $\ve{c}={\rm const}$. Eq. (\ref{2body-area-cartesian})
has two consequences:

\begin{enumerate}
\item The motion is planar. Indeed, the constant vector $\ve{c}$ is
orthogonal to both position vector $\ve{r}$ and velocity vector
$\dot{\ve{r}}$ at any moment of time. The last two vectors define the momentary plane of motion
and since $\ve{c}={\rm const}$ this plane does not change.

\item The area swept out by position vector $\ve{r}$ within an
infinitely small interval of time $dt$ remains constant. Indeed, if at
some moment of time $t$ the position vector is $\ve{r}$ then at time
$t+dt$ the position vector is $\ve{r}+{\dot \ve{r}}\,dt$, where ${\dot
\ve{r}}$ is the velocity vector at time $t$. The area encompassed by
vectors $\ve{r}$ and $\ve{r}+{\dot \ve{r}}\,dt$ can be calculated as
$dA={1\over 2}\, |\ve{r}\times(\ve{r}+{\dot \ve{r}}\,dt)|= {1\over
2}\,|\ve{r}\times{\dot \ve{r}}|\,dt ={1\over 2}\,c\,dt$. Therefore,
$\dot A={1\over 2}\,c={\rm const}$. This is nothing else than the
second Kepler law in differential form (its usual integral form
$\Delta A={1\over 2}\,c\,\Delta t$ immediately follows from the
differential form $\dot A={1\over 2}\,c$). Let us remind that the
standard formulation of the second Kepler's law states that ``a line
joining a planet and the Sun sweeps out equal areas during equal
intervals of time''.

\end{enumerate}

\bigskip
{\small
{\bf Remark.}
We have used only one property of
(\ref{2body-equations-of-motion}) in order to derive (\ref{2body-area-cartesian}):
the property that the force is proportional to $\ve{r}$. The coefficient of
proportionality plays no role here and can be any function of time $t$, position $\ve{r}$,
velocity $\dot{\ve{r}}$. Such forces are called \concept{central forces}.
Motion with any central forces is, therefore, planar and satisfies the second
Kepler's law.
}
\bigskip

Denoting the components of vectors as
$\ve{r}=(x,y,z)$, ${\dot \ve{r}}=(\dot x,\dot y,\dot z)$,
$\ve{c}=(c_x,c_y,c_z)$ one can rewrite (\ref{2body-area-cartesian})
in the form
{
% to produce a non-standard equation number
\addtocounter{equation}{-1}
\catcode`@=11
\renewcommand{\theequation}{\thechapter.\@arabic\c@equation$\,^\prime$}
\catcode`@=12
\begin{eqnarray}
y\,\dot z-\dot y\, z=c_x,
\nonumber\\
z\,\dot x-\dot z\, x=c_y,
\\
x\,\dot y-\dot x\, y=c_z,
\nonumber
\end{eqnarray}
}

Let us now use the fact that the motion is planar and 
re-orient our coordinates in such a way that one of the body's coordinate remain
identically zero. This means that the body remains in the plane $z=0$ of the
coordinate system. Let us then denote the two other coordinates as
$\xi$ and  $\eta$. In these new coordinates Eq. (\ref{2body-area-cartesian})
reads
\begin{equation}\label{2body-area-cartesian-in-plane}
\xi\,\dot\eta-\dot\xi\,\eta=c,
\end{equation}
\noindent
and the equations of motion can be rewritten as
\begin{eqnarray}\label{2body-equations-of-motion-xi-eta}
{\ddot \xi}+\kappa^2\,{\xi\over r^3}=0,
\nonumber\\
{\ddot \eta}+\kappa^2\,{\eta\over r^3}=0
\end{eqnarray}
\noindent
where $r=\sqrt{\xi^2+\eta^2}$.

Let us now multiply the first equation of
(\ref{2body-equations-of-motion-xi-eta}) by $2\,{\dot\xi}$ and the
second one by $2\,{\dot\eta}$, and then add the two resulting
equations to get
\begin{displaymath}
2{\dot\xi}\,{\ddot \xi}+2{\dot\eta}\,{\ddot \eta}=
-\kappa^2\,{2{\dot\xi}\,\xi+2{\dot\eta}\,\eta\over r^3}.
\end{displaymath}
\noindent
Both sides of the latter equation are total time derivatives. Integrating
this equation one gets
\begin{equation}\label{2body-energy-cartesian-in-plane}
{\dot \xi}^2+{\dot \eta}^2=2\kappa^2\,{1\over r}+h,
\end{equation}
\noindent
where $h={\rm const}$ is a constant of integration. 
The validity of (\ref{2body-energy-cartesian-in-plane}) can be checked by
calculating its derivative with respect to time $t$ and comparing it with 
the previous equations.
Quantity $h$ represents one more integral of motion which is called
energy constant. Indeed, the left-hand side of
(\ref{2body-energy-cartesian-in-plane}) is doubled kinetic energy of the
body per unit of mass and the right-hand side is minus doubled
potential energy per unit of mass plus constant $h$.

Let us now introduce polar coordinates $r$ and $u$ instead of 
Cartesian coordinates $\xi$ and $\eta$. Using standard relations
$\xi=r\,\cos u$ and $\eta=r\,\sin u$ which imply, for example,
\begin{eqnarray}
{\dot \xi}={\dot r}\,\cos u - r\,\sin u\,{\dot u},
\nonumber\\
{\dot \eta}={\dot r}\,\sin u + r\,\cos u\,{\dot u}
\nonumber
\end{eqnarray}
\noindent
one gets the integrals of area (\ref{2body-area-cartesian-in-plane}) and
of energy (\ref{2body-energy-cartesian-in-plane}) in polar coordinates
\begin{equation}\label{2body-area-polar}
r^2\,\dot u=c,
\end{equation}
\begin{equation}\label{2body-energy-polar}
{\dot r}^2+r^2\,{\dot u}^2=2\kappa^2\,{1\over r}+h.
\end{equation}
The equations of motion (\ref{2body-equations-of-motion-xi-eta})
can be also expressed in polar coordinates $r$ and $u$. One can show that
the only non-trivial equation reads
\begin{equation}\label{2body-equations-of-motion-polar}
\ddot r-r\,{\dot u}^2+{\kappa^2\over r^2}=0.
\end{equation}

\bigskip
{\small
{\bf Exercise.}
Rewrite the equations of motion (\ref{2body-equations-of-motion-xi-eta})
in polar coordinates $r$ and $u$ explicitly. Show that that they can be
expressed as a sum of derivatives of the integrals (\ref{2body-area-polar})
and (\ref{2body-energy-polar}).
}
\bigskip

\section{Possible Orbits}

{\small {\bf Summary:}\
Conic sections as possible orbits in the two-body problem.
The first Kepler's law. Definitions of the semi-latus rectum, eccentricity
and the argument of pericenter. Apocenter and pericenter. Apsidal line.
Elliptical, parabolic, hyperbolic and rectilinear motions.
}\bigskip

Our aim now is investigate the form of the orbits implies by 
(\ref{2body-area-polar}) and (\ref{2body-energy-polar}).
Since we are interested in the form of the orbits only, we can
eliminate the time variable from the two equations.
Eq. (\ref{2body-area-polar}) implied that $\dot u=c\,r^{-2}$.
Therefore, $\dot r= {d\over dt} r={dr\over du}\,{du\over dt}
={dr\over du}\,c\,r^{-2}$. Substituting this into
(\ref{2body-energy-polar}) one gets
\begin{displaymath}
\left({dr\over du}\right)^2\,{c^2\over r^4}+r^2\,{c^2\over r^4}=
2\kappa^2\,{1\over r}+h
\end{displaymath}
\noindent
and, therefore,
\begin{displaymath}
\left(-{c\over r^2}\right)^2\,
\left({dr\over du}\right)^2=
2\kappa^2\,{1\over r}+h-{c^2\over r^2}.
\end{displaymath}

Now let us consider first the case $c\neq 0$. The previous equations can be
rewritten in the form
\begin{displaymath}
\left(\,{d\sigma\over du}\,\right)^2=h+{\kappa^4\over c^2}-\sigma^2,
\end{displaymath}
\noindent
where
\begin{displaymath}
\sigma={c\over r}-{\kappa^2\over c}.
\end{displaymath}
\noindent
Since the right-hand side of this equation is non-negative (as a square of a real
number $d\sigma/du$) one has also $h+{\kappa^4\over c^2}-\sigma^2\ge0$ or
$h+{\kappa^4\over c^2}\ge\sigma^2$. Considering that $\sigma^2\ge0$ one gets
\begin{displaymath}
h+{\kappa^4\over c^2}\ge0.
\end{displaymath}
\noindent
Therefore, one can designate $\left(h+{\kappa^4\over c^2}\right)^{1/2}=A$, $A\ge0$.

\bigskip
{\small
{\bf Exercise.}
Prove that for any position $\ve{r}$ and velocity ${\dot \ve{r}}$ the
integrals $h$ and $c$ take such values that $h+\kappa^4\,c^{-2}\ge0$.
{\it Hint:} use the definitions of $h$ and $c$ as functions of $\ve{r}$
and ${\dot \ve{r}}$.
}
\bigskip

\noindent
Therefore, one get the differential equation for the orbit
\begin{equation}
\label{dsigma/du-A}
\left(\,{d\sigma\over du}\,\right)^2=A^2-\sigma^2.
\end{equation}
\noindent
Let us first consider the case $A\neq0$. One can consider that $d\sigma/du>0$.
The second case of $d\sigma/du<0$ can be derived from the first one
by setting $u=-u$, that is by mirroring the first case. It is clear, however,
that the orbit in both cases remains the same and it is only the direction in
which the body moves along the orbit which changes. The direction of motion is
not interesting for us for the moment. Therefore, the solution can be written as
\begin{displaymath}
u=\arccos{\sigma\over A}\,+\omega,
\end{displaymath}
\noindent
or
\begin{displaymath}
\sigma=A\cos(u-\omega),
\end{displaymath}
\noindent
$\omega={\rm const}$ being an arbitrary constant.
Taking into account the definitions of $\sigma$ and $A$ one has
\begin{equation}\label{conic-section}
r={p\over 1+e\,\cos(u-\omega)},
\end{equation}
\noindent
where $p>0$ and $e\ge0$ represent two parameters of the orbit defined
through the integrals of motion $h$ and $c$ and parameter $\kappa$:
\begin{equation}\label{p-constants-definition}
p={c^2\over \kappa^2},
\end{equation}
\begin{equation}\label{e-constants-definition}
e=\sqrt{1+h\,{c^2\over\kappa^4}}.
\end{equation}

\bigskip
{\small
{\bf Remark.}
One can see that formally Eq. (\ref{dsigma/du-A}) has one more
solution: $\sigma=\pm A$ which means that $r=p/(1\pm e)$ is constant.
Using the equations of motion in polar coordinates
(\ref{2body-equations-of-motion-polar}) one can see that this is valid
only when $A=0$ (this case is treated below separately). Indeed, a
solution of equations of motion (e.g. a solution of
(\ref{2body-equations-of-motion-xi-eta})) must satisfy also the
corresponding integrals of motion (e.g.
(\ref{2body-area-polar})--(\ref{2body-energy-polar})), but not any
solution satisfying the integrals of motion also satisfy the equations
of motion. That is, the integrals of motion are necessary, but not
sufficient conditions for a function to be a solution of the equations of
motion. Whether a function satisfying of the integrals of motion also
satisfies the equations of motion should be checked explicitly. One can
easily see that (\ref{conic-section}) is really a solution of
(\ref{2body-equations-of-motion-polar}), but $\sigma=\pm A\neq 0$ is
not.
}
\bigskip

\noindent 
Eq. (\ref{conic-section}) shows that the orbit in this case
(we assumed $c\neq0$ and $A\neq0$) is a conic section. The parameter
$p>0$ is called \concept{semi-latus rectum} and $e$ represents the
\concept{eccentricity} of the conic section. From
(\ref{conic-section}) one sees that for $e<1$ (this corresponds to
$h<0$) the orbit is an ellipse, for $e=1$ ($h=0$) a parabola and for
$e>1$ ($h>0$) a hyperbola. This proves the first Kepler's law: the
orbit of every planet is an ellipse with the Sun at one of the two
foci.

For any $e$ the polar angle $u$ can take the value $u=\omega$. In this
case the denominator of (\ref{conic-section}) takes its maximal value
$1+e$. Therefore, the radial distance $r$ is minimal at this point
\begin{displaymath}
r={p\over 1+e}=r_{\rm min}.
\end{displaymath}
\noindent 
The point of the orbit where the distance $r$ takes its
minimal value is called \concept{pericenter} or \concept{periapsis}
(or \concept{perihelion} when motion relative to the Sun is
considered, or \concept{perigee} when motion relative to the Earth is
considered, or \concept{periastron} when motion of a binary star is
considered).  The constant $\omega$ is called \concept{argument of
pericenter}.

For $e<1$, polar angle $u$ can also take the value $u=\omega+\pi$
($\pi=3.14\dots$). Here the distance $r$ takes its maximal value %
\begin{displaymath} r={p\over 1-e}=r_{\rm max}.  \end{displaymath} %
\noindent The point of the elliptic orbit where the distance $r$ takes
its maximal value is called \concept{apocenter} or \concept{apoapsis}
(or \concept{aphel} when motion relative to the Sun is considered, or
\concept{apogee} when motion relative to the Earth is considered, or
\concept{apoastron} when motion of a binary star is considered).
Pericenter and apocenter are called \concept{apsides}. A line
connecting pericenter and apocenter is called \concept{line of
apsides} or \concept{apse line}.

The mean distance of the body calculated as arithmetic mean of the
maximal and minimal values of $r$ is called \concept{semi-major axis}
of the orbit:
\begin{displaymath}
a={1\over 2}\left(r_{\rm min}+r_{\rm max}\right)={p\over 1-e^2}
\end{displaymath}
\noindent
or
\begin{equation}\label{p-definition-a-e}
p=a\,(1-e^2).
\end{equation}
\noindent
Substituting (\ref{p-constants-definition}) and
(\ref{e-constants-definition}) into (\ref{p-definition-a-e}) one gets
the relation between $a$ and the integrals of motion:
\begin{equation}\label{a-constants-definition}
a=-{\kappa^2\over h}.
\end{equation}
\noindent
One can see that $a$ depends only on $\kappa$ and the energy constant $h$,
and not on $c$. Eqs. (\ref{p-definition-a-e}) and
(\ref{a-constants-definition}) represent definition of $a$ for any
non-negative value of $e$ (or for any $h$ and $c$). From
(\ref{a-constants-definition}) it follows that $a$ is infinite for
parabolic motion ($e=1$, $h=0$) and negative for hyperbolic one ($e>1$,
$h>0$).

Let us consider now the two remained cases. First, for $c\neq0$ and $A=0$
the differential equation for the orbit reads
$\left(\,{d\sigma\over du}\,\right)^2=-\sigma^2,$
\noindent
which means that both $\sigma$ and ${d\sigma\over du}$ should be
zero and therefore $\sigma\equiv0$. This means that
\begin{displaymath}
{c\over r}-{\kappa^2\over c}=0
\end{displaymath}
\noindent
and
\begin{equation}\label{r-A=0}
r=p.
\end{equation}
\noindent
This solution coincides with (\ref{conic-section}) for $e=0$ (this agrees also with the
definition of $e$: if $A=0$, one has $h=-\kappa^4/c^2$ and from
(\ref{e-constants-definition}) it follows that $e=0$).

Finally, if $c=0$ from (\ref{2body-area-polar}) one gets
$\dot u=0$ and therefore $u={\rm const}$, which means that the motion is
rectilinear. Substituting this into the energy integral
(\ref{2body-energy-polar}) one gets
\begin{displaymath}
{\dot r}^2=2\kappa^2\,{1\over r}+h.
\end{displaymath}
\noindent
For $h<0$ the motion is bounded since both sides of the latter
equations must be non-negative. For negative $h$ this means that
$r\le-{2\kappa^2\over h}>0$. This is rectilinear motion of elliptical
kind (the elliptical motion with (\ref{conic-section}) with $h<0$ and
$e<1$ is also bounded in space). For non-negative $h$ one can
calculate the velocity of the body for infinite distance $r\to\infty$:
$v_\infty=\lim_{r\to\infty} {\dot r}=\sqrt{h}$. For $h=0$ velocity
goes to zero: $v_\infty=0$.  For $h>0$ the velocity always remains
positive: $v_\infty>0$. These are rectilinear motions of parabolic and
hyperbolic kinds, respectively.

\bigskip
{\small
{\bf Exercise.}
For parabolic case $h=0$, one has ${\dot r}^2=2\kappa^2\,{1\over r}$.
This equation has a simple analytical solution. Find this solution in
its most general form.
}
\bigskip

\section{Orbit in Space}

{\small {\bf Summary:}\ Three Euler angles defining the
orientation of the orbit in space: longitude of the ascending node,
inclination and the argument of pericenter. The rotational matrix
between inertial coordinates in space and the coordinates in the
orbital plane.}\bigskip

Let us now consider the orientation of the orbit in space.  Above we
have seen that the orbit lies in a plane perpendicular to vector
$\ve{c}$. Let us consider two orthogonal Cartesian coordinate systems:
(1) $(x, y, z)$ is some arbitrary inertial reference system where the
equations of motion (\ref{2body-equations-of-motion}) are initially
formulated and (2) $(X, Y, Z)$ is oriented in such a way that
$XY$-plane contains the orbit (that is, axis $Z$ is parallel to vector
$\ve{c}$), and the pericenter of the orbit lies on axis $X$. The
origins of both reference systems coincide and the transformation
between these coordinates is a pure three-dimensional
(time-independent) rotation. The rotation can be parametrized in a
multitude of ways. Historically, it was parametrized by three
Euler-type angles.  Let us first consider the points where the orbit
intersects the $xy$-plane. These points are called nodes. The node at
which the body, in course of its motion, proceeds from the area of
negative $z$ to that of positive $z$ is called \concept{ascending
node}. Let us introduce an intermediate coordinate system $(x_1, y_1,
z_1)$ that is obtained from $(x, y, z)$ by a rotation around axis
$z=z_1$ so that axis $x_1$ contains the ascending node of the orbit:
\begin{equation}
\pmatrix{x_1\cr y_1\cr z_1}
=
\mat{A}_z(\Omega)\,
\pmatrix{x\cr y\cr z},
\end{equation}
\noindent
where $\mat{A}_z$ is the rotational matrix around $z$-axis
\begin{equation}
\mat{A}_z(\alpha)
=
\pmatrix{\cos\alpha&\sin\alpha&0\cr
-\sin\alpha&\cos\alpha&0\cr
0&0&1}.
\end{equation}

\noindent 
Angle $\Omega$ is the called the \concept{longitude of the
ascending node} or simply \concept{longitude of the node}.  One more
intermediate system $(x_2, y_2, z_2)$ is obtained from $(x_1, y_1,
z_1)$ by a rotation around axis $x_1=x_2$ so that the direction of
axis $z_2$ coincides with vector $\ve{c}$:
\begin{equation}
\pmatrix{x_2\cr y_2\cr z_2}
= \mat{A}_x(i)\,\pmatrix{x_1\cr y_1\cr z_1}
= \mat{A}_x(i)\,\mat{A}_z(\Omega)\,\pmatrix{x\cr y\cr z},
\end{equation}
\noindent
where $\mat{A}_x$ is the rotational matrix around $x$-axis
\begin{equation}
\mat{A}_x(\alpha)=\pmatrix{1&0&0\cr
0&\cos\alpha&\sin\alpha\cr
0&-\sin\alpha&\cos\alpha\cr}.
\end{equation}
Angle $i$ is called \concept{inclination}. Two angles -- longitude of
the node $\Omega$ and inclination $i$ -- fully define the orientation
of the orbital plane in space. The orbit lies in the $x_2y_2$-plane. 
Coordinates $(x_2, y_2)$ coincides with coordinates $(\xi,\eta)$ used above.
The last step is to define the orientation of the orbit
in the orbital plane. This is done by using argument of pericenter
$\omega$. The final coordinate system $(X, Y, Z)$ 
can be obtained from $(x_2, y_2, z_2)$ by a rotation around axis $z_2$:
\begin{equation}
\pmatrix{X\cr Y\cr Z}=
\mat{A}_z(\omega)\,\pmatrix{x_2\cr y_2\cr z_2}=
\mat{A}_z(\omega)\,\mat{A}_x(i)\,\mat{A}_z(\Omega)\,
\pmatrix{x\cr y\cr z}\,.
\end{equation}
In the following the inverse transformation plays an important role:
\begin{equation}
\label{XYZ2xyz}
\pmatrix{x\cr y\cr z}=\mat{P}\pmatrix{X\cr Y\cr Z}\,,
\end{equation}
\begin{equation}
\mat{P}=\mat{A}^T_z(\Omega)\,\mat{A}^T_x(i)\,\mat{A}^T_z(\omega)\,.
\end{equation}
\noindent
Here superscript $T$ denotes the transpose of the corresponding matrix.
Note that for any rotational matrix $\mat{R}^T=\mat{R}^{-1}$. 
Explicitly one has:
{
\renewcommand{\mathindent}{0cm}
\begin{eqnarray}\label{P-matrix}
\mat{P}&=&
\left(
\begin{array}{rrr}
\cos \Omega \,\cos \omega  - \sin \Omega \, \cos i\,\sin \omega &
-\cos \Omega \, \sin \omega  - \sin \Omega \,\cos i\,\cos\omega &
\sin \Omega \,\sin i
\\
\quad\sin \Omega \,\cos \omega  + \cos \Omega \,\cos i\, \sin \omega &
\quad- \sin \Omega \,\sin \omega  + \cos \Omega \,\cos i\,\cos\omega &
\quad- \cos \Omega \,\sin i
\\
\sin i\,\sin \omega &
\sin i\,\cos\omega &
\cos i
\end{array}
\ \right)\,.
\nonumber\\
\end{eqnarray}
}

\begin{figure}
\begin{center}
\resizebox{10.00cm}{!}{\includegraphics{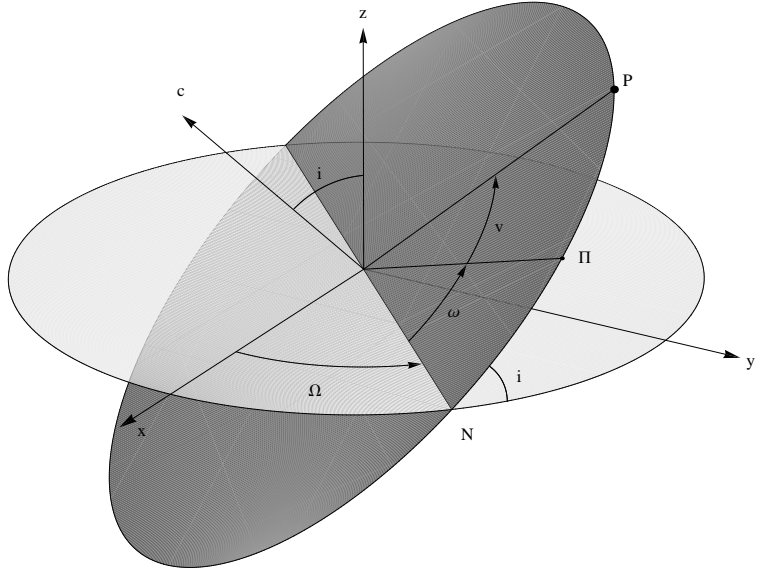}}
\end{center}
\caption[Orbit in space]{\small Orbit in space.
The picture shows the definition of three Euler-like angles 
$\Omega$, $i$ and $\omega$
that fully describe the orientation of the orbit of a two-body problem in space.
$N$ is the ascending node, $\Pi$ is the pericenter, $P$ is the position of the body.
One more angle (true anomaly $v$) defines the position of the body 
on the orbit.
\label{Fig-Orbit-in-space}}
\end{figure}

\section{Kepler Equation}

{\small {\bf Summary:}\
True Anomaly. Kepler equation in true anomaly. Eccentric anomaly.
Various relations between the true and eccentric anomaly.
Kepler equation in eccentric anomaly. Mean anomaly.
The period of motion and the third Kepler law.
}\bigskip

The \concept{true anomaly} $v$ is defined as $v=u-\omega$. Since $\omega={\rm const}$
one has $\dot v=\dot u$. Therefore, integral of areas
(\ref{2body-area-polar}) can be written as
\begin{displaymath}
\dot v=c=\kappa\,\sqrt{p}\,{1\over r^2}.
\end{displaymath}
\noindent
Using (\ref{conic-section}) one gets
\begin{equation}\label{Kepler-in-true-anomaly}
{dv\over (1+e\,\cos v)^2}=\kappa\,p^{-3/2}\,dt\,,
\end{equation}
\noindent
which after integration gives the Kepler equation in true anomaly
\begin{displaymath}
\kappa\,p^{-3/2}\,t=\int {dv\over (1+e\,\cos v)^2}+{\rm const}\,.
\end{displaymath}

In this Section from now on we consider only the case of elliptical
motion with $e<1$. The integral above cannot be computed
analytically. In order to simplify the computations one introduces the
so-called {\it eccentric anomaly}. The eccentric anomaly $E$ is
defined by its relation to the true anomaly $v$:
\begin{equation}\label{cos-v-cos-E}
\cos v={\cos E-e\over1-e\,\cos E},
\end{equation}
\noindent
It is easy to calculate $\sin^2 v$ from this equation. Choosing the
sign so that $E=v$ for $e=0$ one gets
\begin{equation}\label{sin-v-E}
\sin v={\sqrt{1-e^2}\,\sin E\over1-e\,\cos E}.
\end{equation}
\noindent
Solving (\ref{cos-v-cos-E}) for $\cos E$ one gets
\begin{equation}\label{cos-E-cos-v}
\cos E={\cos v+e\over1+e\,\cos v},
\end{equation}
\noindent
so that
\begin{equation}\label{1-e*cosE}
1-e\,\cos E={1-e^2\over1+e\,\cos v},
\end{equation}
\noindent
which together with (\ref{sin-v-E}) gives
\begin{equation}\label{sin-E-v}
\sin E={\sqrt{1-e^2}\,\sin v\over1+e\,\cos v}.
\end{equation}
\noindent
It is easy to see that (\ref{cos-E-cos-v}) and (\ref{sin-E-v})
can be obtained from (\ref{cos-v-cos-E}) and (\ref{sin-v-E}) by
making the substitution $(E,v,e)\,\longrightarrow\,(v,E,-e)$.
From (\ref{cos-v-cos-E}) and (\ref{sin-v-E}) one can derive a relation
between $v$ and $E$ which is convenient for numerical calculations
\begin{equation}\label{tan-v-E}
\tan {v\over 2}=\sqrt{1+e\over 1-e}\,\tan {E\over 2}.
\end{equation}
\noindent
From this equation it is easy to see that $E$ and $v$ are equal at apsides.

Now let us express the Kepler equation (\ref{Kepler-in-true-anomaly})
in terms of the eccentric anomaly. Taking a derivative of (\ref{1-e*cosE})
one gets
\begin{displaymath}
\sin E\ dE={(1-e^2)\,\sin v\over(1+e\,\cos v)^2}\,dv,
\end{displaymath}
\noindent
which together with (\ref{sin-v-E}) gives
\begin{displaymath}
{dv\over(1+e\,\cos v)^2}=(1-e^2)^{-3/2}\,(1-e\,\cos E)\,dE.
\end{displaymath}
\noindent
Substituting this into (\ref{Kepler-in-true-anomaly}) one gets

\begin{equation}\label{Kepler-in-E-diff}
(1-e\,\cos E)\,dE=\kappa\,a^{-3/2}\,dt
\end{equation}

\noindent
or after integrating

\begin{equation}\label{Kepler-in-E-int}
E-e\,\sin E=M,
\end{equation}

\noindent
where $M=n(t-T)$ is the mean anomaly, $n=\kappa\,a^{-3/2}$ is the mean
motion, and $T$ is an integration constant representing the moment time
at which $M=0$. It is easy to see from (\ref{Kepler-in-E-int}) and
(\ref{tan-v-E}) that for $t=T$ one has $M=E=v=0$ and the body is
situated in its pericenter. However, for $e=0$ the orbit is circular so
that any point can be declared to be pericenter. That is why, the
definition of the mean anomaly is often taken to be $M=M_0+n(t-t_0)$,
where $M_0$ is the value of the mean anomaly at some moment $t=t_0$.

The period of motion $P$ an be defined as a time interval between two
successive pericenters ($M=0$ and $M=2\pi$). Then it is clear that
$P={2\pi\over n}$, which can be re-written as
\begin{displaymath}
{P^2\over a^3}={4\pi\over\kappa^2}={\rm const}.
\end{displaymath}
\noindent
This is the third Kepler's law: the square of the
orbital period of a planet is directly proportional to the cube of the
semi-major axis of its orbit. However, the constant $\kappa$ depends
on the masses of both bodies of the two-body problem. Considering the
motion of two planets relative to the Sun in the framework of the 
two-body problem we have two different constants $\kappa_i$ entering the
corresponding equations of motion of each of the two planets
$\kappa^2_i=G\,(m_{\rm Sun}+m_i)$. Hence, one gets the third Kepler's
law in its correct form
\begin{displaymath}
\left({P_1\over P_2}\right)^2=\left({a_1\over a_2}\right)^3\,
{m_{\rm Sun}+m_1\over m_{\rm Sun}+m_2}
\end{displaymath}
\noindent
In the Solar system $m_i/m_{\rm Sun}<10^{-3}$ and the
last factor is almost unity.

\bigskip

Using the eccentric anomaly $E$ it is easy to calculate the position
and velocity of the object in two-body motion. For the position one
has (here we use the known expressions of the true anomaly $v$ and
relation (\ref{cos-v-cos-E})--(\ref{sin-v-E})):
\begin{eqnarray}
\label{r-E}
r&=&{p\over 1+e\,\cos v}=a\,(1-e\,\cos E),\\
\label{X-E}
X&=&r\,\cos v=a\,(\cos E-e),\\
\label{Y-E}
Y&=&r\,\sin v=a\,\sqrt{1-e^2}\sin E.
\end{eqnarray}
\noindent
Differentiating (\ref{X-E})--(\ref{Y-E}) with respect to time $t$
and using that
\begin{equation}\label{E-dot}
\dot E={n\over 1-e\,\cos E}
\end{equation}
\noindent
(the latter equation can be derived by differentiating the Kepler
equation (\ref{Kepler-in-E-int})) one gets 
\begin{eqnarray}
\label{dotX-E}
\dot X&=&-{a\,n\,\sin E\over 1-e\,\cos E},\\
\label{dotY-E}
\dot Y&=&{a\,\sqrt{1-e^2}\,n\,\cos E\over 1-e\,\cos E}.
\end{eqnarray}
\noindent
We have now all formulas that are necessary to compute the position
and velocity of a body in two-body motion. These calculations can be
performed if eccentric anomaly $E$ is known as a function of time $t$.
The relation between $E$ and $t$ is given by the Kepler equation
(\ref{Kepler-in-E-int}).  However, the latter equation is transcendent
and cannot be solved algebraically. Let us turn to the analysis of the
Kepler equation.

\section{Solving the Kepler Equation}

{\small {\bf Summary:}\
Existence and uniqueness of the solution. Iterative solution.
Newtonian solution.
}\bigskip

Let us confine ourselves by the case of elliptic motion with $0\le e<1$.
The Kepler equation $E-e\sin E=M$ can be considered as an implicit
function $E(M)$ or as explicit function $M(E)$
($e$ being a parameter in both cases).
\begin{enumerate}
\item[1)] $E(M)$ is a continuous function as an inverse function to a continuous function
$M(E)$.
\item[2)] Since $|E-M|=e\,|\sin E|<1$ one has
\begin{eqnarray}
\lim_{E\to+\infty}M(E)&=&+\infty,
\nonumber\\
\lim_{E\to-\infty}M(E)&=&-\infty.
\nonumber
\end{eqnarray}
\end{enumerate}
\noindent
From properties 1 and 2 follows that for any $M$ there exists at least
one $E$ such that the Kepler equation is satisfied.
\begin{enumerate}
\item[3)]  Since $\displaystyle{dM\over dE}=1-e\,\cos E>0$, $M(E)$ is monotone, and
therefore, $E(M)$ is also monotone (as an inverse function of $M(E)$).
\end{enumerate}
\noindent
From these three properties it follows that for any $M$ there exists
only one $E$ such that $E-e\,\sin E=M$.

Let us now consider how one can solve the Kepler equation.
Generally we have a transcendent equation
\begin{equation}
f(x)=0,
\nonumber
\end{equation}
\noindent
that should be solved numerically. Moreover, we just have proved that
in case of the Kepler equation one has only one solution for any $0\le
e<1$ and any $M$. Many numerical methods to find the solution are
known. Let us consider two simplest methods which, however, are
sufficient in many cases.
\begin{enumerate}
\item[I.] Iterations

The method consists in starting with some initial value for
$x$ (say $x_0$) and iterating the formula $x_{i+1}=x_i-f(x_i)$ until
the subsequent values of $x$ ($x_i$ and $x_{i+1}$) are close enough to
each other: if $|x_{i+1}-x_i|<\varepsilon$, then $x_{i+1}$ is a solution
of $f(x)=0$ such that $|f(x_i)|<\varepsilon$. Convergence of the
iterative sequence is guaranteed if the derivative $f^\prime(x)$ of
$f(x)$ satisfies the inequality $|f^\prime(x)-1|<1$ (Schwarz, 1993).
One can
easily show that this inequality holds in case of the Kepler equation with $e<1$.
In general, the convergence also depends on the starting point $x_0$.

For the Kepler equation one gets $f(E)=E-e\,\sin E-M$ and the algorithm
can be written as
\begin{eqnarray}
E_0&=&M,
\nonumber\\
E_{i+1}&=&M+e\,\sin E_i, \quad i=0,1,\dots
\nonumber
\end{eqnarray}
\noindent
One can prove that the initial condition $E_0=M$ guarantees that the
iterations converge for any $e$ and $M$.

\bigskip

\item[II.] Newton's method

Another well-known method is the Newton's (or Newton-Raphson) one.
Again starting from some initial value for the root's estimate $x_0$
one iterates $x_{i+1}=x_i-f(x_i)/f^\prime(x_i)$, where
$f^\prime(x)$ is the derivative of $f(x)$.  Again if
$|x_{i+1}-x_i|<\varepsilon$, then $x_i$ is a solution of $f(x)=0$ with
the corresponding accuracy. Convergence of the iterative sequence is
guaranteed if the initial guess $x_0$ lies close enough to the root
(Section 9.4 of Press {\it et al.} 1992).
%\citep[Section 9.4]{Press:Teukolsky:Vetterling:Flannery:1992}

For the Kepler equation this implies the scheme
\begin{eqnarray}
E_0&=&M,
\nonumber\\
E_{i+1}&=&\displaystyle{M+e\,\left(\sin E_i-E_i\,\cos E_i\right)\over 1-e\,\cos E_i},
\quad i=0,1,\dots
\nonumber
\end{eqnarray}
\noindent
Again one can prove that the initial condition $E_0=M$ guarantees that
the iterations converge for any $e$ and $M$. It is well known that if
the initial guess of the rule is good enough, the Newton's method
converges much faster than the iteration method.

\end{enumerate}

\section{Hyperbolic and Parabolic Motion}

{\small {\bf Summary:}\
The eccentric anomaly and the Kepler equation for the hyperbolic
motion. Explicit solution for the parabolic motion.
}\bigskip

When we introduced the eccentric anomaly $E$ and discussed the Kepler
equation above we have concentrated on elliptic motion with eccentricity
$0\le e<1$. Let us now consider two other cases: hyperbolic motion with
$e>1$ and parabolic one with $e=1$. The formula for the form of the
orbit
\begin{displaymath}
r={p\over 1+e\,\cos v}
\end{displaymath}
\noindent
remains valid for any $e$. Let us first consider the case of {\it
hyperbolic motion} with $e>1$. The semi-latus rectum is non-negative
$p=a(1-e^2)={c^2\over \kappa^2}>0$ (the case $c=0$ leads to the
rectilinear motion, has been considered above and will not be
considered here). But $e=\sqrt{1+h\,{c^2\over \kappa^4}}$ and,
therefore, $e>1$ implies $h>0$, i.e. the total energy of the two-body
system is positive. Since $a=-{\kappa^2\over h}$, $h>0$ implies in turn
that the semi-major axis is negative $a<0$.

If one takes the transformations (\ref{cos-E-cos-v}) and
(\ref{sin-E-v}) relating sine and cosine of the eccentric one $E$ to
the sine and cosine of the true anomaly $v$, it is easy to see that
$\sin E$ is imaginary and $\cos E$ is real, but can exceed unity.
Therefore, for hyperbolic motion $E$ is imaginary. One can continue in
this way and work directly with complex numbers, but it is not
convenient. We, therefore, try to make all our equations real again (as
it is in the case of elliptical motion). To this end we define
a new anomaly instead of $E$. Since
\begin{displaymath}
e^{\mi x}=\cos x+\mi\sin x,
\end{displaymath}
\noindent
$\mi$ being imaginary unit, one has
\begin{displaymath}
\cos x={1\over 2}\,\left(e^{\mi x}+e^{-\mi x}\right),
\quad
\sin x={1\over 2\mi}\,\left(e^{\mi x}-e^{-\mi x}\right),
\end{displaymath}
\noindent
and for $x=\mi\,y$ one gets
\begin{eqnarray}
\cos \mi\,y&=&\cosh y,
\nonumber\\
\sin \mi\,y&=&\mi\,\sinh y.
\nonumber
\end{eqnarray}
\noindent
Using these formulas one can introduce a new anomaly $H$ defined as
$H=\mi E$, so that
\begin{eqnarray}
\cos E&=&\cosh H,
\nonumber\\
\sin E&=&-\mi\sinh H.
\nonumber
\end{eqnarray}

\noindent
This allows one to re-write Eq. (\ref{r-E})--(\ref{Y-E}) as
\begin{eqnarray}
\label{r-H}
r&=&{p\over 1+e\,\cos v}=a\,(1-e\,\cos E)=|a|\,(e\,\cosh H-1),\\
\label{X-H}
X&=&r\,\cos v=a\,(\cos E-e)=|a|\,(e-\cosh H),\\
\label{Y-H}
Y&=&r\,\sin v=a\,\sqrt{1-e^2}\sin E=|a|\,\sqrt{e^2-1}\,\sinh H,
\end{eqnarray}

\noindent
and the Kepler equation (\ref{Kepler-in-E-int}) as

\begin{equation}\label{Kepler-hyperbolic}
e\,\sinh H-H=M_{\rm hyp},
\end{equation}

\begin{equation}\label{hyperbolic-M}
M_{\rm hyp}=\kappa\,|a|^{-3/2}\,(t-T).
\end{equation}

\noindent
The signs in (\ref{Y-H}) and (\ref{Kepler-hyperbolic}) are chosen so
that the body moves in the positive direction of the $Y$-axis for
$t=T$. Eq. (\ref{Kepler-hyperbolic}) is the Kepler equation for
hyperbolic motion. It has only one solution for any value of mean
anomaly $M_{\rm hyp}$.

Let us now consider the simple case of {\it parabolic motion}. Parabolic
motion corresponds to $e=1$ and this implies that the total energy of
the system is zero: $h=0$. The equation for the form of the orbit can be
simplified to
\begin{displaymath}
r={p\over 1+\cos v}=q\,(1+\sigma^2),
\end{displaymath}
\noindent
where $\sigma=\tan {v\over 2}$ and $q=p/2$ is the perihelion distance of the
parabolic orbit (since in the perihelion one has $v=0$ and $r=p/2$). From the
definition of $\sigma$ one has
\begin{displaymath}
  \sin v={2\sigma\over 1+\sigma^2},\quad
  \cos v={1-\sigma^2\over 1+\sigma^2}.
\end{displaymath}
\noindent
Therefore, the distance and coordinates of the body on a parabolic orbit
can be written as
\begin{eqnarray}
\label{r-sigma}
r&=&q\,(1+\sigma^2),\\
\label{X-sigma}
X&=&r\,\cos v=q\,(1-\sigma^2),\\
\label{Y-sigma}
Y&=&r\,\sin v=2\,q\,\sigma.
\end{eqnarray}
\noindent
Now, using the integral of angular momentum one gets
\begin{displaymath}
r^2\dot v=c=\kappa\,\sqrt{p}=\kappa\,\sqrt{2\,q}
\end{displaymath}
\noindent
and, therefore,
\begin{equation}\label{parabolic-dot-v}
\dot v=\kappa\,\sqrt{2}\,q^{-3/2}\,(1+\sigma^2)^{-2}.
\end{equation}
\noindent
Since $\sigma=\tan {v\over 2}$ we have
\begin{displaymath}
{d\sigma\over dv}={1\over 2}\,(1+\sigma^2).
\end{displaymath}
\noindent
Integrating this equation
and using (\ref{parabolic-dot-v}) one gets
\begin{equation}\label{Kepler-parabolic}
{1\over 3}\sigma^3+\sigma=M_{\rm par},
\end{equation}
\begin{equation}\label{parabolic-M}
M_{\rm par}={\kappa\over\sqrt{2}}\,q^{-3/2}\,(t-T).
\end{equation}
\noindent
Eq. (\ref{Kepler-parabolic}) is the Kepler equation for parabolic motion.
This equation can be solved analytically using, e.g., the well-known Kardan
formulas. Indeed, the equation has only one real solution for any value of $M_{\rm par}$
\begin{eqnarray}\label{Kepler-parabolic-solution}
\sigma&=&{1\over2}\,Q^{1/3}-2\,Q^{-1/3},
\nonumber\\
Q&=&12\,M_{\rm par}+4\,\sqrt{4+9\,M_{\rm par}^2}.
\end{eqnarray}
\noindent
Note that for any $M_{\rm par}$ value of $Q$ remains positive.
This means that the parabolic motion can be represented by an explicit
analytical formula (this can be also done for
circular motion with $e=0$ and for rectilinear motion of parabolic type
with $c=0$ and $h=0$).

\bigskip
{\small
{\bf Exercise.}
Write the explicit formulas for the coordinates for the case $e=0$.
}
\bigskip

\section{Relation between Position, Velocity and the Kepler Elements}
\label{Section-2-body-summary}

{\small {\bf Summary:}\
Calculation of position and velocity from the Kepler elements.
Calculation of the Kepler elements from the position and velocity.
Orbit determination (an overview).
}\bigskip

We have seen above that there are two equivalent ways to represent a
particular two body motion: (1) to specify the initial conditions for
the equation of motion, i.e. the position and velocity vectors
$\ve{r}=(x,y,z)$ and ${\dot \ve{r}}=(\dot x,\dot y,\dot z)$ together
with the corresponding moment of time $t_0$ and the parameter $\kappa$,
and (2) to fix the whole set of the six Kepler elements $a$, $e$, $i$,
$\omega$, $\Omega$, $M_0=M(t_0)$ again together with the moment of time
$t_0$ for which the mean anomaly $M_0$ is supposed to be known and the
parameter $\kappa$. Very often in the practical calculation one wants
to switch between these two representations, that is to transform the
position and velocity into the corresponding Kepler elements or vice
verse. Here we give the set of formulas enabling one to perform these
transformations for the case of elliptic motion.

The transformation from the Kepler elements to the position and
velocity vectors can be done in the following way:

\begin{enumerate}
\item calculate mean motion as $n=\kappa\,a^{-3/2}$ and mean anomaly as
$M=n\,(t-t_0)+M_0$ (here the position and velocity vectors can be calculated
for any arbitrary moments of time $t$, not necessarily for the moment $t_0$ for
which the mean anomaly $M_0$ is specified),

\item calculate eccentric anomaly $E$ from $E-e\sin E=M$,
\item calculate position and velocity in the orbital plane:
\begin{eqnarray}
X&=&a\,(\cos E-e),
\nonumber\\
Y&=&a\,\sqrt{1-e^2}\,\sin E,
\nonumber\\
\dot X&=&-{a\,n\,\sin E\over 1-e\,\cos E},
\nonumber\\
\dot Y&=&{a\,n\,\sqrt{1-e^2}\,\cos E\over 1-e\,\cos E},
\nonumber
\end{eqnarray}
\item calculate the position and velocity vectors in space as
\begin{eqnarray}
\pmatrix{x\cr y\cr z}&=&P\,\pmatrix{X\cr Y\cr 0},
\nonumber\\
\nonumber\\
\nonumber\\
\pmatrix{\dot x\cr \dot y\cr \dot z}&=&P\,\pmatrix{\dot X\cr \dot Y\cr 0},
\nonumber
\end{eqnarray}
\noindent 
where $P=\mat{A}^T_z(\Omega)\,\mat{A}^T_x(i)\,\mat{A}^T_z(\omega)$ defined by (\ref{P-matrix}).
\end{enumerate}

The transformation from the position and velocity vectors to the Kepler
elements is a bit more complicated and can be done as follows:

\begin{enumerate}
\item from the integrals of the areas $\ve{r}\times{\dot \ve{r}}=\ve{c}$
one gets $\ve{c}=(c_x,c_y,c_z)$
\begin{eqnarray}
c_x&=&y\,\dot z-\dot y\, z,
\nonumber\\
c_y&=&z\,\dot x-\dot z\, x,
\nonumber\\
c_z&=&x\,\dot y-\dot x\, y,
\nonumber\\
c&=&|\ve{c}|=\sqrt{c_x^2+c_y^2+c_z^2}.
\nonumber
\end{eqnarray}
\noindent
Then the semi-latus rectum can be calculated as
\begin{equation}
p={c^2\over \kappa^2},
\nonumber
\end{equation}
\noindent
and from
\begin{eqnarray}
\pmatrix{c_x\cr c_y\cr c_z}=\mat{A}^T_z(\Omega)\,\mat{A}^T_x(i)\,\pmatrix{0\cr 0\cr c}
=
\left(\begin{array}{rrr}
c\sin i\sin\Omega\\ -c\sin i\cos\Omega\\ c\,\cos i
\end{array}
\right)
\nonumber
\end{eqnarray}
\noindent
which gives us three equations. The third equation
can be written as
\begin{equation}
\cos i = {c_z\over c}
\end{equation}
\noindent
and since $0\le i\le\pi$ this one equation is sufficient to calculate
the inclination $i$. The two other equations read
\begin{eqnarray}
\sin\Omega&=&{c_x\over\sqrt{c_x^2+c_y^2}},
\nonumber\\
\cos\Omega&=&-{c_y\over\sqrt{c_x^2+c_y^2}}
\end{eqnarray}
\noindent
and allow one to calculate $\Omega$. Note that if $c_x^2+c_y^2=0$, the
inclination $i=0$ and $\Omega$ is not defined.
\item From Eq. (\ref{conic-section}) and $v=u-\omega$ one gets two equations
\begin{eqnarray}
e\,\cos v&=&{p\over r}-1,
\nonumber\\
e\,\sin v&=&{\sqrt{p}\over \kappa}\,{\ve{r}\cdot{\dot \ve{r}}\over r}
\end{eqnarray}
\noindent
which can be used to calculate both the eccentricity $e$ and the true
anomaly $v$. Then using (\ref{tan-v-E}) one can calculate the eccentric
anomaly $E$, and from (\ref{Kepler-in-E-int}) the mean anomaly $M$. All
these values of anomalies $v$, $E$ and $M$ correspond to time $t_0$ for
which the position and velocity of the body is specified. Finally, from
$p$ and $e$ it is easy to calculate the semi-major axis as
$a=p\,{(1-e^2)}^{-1}$;
\item From
\begin{eqnarray}
\pmatrix{x\cr y\cr z}=\mat{A}^T_z(\Omega)\,\mat{A}^T_x(i)\,
\pmatrix{r\cos (v+\omega)\cr r\sin (v+\omega)\cr 0}
\nonumber
\end{eqnarray}
\noindent
one gets
\begin{eqnarray}
\label{cos-sin-v+omega}
\cos (v+\omega)&=&{x\over r}\,\cos\Omega+{y\over r}\,\sin\Omega,
\nonumber\\
\sin (v+\omega)&=&\left(-{x\over r}\,\sin\Omega+{y\over r}\,\cos\Omega\right)\,
\cos i+{z\over r}\,\sin i.
\end{eqnarray}
\noindent
From these two equations one calculates the angle $v+\omega$ and since
$v$ is known, the argument of perihelion $\omega$.
\end{enumerate}

\section{Series Expansions in Two-Body Problem}

{\small {\bf Summary:}\
Series in powers of time. Fourier series in multiples of the mean anomaly.
Series in powers of the eccentricity.
}\bigskip

As we have seen above a fully analytical solution of the two-body
problem is impossible: one has a transcendent Kepler equation cannot be
solved analytically. The only possibility to get an analytical solution
is to use some kind of expansions. Below we consider three types of
expansions which are widely used in celestial mechanics.

\subsection{Taylor expansions in powers of time}
% powers of tau

The first kind of expansion is the Taylor expansion in powers of time.
Let us consider the positional vector $\ve{r}$ and expand it into
Taylor series
\begin{equation}
\label{ve-r-Taylor}
\ve{r}(t+\tau)=\sum_{k=0}^\infty {1\over k!}\,\ve{r}^{(k)}(t)\,\tau^k,
\end{equation}
\noindent
where $\ve{r}^{(k)}$ are the derivatives of $\ve{r}$ of order $k$.
For $k=0$ and $k=1$ they represent the initial conditions for the motion
$\ve{r}^{(0)}(t)=\ve{r}(t)$ and $\ve{r}^{(1)}(t)={\dot \ve{r}}(t)$.
Using the equation of motion
\begin{displaymath}
{\ddot \ve{r}}=-\kappa^2\,{\ve{r}\over r^3}
\end{displaymath}
\noindent
it is clear that the higher derivatives for $k\ge2$ can be calculated
in terms of $\ve{r}$ and ${\dot \ve{r}}$. E.g.,
\begin{eqnarray}
\ve{r}^{(2)}&=&{\ddot \ve{r}}=-\kappa^2\,S^{-3/2}\,\ve{r},
\nonumber\\[3mm]
\ve{r}^{(3)}&=&{3\over 2}\,\kappa^2\,S^{-5/2}\,{\dot S}\,\ve{r}
-\kappa^2\,S^{-3/2}\,{\dot \ve{r}},
\end{eqnarray}
\noindent
and so on. Here $S=r^2$, ${\dot S}=2\,\ve{r}\cdot\dot{\ve{r}}$, and
the second- and higher-order derivatives of $S$
appearing in $\ve{r}^{(k)}$ for $k\ge4$ can be calculated using
\begin{displaymath}
\ddot S=2\,h+2\kappa^2\,S^{-1/2},
\end{displaymath}
\noindent
where $h={\dot \ve{r}}\cdot{\dot \ve{r}}-{2\kappa^2\over r}$ is the energy
integral. Therefore, it is clear that $\ve{r}(t+\tau)$ can be
represented as a linear combination of $\ve{r}(t)$ and ${\dot
\ve{r}}(t)$
\begin{eqnarray}
\label{F-G}
\ve{r}(t+\tau)=F\,\ve{r}(t)+G\,{\dot \ve{r}}(t),
\end{eqnarray}
\noindent
while the functions $F$ and $G$ can be expanded in their Taylor series
in powers of $\tau$:
\begin{eqnarray}
\label{F-expand}
F=\sum_{k=0}^\infty {1\over k!}\,F_k\,\tau^k,
\nonumber\\[3mm]
\label{G-expand}
G=\sum_{k=0}^\infty {1\over k!}\,G_k\,\tau^k.
\end{eqnarray}
\noindent
The coefficients $F_k$ and $G_k$ are functions of $\kappa$, $h$, $S$
and $\dot S$ only, and, therefore, can be calculated from the initial
conditions $\ve{r}(t)$ and ${\dot \ve{r}}(t)$. Comparing
(\ref{F-G})--(\ref{G-expand}) to (\ref{ve-r-Taylor}) one gets for any
$k$
\begin{equation}
\label{ve-r-k-F-G}
\ve{r}^{(k)}= F_k\,\ve{r}+G_k\,{\dot \ve{r}}.
\end{equation}
\noindent
Taking the derivative of (\ref{ve-r-k-F-G}) and comparing
again with (\ref{ve-r-k-F-G}) written for $k+1$
\begin{eqnarray}
\label{ve-r-k-F-G-k+1}
\ve{r}^{(k+1)}= F_{k+1}\,\ve{r}+G_{k+1}\,{\dot \ve{r}}
\nonumber
\end{eqnarray}
\noindent
one gets the recursive formulas for
$F_k$ and $G_k$:
\begin{eqnarray}
\label{F-recursive}
F_{k+1}=\dot F_k-\kappa^2\,S^{-3/2}\,G_k,
\\[3mm]
\label{G-recursive}
G_{k+1}=F_k+\dot G_k.
\end{eqnarray}
\noindent
The initial conditions for (\ref{F-recursive})--(\ref{G-recursive})
can be derived by considering the zero-order expansion
$\ve{r}(t+\tau)=\ve{r}(t)+{\cal O}(\tau)$:
\begin{eqnarray}
\label{F0}
F_0=1,
\\
\label{G0}
G_0=0.
\end{eqnarray}
\noindent
Using (\ref{F-recursive})--(\ref{G-recursive}) with (\ref{F0})--(\ref{G0})
one gets, for example,
\begin{eqnarray}
F_1&=&0,
\nonumber\\
G_1&=&1,
\nonumber\\
F_2&=&-\kappa^2\,S^{-3/2},
\nonumber\\
G_2&=&0,
\nonumber\\
\dots
\end{eqnarray}
A detailed analysis of the two-body function by means of the complex
analysis shows that the convergence of the derived series is guaranteed
only for $|\tau|$ smaller than some limit depending on parameters of
motion:
\begin{equation}
\label{tau-limit}
|\tau|<{1\over \kappa}\,q^{3/2}\,\alpha(e),
\end{equation}
\noindent
where $q$ is the perihelion distance, $e$ is the eccentricity and
\begin{eqnarray}
\alpha(e)=\left[
\begin{array}{ll}
(1-e)^{-3/2}\,\left(\log\displaystyle{1+\sqrt{1-e^2}\over e}-\sqrt{1-e^2}\right),&e\le1
\\[5mm]
(e-1)^{-3/2}\,\left(\sqrt{e^2-1}-\arctan\sqrt{e^2-1}\right),&e>1
\end{array}
\right.
\end{eqnarray}
\noindent
Note that $\alpha(e)$ is a continuous monotone function for $e\ge0$
and
\begin{eqnarray}
\lim_{e\to 0}\alpha(e)&=&\infty,
\nonumber\\
\lim_{e\to 1}\alpha(e)&=&{2\sqrt2\over 3},
\nonumber\\
\lim_{e\to \infty}\alpha(e)&=&0.
\nonumber
\end{eqnarray}
\noindent
This means that the convergence is guaranteed for any $\tau$ if and
only if the eccentricity of the orbit is zero and that the higher the
eccentricity is the lower is the maximal $\tau$ for which the series in
powers of time converge.

\subsection{Fourier expansions in multiples of the mean anomaly}
% Fourier series

It is well known that any continuous complex function $f(x)$ of a real
argument $x$ with a period of $2\pi$ (i.e. $f(x+2\pi)=f(x)$ for any
$x$) can be expanded into Fourier series
\begin{eqnarray}
\label{f(x)-Fourier}
f(x)&=&\sum_{k=-\infty}^{+\infty}\,f_k\,e^{\mi\,k\,x},
\nonumber\\
f_k&=&{1\over 2\pi}\,\int_0^{2\pi}f(x)\,e^{-\mi\,k\,x}\,dx,
\end{eqnarray}
\noindent
which converges for any $x$. This kind of expansions can also be
applied to the two-body problem. If $f(x)$ has additional properties
(i.e., real or odd) formula (\ref{f(x)-Fourier}) can be simplified.
A nice overview of all special cases can be found in Chapter 12
of Press {\it et al.} (1992).
%\citet{Press:Teukolsky:Vetterling:Flannery:1992}

Let us consider function $f(E)=e\,\sin E$. This function is obviously
real and odd ($f(-E)=-f(E)$). Therefore, the Fourier expansion can be
simplified to be
\begin{eqnarray}
\label{E-Fourier-M}
E&=&M+e\,\sin E=M+\sum_{k=1}^\infty a_k \sin k M,
\\[5mm]
\label{ak-E-Fourier-M}
a_k&=&{2\over \pi}\,\int_0^\pi e\,\sin E\,\sin kM\,dM={2\over k}\,J_k(k\,e),
\end{eqnarray}
\noindent
where $J_n(x)$ are the \concept{Bessel functions of the first kind} defined as
\begin{eqnarray}
\label{Bessel-J-n}
J_n(x)={1\over \pi}\,\int_0^\pi \cos\left(n\,\theta-x\,\sin\theta\right)\,d\theta.
\end{eqnarray}
Many properties of $J_n(x)$ can be found e.g. in Chapter 9 of 
Abramowitz \& Stegun (1965).

\bigskip

{\small
{\bf Exercise.}
Prove the second equality in (\ref{ak-E-Fourier-M}) by taking the integral
by parts.
}
\bigskip

\noindent
Therefore,
\begin{eqnarray}
\label{E-Fourier-M-final}
E&=&M+\sum_{k=1}^\infty {2\over k}\,J_k(k\,e)\,\sin k M.
\end{eqnarray}
\noindent
To give one more example let us note that
\begin{displaymath}
{a\over r}={1\over n}\, \dot E\,.
\end{displaymath}
\noindent
Therefore, taking a derivative of (\ref{E-Fourier-M-final}) one gets
\begin{eqnarray}
\label{a-over r-Fourier-M}
{a\over r}&=&1+\sum_{k=1}^\infty 2\,J_k(k\,e)\,\cos k M.
\end{eqnarray}
\noindent
In general one has
\begin{eqnarray}
\label{Hansen}
\left({r\over a}\right)^n\,e^{\mi\,m\,v}&=&\sum_{k=-\infty}^\infty
X_k^{n,m}(e)\,e^{\mi\,k\,M},
\end{eqnarray}
\noindent
where $X_k^{n,m}(e)$ are a three-parametric family of functions called
Hansen coefficients.

\subsection{Taylor expansions in powers of the eccentricity}
% powers of e

The third kind of expansions are series in powers of eccentricity $e$.
Let us consider these series for the example of the eccentric anomaly.
Re-writing the Kepler equation in the form
\begin{equation}
\label{for-expansion-of-E}
E=M+e\sin E,
\end{equation}
\noindent
one has iteratively
\begin{eqnarray}
E&=&M+{\cal O}(e),
\nonumber\\
E&=&M+e\,\sin(M+{\cal O}(e))=M+e\,\sin M+{\cal O}(e^2),
\nonumber\\
E&=&M+e\,\sin(M+e\,\sin M+{\cal O}(e^2))=M+e\,\sin M+{1\over 2}e^2\,\sin 2M+{\cal O}(e^3),
\nonumber\\
&&\dots
\nonumber
\end{eqnarray}
\noindent
Here we used the expansion $\sin(M+e\,\sin M)= \sin M+{1\over 2}e\,\sin
2M+{\cal O}(e^2)$. Note that at each step of the iteration the
expansion for $E$ derived on the previous step in substituted under
sinus in the right-hand side of (\ref{for-expansion-of-E}) and the
sinus is expanded in powers of $e$ to the corresponding order. In
general one can write
\begin{equation}
E=M+\sum_{k=1}^{\infty}a_k(M)\,e^k,
\end{equation}
\noindent
and the coefficients $a_1$ and $a_2$ have been explicitly calculated
above. Further coefficients can be calculated by the same iterative
scheme. The series in powers of $e$ converge for all $e$ lower than the
so-called Laplace limit:
\begin{equation}
0\le e < e^*=0.6627434193492\dots
\end{equation}

%\section{Astronomical units of measurements}
%
%{\small {\bf Summary:}\
%The SI units for mass, distance and time. The astronomical
%units of measurements. The relations between the SI and astronomical units.
%}\bigskip

\chapter{The N-body problem}

\section{Equations of motion}

{\small {\bf Summary:}\
Equations of motion of the N-body problem. Gravitational potential.
}\bigskip

Let us consider N bodies having positions $\ve{\rho}_i$, $i=1, \dots, N$  in an inertial
reference system and characterized by their masses $m_i$. Here index
$i$ enumerates the bodies. Introducing the position of body $j$ with respect 
to body $i$ as $\ve{\rho}_{ij}=\ve{\rho}_j-\ve{\rho}_i$ one gets the equations
of motion of such a system
\begin{eqnarray}
\label{N-body-eqm-with-mass}
m_i\ddot{\ve{\rho}}_i=\sum_{j=1,\ j\neq i}^N\,G\,{m_i\,m_j\over\rho_{ij}^3}
\,\ve{\rho}_{ij}
\end{eqnarray}
\noindent
or 
\begin{equation}
\label{N-body-eqm}
\ddot{\ve{\rho}}_i=\sum_{j=1,\ j\neq i}^N\,G\,{\,m_j\over\rho_{ij}^3}
\,\ve{\rho}_{ij}\,.
\end{equation}
\noindent
These equations can also be be written in another form: 
\begin{equation}
\label{N-body-eqm-gradient}
m_i\ddot{\ve{\rho}}_i=\grad_i\,U\,
\end{equation}
\noindent
where $\grad_i\,U$ is the vector of partial derivatives of $U$ 
with respect to the components of $\ve{\rho}_i$. Denoting
$\ve{\rho}_i=(x_i,y_i,z_i)$ for any function $f$ one defines $\grad_i\,f$ as a vector
with the following components
\begin{equation}
\label{def-grad}
\grad_i\,f=\left(
{\partial \over \partial x_i}\,f\,,
{\partial \over \partial y_i}\,f\,,
{\partial \over \partial z_i}\,f
\right)\,.
\end{equation}
\noindent
The potential $U$ of $N$ gravitating bodies reads
\begin{equation}
U={1\over 2}\,\sum_{i=1}^N\sum_{j=1,\ j\neq i}^N {G\,m_i\,m_j\over\rho_{ij}}
=\sum_{i=1}^N\sum_{j=1}^{i-1} {G\,m_i\,m_j\over\rho_{ij}}\,.
\end{equation}
Since gradient of $U$ can be written as
$$
\grad_k\,U=\sum_{j=1,\ j\neq k}^N\,G\,m_k\,m_j\,\grad_k\,{1\over \rho_{kj}}=
\sum_{j=1,\ j\neq k}^N {G\,m_k\,m_j\over\rho_{kj}^3}\,\ve{\rho}_{kj}
$$
(the last equality uses that $\displaystyle{\grad_k{1\over
\rho_{kj}}={\ve{\rho}_{kj}\over \rho_{kj}^3}}$), 
it can be seen that (\ref{N-body-eqm-gradient}) really holds.

\section{Classical integrals of the $N$-body motion}

{\small {\bf Summary:}\
Center of mass integral in the $N$-body.
Integral of angular momentum in the $N$-body problem.
Integral of energy in the $N$-body problem.
}\bigskip

The equations of motion of the $N$-body problem possess similar 10 integrals
of motion that we already discussed for the two-body problem.
Summing up the equations (\ref{N-body-eqm-with-mass}) one sees that
$$
\sum_{i=1}^N m_i\,\ddot{\ve{\rho}}_i=0
$$
\noindent
Since the masses $m_i$ are constant this leads to
\begin{equation}
\label{N-body-CoM-A}
\sum_{i=1}^N m_i\,\dot{\ve{\rho}}_i=\ve{A}={\rm const}
\end{equation}
\noindent
and
\begin{equation}
\label{N-body-CoM-AB}
\sum_{i=1}^N m_i\,\ve{\rho}_i=\ve{A} t+ \ve{B},\qquad \ve{B}={\rm const}\,.
\end{equation}
\noindent
Components of $\ve{A}$ and $\ve{B}$ are six center of mass integrals in the $N$-body problem. 
These are fully analogous to (\ref{2body-integral1})--(\ref{2body-cm-position}).
The position of the center of mass of the $N$-body system obviously read
$$
\ve{R}={\sum_{i=1}^N m_i\,\ve{\rho}_i\over\sum_{i=1}^N m_i}
$$

Eq. (\ref{N-body-eqm-with-mass}) also implies
$$
\sum_{i=1}^N m_i\,\ve{\rho}_i\times\ddot{\ve{\rho}}_i=0.
$$
\noindent
Integrating one gets three more integrals -- integrals of angular momentum:
\begin{equation}
\sum_{i=1}^N m_i\,\ve{\rho}_i\times\dot{\ve{\rho}}_i=\ve{C}={\rm const}.
\end{equation}
\noindent
The plane perpendicular to vector $\ve{C}$ remains time-independent
(since $\ve{C}$ is constant). This plane is called \concept{invariant
plane} of the N-body system or \concept{Laplace plane}. In the Solar system 
the invariant plane lies close to the orbital plane of Jupiter.

Finally, summing up scalar products of each of the equations
(\ref{N-body-eqm-gradient}) with $\dot{\ve{\rho}}_i$ one gets
\begin{equation}
\label{for-energy-integral}
\sum_{i=1}^N m_i\,\dot{\ve{\rho}}_i\cdot\ddot{\ve{\rho}}_i
=\sum_{i=1}^N\grad_i\,U\cdot\,\dot{\ve{\rho}}_i={dU\over dt}
\end{equation}
\noindent
On the other hand, the left-hand side of this equation can be written as
$$
\sum_{i=1}^N m_i\,\dot{\ve{\rho}}_i\cdot\ddot{\ve{\rho}}_i
={dT\over dt},
$$
\noindent
where
$$
T={1\over 2}\,\sum_{i=1}^N m_i\,\dot{\ve{\rho}}_i\cdot\dot{\ve{\rho}}_i.
$$
\noindent
Since both sides of (\ref{for-energy-integral}) represent full derivatives
one can integrate the equation to get the integral of energy in the $N$-body problem
\begin{equation}
{1\over 2}\,\sum_{i=1}^N m_i\,\dot{\ve{\rho}}_i^2
-{1\over 2}\,\sum_{i=1}^N\sum_{j=1,\ j\neq i}^N {G\,m_i\,m_j\over\rho_{ij}}
=H={\rm const}\,,
\end{equation}

\noindent
where $H$ is the total (mechanical) energy of the system of $N$ gravitating bodies.

These 10 integrals can be used to decrease the order of the system
(\ref{N-body-eqm}) or to check the accuracy of numerical
integrations. One often uses barycentric coordinates of the $N$-body
system in which $\ve{A}=0$ and $\ve{B}=0$. In this case
(\ref{N-body-CoM-A})--(\ref{N-body-CoM-AB}) can be used to compute the
position and velocity of one arbitrary body if the positions and
velocities of other $N-1$ bodies are known. This procedure can be used
to compute initial conditions satisfying
(\ref{N-body-CoM-A})--(\ref{N-body-CoM-AB}) with $\ve{A}=0$ and
$\ve{B}=0$.  Alternatively, one body can be completely eliminated from
the integration, so that at each moment of time position and velocity
for that body are calculated using
(\ref{N-body-CoM-A})--(\ref{N-body-CoM-AB}) with $\ve{A}=0$ and
$\ve{B}=0$, and the corresponding equation is excluded from
(\ref{N-body-eqm}) or (\ref{N-body-eqm-with-mass}).  Four remaining
integrals are usually used to check the accuracy of the integration,
the integral of energy is especially sensitive to numerical errors of
usual (non-symplectic) integrators.

\section{The disturbing function}
\label{section-the-disturbing-function}

{\small {\bf Summary:}\
Planetary motion as perturbed two-body motion.
The planetary disturbing function.
}\bigskip

If the mass of one body is much larger than other masses in the
$N$-body system it is sometimes advantageous to write the equations of
motion in a non-inertial reference system centered on that dominating
body. In solar system the Sun is obviously dominating, having the mass
about 1000 times larger than the planets.

Let us consider a system of $N+1$ body numbered from $0$ to $N$. Suppose that
the mass of body $0$ is much larger than the masses of all other bodies
$m_i\ll m_0$ for $i=1,\dots,N$. The equations of motion
(\ref{N-body-eqm}) for bodies $0$ and $i$ can be written as
\begin{eqnarray}
\label{eqm-preparation}
\ddot{\ve{\rho}}_0&=&\sum_{j=1}^N {G\,m_j\,\ve{\rho}_{0j}\over \rho_{0j}^3}
={G\,m_i\,\ve{\rho}_{0i}\over \rho_{0i}^3}+
\sum_{j=1,\ j\neq i}^N {G\,m_j\,\ve{\rho}_{0j}\over \rho_{0j}^3},
\nonumber \\
\ddot{\ve{\rho}}_i&=&\sum_{j=0,\ j\neq i}^N {G\,m_j\,\ve{\rho}_{ij}\over \rho_{ij}^3}
={G\,m_0\,\ve{\rho}_{i0}\over \rho_{i0}^3}+
\sum_{j=1,\ j\neq i}^N {G\,m_j\,\ve{\rho}_{ij}\over \rho_{ij}^3}\,.
\end{eqnarray}
\noindent
Let us designate the position of body $i$ relative to body $0$ as
$\ve{r}_i\equiv \ve{\rho}_{0i}=\ve{\rho}_{i}-\ve{\rho}_{0}$. Then 
subtracting two equations in (\ref{eqm-preparation}) one has
\begin{eqnarray}
\ddot{\ve{r}}_i=\ddot{\ve{\rho}}_{i}-\ddot{\ve{\rho}}_{0}=
-{G\,m_0\,\ve{r}_i\over r_i^3}-{G\,m_i\,\ve{r}_i\over r_i^3}
+\sum_{j=1,\ j\neq i}^N \left( {G\,m_j\,\ve{r}_{ij}\over r_{ij}^3}
-{G\,m_j\,\ve{r}_j\over r_j^3}\right),
\end{eqnarray}
\noindent
where $\ve{r}_{ij}\equiv \ve{r}_{j}-\ve{r}_{i}=\ve{\rho}_{ij}$. Finally,
the equations of motion of body $i$ with respect to body $0$ can be written as
\begin{equation}
\label{eqm-N-body-relative}
\ddot{\ve{r}}_i+{G\,(m_0+m_i)\,\ve{r}_i\over r_i^3}
=\sum_{j=1,\ j\neq i}^N G\,m_j\left( {\ve{r}_{ij}\over r_{ij}^3}
-{\ve{r}_j\over r_j^3}\right).
\end{equation}
\noindent
This equation coincide with the equations
of motion (\ref{2body-equations-of-motion}) of two-body problem
if the right-hand side is zero (e.g., for $m_j=0$, $1\le j\le N$, $j\neq i$).
The right-hand side can be considered as a perturbation of two-body motion.
The same equations of motion can be rewritten in the form
\begin{eqnarray}
&&\ddot{\ve{r}}_i+{G\,(m_0+m_i)\,\ve{r}_i\over r_i^3}
={\rm grad}_i R\,,
\\
\label{perturbation-potential-N-body}
&&R=\sum_{j=1,\ j\neq i}^N G\,m_j\,
\left({1\over r_{ij}}-{\ve{r}_j\cdot\ve{r}_i\over r_j^3}\right)\,.
\end{eqnarray}
\noindent
These equations can be directly integrated numerically or analyzed
analytically to obtain the motion of planets and minor bodies with
respect to the Sun. It is also clear that if we have only two bodies,
$R$ vanishes and the remaining equations of motion describe two-body
problem. Therefore, the forces coming from $R$ can be considered as
perturbations of two-body problem. These perturbations are small in the case when
$m_0+m_i\gg m_j$ and the heliocentric motion of body is close to the
solution of two-body problem. For this reason, $R$ is called
planetary \concept{disturbing function}. The idea to treat any motion of a
dynamical system as a perturbation of some known motion of a simplified
dynamical system is natural and widely used in my areas of physics and
astronomy. In case of dynamics of celestial bodies a suitable
simplification is the two-body motion which is simple and given by
analytical formulas. As perturbations one can consider not only
$N$-body forces as given above, but also non-gravitational forces,
relativistic forces, etc.
 
The problem of motion of N-bodies is a very complicated problem. Since
its formulation the $N$-body problem has led to many new branches in
mathematics. Here let us only mention the Kolmogorov-Arnold-Moser
(KAM) theory that proves the existence of stable quasi-periodic
motions in the $N$-body problem. A review of mathematical results known
in the area of the $N$-body problem is given in the encyclopedic book of
Arnold, Kozlov \& Neishtadt (1997).
 
The main practical tool to solve the equations of the $N$-body problem
is numerical integration. One can distinguish three different modes of
these numerical integrations. First mode is integrations for
relatively short time span and with highest possible accuracy. This
sort of solutions is used for the solar system ephemerides and space
navigation. Some aspects of these high-accuracy integrations are
discussed in Section \ref{Section-ephemerides} below. 
Second sort of integrations are
integrations of a few bodies over very long periods of time with the
goal to investigate long-term dynamics of the motion of the major and
minor bodies of the solar system or exoplanetary systems. Usually one
considers a subset of the major planets and the Sun as gravitating
bodies and investigates long-term motion of this system or the
long-term dynamics of massless asteroids. For this sort of solution,
it is important to have correct phase portrait of the motion and not
necessarily high accuracy of individual orbits. Besides that, usually
the initial conditions of the problem are such that no close
encounters between massive bodies should be treated. Symplectic
integrators are often used for these integrations because of their
nice geometrical properties (e.g., the symplectic integrators do not
change the integral of energy). Resonances of various nature play crucial
role for such studies and are responsible for the existence of chaotic
motions. A good account of recent efforts in this area can be
found in Murray \& Dermott (1999) and Morbidelli (2002). Third kind of
numerical integrations is integrations with arbitrary initial
conditions that do not exclude close encounters between gravitating
bodies. Even for small $N$ numerical integrations of (\ref{N-body-eqm})
in this
general case is not easy, e.g. because of possible close encounters of
the bodies which make the result of integration extremely sensitive to
small numerical errors. During last half of a century significant
efforts have been made to improve the stability and reliability of
such numerical simulations. This includes both analytical change of
variables known as ``regularisation'' and clever tricks in the
numerical codes. An exhaustive review of these efforts can be found in
Aarseth (2003). To increase the performance and make it possible to
integrate the $N$-body problem for large $N$ special-purpose hardware
GRAPE has been created on which special parallel $N$-body code can be
run. Nowadays, direct integrations of the $N$-body problem are possible
with $N$ up to several millions. This makes it possible to use these
$N$-body simulations to investigate the dynamics of stellar clusters and
galaxies (Aarseth, Tout \& Mardling, 2008).
 
\section{Overview of the three-body problem }
 
\begin{figure}
\begin{center}
\resizebox{!}{8.0cm}{\includegraphics{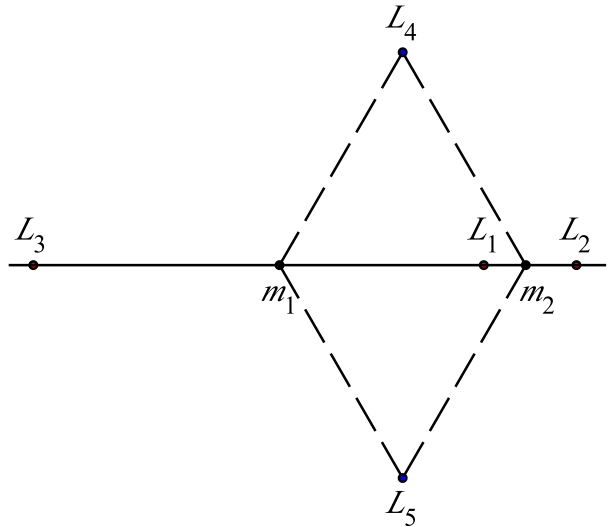}}
\end{center}
\caption[Lagrange points 
in the three-body problem]{\small Lagrange points $L_1$, $L_2$, $\dots$, $L_5$  
in the three-body problem.
Two masses labeled with $m_1$ and $m_2$ are at the indicated positions. The third
body has mass $m_3$. The positions of the Lagrange points depend only on
the mass ratios between $m_1$, $m_2$, and $m_3$. For this plot masses with
$m_2/m_1=m_3/m_1=0.1$ were used.
\label{Figure-1-celmech}}
\end{figure}

Special cases of the $N$-body problem are the two- and three-body
problems. The two-body problem is the basis of all practical
computations of the motion of celestial bodies and has been considered
above.  The three-body problem also has important practical
applications. The real motion of the Moon is much better described by
the three-body system Sun-Earth-Moon than by the two-body problem
Earth-Moon. The motion of asteroids and comets can often be
approximated by the system Sun-Jupiter-asteroid. The two-body problem
can be solved completely. Already the three-body problem is so much
complicated that, in general case, it cannot be solved in analytically
closed form. The motion of three attracting bodies already contains
most of the difficulties of the general $N$-body problem. However,
many theoretical results describing solutions of the three-body
problem have been found. For example, all possible final motions (motions at
$t\to\pm\infty$) are known. Also many classes of periodic orbits were
found. The three-body problem has five important special solutions
called \concept{Lagrange solutions}. These are points of dynamical equilibrium:
all three masses remain in one plane and have in that plane a
Keplerian orbit (being a conic section) with the same focus and with
the same eccentricity. Therefore, in this case the motion of each body
is effectively described by the equations of the two-body problem. The
geometrical form of the three-body configuration (i.e. the ratio of
mutual distances between the bodies) remains constant, but the scale
can change and the figure can rotate. In a reference system where
positions of arbitrary two bodies are fixed there are five points
where the third body can be placed (see Figure \ref{Figure-1-celmech}).
The three bodies either are always situated on a straight line (three
rectilinear Lagrange solutions $L_1$, $L_2$, and $L_3$) or remain at
the vertices of an equilateral triangle (two triangle Lagrange
solutions $L_4$ and $L_5$). Three rectilinear solutions were first
discovered by Leonhard Euler (1707 -- 1783) and are sometimes called Euler's
solutions.

A simplified version of the three-body problem -- the so-called
restricted three-body problem -- is often considered. In the
restricted three-body problem the mass of one of the bodies is
considered to be negligibly small, so that the other two bodies can be
described by the two-body problem and the body with negligible mass
moves in the given field of two bodies with given Keplerian
motion. Clearly, in many practical situations the mass of the third
body can indeed be neglected (e.g., for the motion of minor bodies or
spacecrafts in the field of the Sun and one of the
planets). Sometimes, it is further assumed that the motion of all
three bodies is co-planar (``planar restricted three-body problem'')
and/or that the orbit of the two massive bodies is circular
(``circular restricted three-body problem''). The five Lagrange
solutions do exist also in these restricted versions of the three-body
problem. In the circular restricted three-body problem the five
configurations remain constant in the reference system co-rotating
with the two massive bodies. These points are called libration or
equilibrium points.  Oscillatory (librational) motion around these
points has been investigated in detail. In the linear approximation,
such librational orbits around $L_4$ and $L_5$ in the circular
restricted three-body problem are stable provided that the ratio of
the masses of two massive bodies is less than
$1/2-\sqrt{23/108}\approx0.03852$.  The orbits around $L_1$, $L_2$,
and $L_3$ are unstable.  Interestingly, librational motions around
$L_4$ and $L_5$ are realized in the Solar system.  E.g. the asteroid
family called Trojans has orbits around $L_4$ and $L_5$ of the system
Sun-Jupiter-asteroid. These librational orbits are stable since the
ratio of the masses of Jupiter and the Sun is about $10^{-3}$ which is much
smaller than the limit given above. The rectilinear Lagrange
points have also practical applications. Librational orbits around
these points -- the so-called \concept{Lissajous orbits} -- are very attractive
for scientific space missions. Lissajous orbits around $L_1$ and $L_2$
of the system Sun-Earth-spacecraft are used for such space missions as
WMAP, Planck, Herschel, SOHO, Gaia and James Webb Space
Telescope. Points $L_1$ and $L_2$ of the system Sun-Earth-spacecraft
are situated on the line Sun-Earth at the distance of about 1.5
million kilometers from the Earth (see Figure \ref{Figure-1-celmech}).
Although the Lissajous orbits are unstable, the maneuvers needed to
maintain these orbits are simple and require very limited amount of
fuel. On the other side, placing a spacecraft on an orbit around $L_1$
or $L_2$ guarantees almost uninterrupted observations of celestial
objects, good thermal stability of the instruments, and optimal
distance from the Earth (too far for the disturbing influence of the
Earth's figure and atmosphere, and close enough for high-speed
communications).

One more important result in the circular restricted three-body
problem is the existence of an additional integral of motion called
the \concept{Jacobi integral}. This integral can be used to recognize, e,
g. comets even after close encounters with planets. This is the
so-called Tisserand criterion: the Jacobi integral should remain the
same before and after the encounter even if the heliocentric orbital
elements of the comet have substantially changed. The value of the
Jacobi integral also defines (via the so-called \concept{Hill's surfaces of
zero velocity}) spatial region in which the massless body must be
found. The details on the Jacobi integral can be found, e.g. in the
book of Roy (2005).
 
Finally, let us note that although the $N$-body problem in general and
three-body problem in particular are one of the oldest problems in
astronomy, new results in this area continue to appear. Good example
here is a remarkable figure-eight periodic solution of the three-body
problem discovered by Chenciner \& Montgomery (2000).
 
\section{Planetary ephemerides}
\label{Section-ephemerides}

{\small {\bf Summary:}\
Modern ephemerides: dynamical models, observations, representations.
}\bigskip

Modern ephemerides of the solar system bodies are numerical solutions
obtained by numerical integration of the differential equations of
motion and by fitting the initial conditions of these integrations and
other parameters of the force model to observational data.
 
The equations of motion used here are the $N$-body equations discussed
above augmented by a number of smaller forces. These forces include
relativistic $N$-body forces (the so-called \concept{Einstein-Infeld-Hoffmann
equations}), Newtonian forces due to asteroids, the effects of the
figures (non-sphericity) of the Earth, Moon and the Sun as well as
some non-gravitational forces. For the Sun it is sufficient to
consider the effect of the second zonal harmonics $J_2^\odot$.  The
zonal harmonics $J_n$ of the Earth and the Moon are usually used up to
$n\le4$. Mostly one needs forces coming from the interaction of these
zonal harmonics with other bodies modeled as point masses. The
dynamics of the Earth-Moon system requires even more detailed modeling
since the translational motion of the Earth and the Moon are coupled
with their rotational motions and deformations in a tricky way. For
the Moon even more subtle effects due to tesseral harmonics $C_{nk}$
and $S_{nk}$ again with $n\le4$ should be taken into account. Tidal
deformations of the Earth's gravitational field influence the
translational motion of the Moon and should be taken into
account. Rotational motion of the Earth is well known and obtained
from dedicated observations by the International Earth Rotation and
Reference Systems Service (IERS). These results are good enough and usually
taken for granted for the solar system ephemerides.  Rotational motion
of the Moon is often called physical libration and is an important
part of the process of construction of solar system
ephemerides. Physical libration is modeled as rotational motion of a
solid body with tidal and rotational distortions, including both
elastic and dissipational effects. A discussion of all these forces
coming from the non-point-like structure of gravitating bodies can be
found in Standish \& Williams (2010).
 
Asteroids play an important role for high-accuracy modeling of the
motion of the inner solar system, the motion of Mars being especially
sensitive to the quality of the model of asteroids. Since masses of
asteroids are poorly known for most of them, the modeling is not
trivial. Usually, asteroids are treated in three different ways. A
number of ``big'' asteroids are integrated together with the major
planets, the Moon and the Sun. For these ``big'' asteroids the masses
are estimated from the same observational data that are used to fit
the ephemeris. Among these ``big'' asteroids are always ``the big three''
-- Ceres, Pallas and Vesta -- and, sometimes, up to several ten
asteroids which influence the motion of Mars more than other
asteroids. For some hundred asteroids their masses are estimated using
their taxonomic (spectroscopic) classes and their estimated radii that
are determined by photometry, radar data or observations of stellar
occultations by asteroids. For each of the three taxonomic classes Ð C
(carbonaceous chondrite), S (stony) and M (iron) Ð the mean density is
determined as a part of the ephemeris construction. The cumulative
effect of other asteroids is sometimes empirically modeled by a
homogeneous massive ring in the plane of ecliptic. The mass of the
ring and its radius are again estimated from the same data that are
used for the ephemeris (Pitjeva, 2007; Kuchynka et al. 2010).
 
Since the equations of motion are ordinary differential equations any
method for numerical integration of ordinary differential equations
can be used to solve them. A very good practical overview of
numerical integration methods is given in Chapter 4 of Montenbruck \&
Gill (2000). Even more details can be found in Chapter 7 of Beutler
(2005, Part I). In practice, for planetary motion, one uses either
multistep Adams (predictor-corrector) methods (Standish \& Williams,
2008; Fienga et al., 2008) or the Everhart integrator (Everhart, 1985;
Pitjeva, 2005). The latter is a special sort of implicit Runge-Kutta
integrators. Numerical round-off errors are an important issue for the
integrations of planetary ephemerides. Usual double precision (64 bit)
arithmetic is not sufficient to achieve the goal accuracy and one
often uses quadruple precision (128 bit) arithmetic. Since the
beginning of the 1970s the JPL ephemeris team uses the variable
stepsize, variable order multistep Adams integrators called DIVA/QIVA
(Krogh, 2004). Fienga at al. (2008) have shown that only a few
arithmetical operations in the classical Adams integrator of order 12
must be performed with quadruple precision to achieve an acceptable
accuracy over longer integration intervals. This substantially
increases the performance of numerical integrations.
 
Observational data used for planetary ephemerides include radar
observations of earth-like planets, radar and Doppler observations of
spacecrafts (especially planetary orbiters), VLBI observations of
spacecrafts relative to some reference quasars, Lunar Laser Ranging
data, and, finally, optical positional observations of major planets
and their satellites (especially important for outer planets with very
few radiometric observations).
 
A total of 250 parameters are routinely fitted for the construction of
planetary ephemerides. These parameters include initial positions and
velocities of the planets and some of their satellites, the
orientation of the frame with respect to the ICRF, the value of
Astronomical Unit in meters (or the mass parameter $GM_\odot$ of the Sun), the
parameters of the model for asteroids (see above), various parameters
describing rotational and translational motion of the Earth-Moon
system, various parameters used in the reduction of observational data
(phase corrections for planetary disk observations, corrections to
precession and equinox drift, locations of various relevant sites on
the Earth and other bodies, parameters of the solar corona, parameters
describing the geometrical figures of Mercury, Venus and Mars,
etc.). Useful discussion of various models used for data modeling is
given by Moyer (2003) and Standish \& Williams (2010). The masses of
the major planets can be also fitted from the same data, but, when
available, they are taken from the special solutions for the data of
planetary orbiters. However, the masses of the Earth and the Moon are
often determined in the process of construction of planetary
ephemerides.

Modern ephemerides are represented in the form of Chebyshev
polynomials. The details of the representation can vary from one
ephemeris to another, but the principles are the same: each scalar
quantity is represented by a set of polynomials $p_i$ of the form
 
\begin{equation}
\label{Chebyshev-representation}
p_i(t)=\sum_{k=0}^{N_i}a_k^{(i)}T_k(x),\quad x={2t-t_i-t_{i+1}\over t_{i+1}-t_i},
\end{equation}

\noindent 
where $T_k(x)$ are the Chebyshev polynomials of the first
kind (given by the recurrent relations $T_0=1$, $T_1=x$, and
$T_{k+1}=2xT_k-T_{k-1}$), and coefficients $a_k^{(i)}$ are real
numbers. Each polynomial $p_i(t)$ is valid for some interval of
time $t_i\le t\le t_{i+1}$ (so that $-1\le x\le 1$). 
The representation (\ref{Chebyshev-representation})
is close to the optimal uniform
approximation of a function by polynomials of given order (Press et
al., 2007, Section 5.8), and, thus, gives nearly optimal
representation of a function using given number of free
parameters. The orders of polynomials $N_i$ are usually the same for all
time intervals, but do depend on the quantity to be
represented. Sometimes (e.g. for the JPL ephemerides) one polynomial
represents both the position and the velocity of a body. The velocity
can be then calculated as a derivative of (\ref{Chebyshev-representation}):

\begin{equation}
\label{Chebyshev-representation-derivative}
{d\over dt}\,p_i(t)={2\over t_{i+1}-t_i}\sum_{k=1}^{N_i}k a_k^{(i)}U_k(x),
\end{equation}
 
\noindent 
where $U_k$ are the Chebyshev polynomials of the second kind
(given by the recurrent relations $U_0=1$, $U_1=2x$, and
$U_{k+1}=2xU_k-U_{k-1}$).  At the boundaries $t_i$ of the time
intervals, the polynomials $p_i$ must satisfy conditions like
$p_{i-1}(t_i)=p_i(t_i)$, so that the approximating function is continuous.
One can also imply additional constraints ${d\over
dt}\,p_{i-1}(t_i)={d\over dt}\,p_i(t_i)$ for the derivatives to make
the approximating function continuously differentiable. An efficient technique to
compute the coefficients $a_k^{(i)}$ starting from values of the
quantity to be represented is described by Newhall~(1989).

There are three sources of modern planetary ephemerides: Jet
Propulsion Laboratory (JPL, Pasadena, USA; DE ephemerides), Institut
de M\'echanique C\'eleste et de Calcul des \'Eph\'em\'erides
(IMCCE, Paris Observatory, France; INPOP ephemerides) and Institute of
Applied Astronomy (IAA, St.Petersburg, Russia; EPM ephemerides). All
of them are available from the Internet:
 
\begin{itemize}
 \item[--] \url{http://ssd.jpl.nasa.gov/?planet_eph_export} 
for the DE ephemerides, 
\item[--] \url{http://www.imcce.fr/fr/presentation/equipes/ASD/inpop/} 
for the INPOP, and 
\item[--] \url{ftp://quasar.ipa.nw.ru/incoming/EPM2004/} for the EPM ephemerides.  
\end{itemize}

Different versions of the ephemerides have different intervals of
validity, but typically these are several hundred years around the
year 2000. Longest readily available ephemerides are valid for a time
span of 6000 years. Further details on these ephemerides can be found
in Standish \& Williams (2010), Folkner (2010), Fienga et al. (2008),
and Pitjeva (2005), respectively.
 
Let us also mention that for the lower-accuracy applications,
semi-analytical theories of planetary motion called VSOP are available
(Bretagnon, Francou, 1988; Moisson, Bretagnon, 2001). The
semi-analytical theories are given in the form of Poisson series
$\sum_k c_k t^{n_k} \cos(a_k t+b_k)$, where $a_k$, $b_k$, and $c_k$
are real numbers, and $n_k$ is the integer power of time $t$.
Formally any value of time can be substituted into such series, but
the theory is meaningful only for several thousand years around the year
2000. The VSOP ephemeris contains only major planets and the 
Earth-Moon barycenter. The best semi-analytical theory of motion
of the Moon with respect to the Earth is called ELP82. 
This theory can also be used in low-accuracy applications.

\chapter{Elements of the Perturbation Theory}

\section{The method of the variation of constants}
\label{Section-osculating-elements-general}

{\small {\bf Summary:}\
The variation of constants as a method to solve differential equations.
Instantaneous elements. Osculating elements.
}\bigskip

The equations of motion of two-body problem considered above in great detail
read

$$\ddot{\ve{r}}+\kappa^2\,{\ve{r}\over r^3}=0.$$

\noindent
The simplicity of the two-body motion and the fact that many practical
problems of celestial mechanics are sufficiently close to two-body
motion make it practical to use two-body motion as zero approximation
to the motion in more realistic cases and treat the difference by the
usual perturbative approach. Special technique for the motion of
celestial bodies is called osculating elements. The solution of the
two-body problem discussed above can be symbolically written as

\begin{equation}
\label{r-Kepler}
\ve{r}=\ve{f}(t,e_1,e_2,e_3,e_4,e_5,e_6),
\qquad e_i={\rm const},\ i=1\dots6,
\end{equation}

\noindent where $e_i$ are six Keplerian elements: semi-major axis $a$,
eccentricity $e$, inclination $i$, argument of pericenter $\omega$ and longitude of the
node $\Omega$. In general case of arbitrary additional forces it is always
possible to write the equations of motion of a body as

\begin{equation}
\label{eqm-perturbed-general}
\ddot{\ve{r}}+\kappa^2\,{\ve{r}\over r^3}=\ve{F},
\end{equation}

\noindent 
where $\ve{F}=\ve{F}(t,\ve{r},\dot{\ve{r}},\dots)$ is arbitrary force
depending in general on the position and velocity of the body under
study, time and any other parameters. One example of such a
disturbing force is given by (\ref{eqm-N-body-relative}) for the
$N$-body problem. The general idea is to use the same functional form
for the solution of (\ref{eqm-perturbed-general}) as we had for the two-body
motion, but with constants (former Kepler elements) being time-dependent:

\begin{equation}
\label{f-Kepler-osculating}
\ve{r}=\ve{f}(t,e_1,e_2,e_3,e_4,e_5,e_6),
\qquad e_i=e_i(t),\ i=1\dots6.
\end{equation}

\noindent 
This is always possible since $\ve{F}$ has three degrees of freedom
(three arbitrary components) and representation
(\ref{f-Kepler-osculating}) involves six arbitrary functions of time.
Let us stress the following. Eq. (\ref{f-Kepler-osculating}) means
that if elements $e_i(t)$ are given for some $t$, the position
$\ve{r}$ of the body under study for that $t$ can be computed using usual formulas
of two-body problem as summarized in Section
\ref{Section-2-body-summary}. The idea of (\ref{f-Kepler-osculating})
is closely related to the idea of the method of \concept{variation of
constants}, also known as \concept{variation of parameters}, developed by
Joseph Louis Lagrange. This method is a general method to solve
inhomogeneous linear ordinary differential equations.

As mentioned above the representation in (\ref{f-Kepler-osculating})
has three ``redundant'' degrees of freedom. These three degrees of
freedom can be used to make it possible to compute not only position
$\ve{r}$, but also velocity $\dot{\ve{r}}$ from the solution of
(\ref{eqm-perturbed-general}) using standard formulas of the two-body
problem summarized in Section \ref{Section-2-body-summary}. This can
be done if the elements $e_i(t)$ satisfy the following condition

\begin{equation}
\label{osculating-condition}
\sum_{i=1}^6 {\partial\ve{f}\over \partial e_i}\,\dot e_i=0\,.
\end{equation}

\noindent
Indeed, in general case, time derivative of $\ve{r}$ given by 
(\ref{f-Kepler-osculating}) reads

\begin{equation}
\dot{\ve{r}}={\partial\ve{f}\over \partial t}
+\sum_{i=1}^6 {\partial\ve{f}\over \partial e_i}\,\dot e_i\,.
\end{equation}

\noindent
Therefore, condition (\ref{osculating-condition}) 
guarantees that the time derivative of (\ref{f-Kepler-osculating})
is given by the
partial derivative of $\ve{f}$ with respect to time

\begin{eqnarray}
\label{dot-r-osculating}
%\ve{r}&=&\ve{f}(t,e_1,e_2,e_3,e_4,e_5,e_6),
%\nonumber\\
\dot{\ve{r}}&=&{\partial\over \partial t}
\,\ve{f}(t,e_1,e_2,e_3,e_4,e_5,e_6)\,.
\end{eqnarray}

\noindent 
This means that velocity $\dot{\ve{r}}$ can be calculated by the standard
formulas of two-body problem (indeed, (\ref{dot-r-osculating}) coincided with
the derivative of (\ref{r-Kepler} with constant $e_i$ that represent the usual solution
of the two-body problem). The elements having these properties
are called \concept{osculating elements}. The osculating elements are
in general functions of time. To compute position and velocity at any given
moment one first has to calculate the values of the six osculating
elements for this moment of time and then use the standard equations
summarized in Section \ref{Section-2-body-summary}.

Let us stress that, with osculating elements, not only vectors of
position $\ve{r}$ and velocity $\dot{\ve{r}}$ can be computed using
formulas of the two-body problem, but also any functions of these two
vectors. Let us give an explicit example here. For a given moment of
time $t$ the absolute value of $\ve{r}$ is given as
$r(t)=a(t)\,(1-e(t)\,\cos E(t))$.  Here $a(t)$ and $e(t)$ are
osculating semi-major axes and eccentricity. Osculating eccentric
anomaly $E(t)$ can be computed from Kepler equation $E-e(t) \sin E=M$,
where again osculating eccentricity $e(t)$ should be used. The
derivative of $r$ can be computed $\dot r=\dot E\,a\,e\,\cos E
=a\,e\,\cos E\,{\kappa\,a^{-3/2}\over 1-e\,\cos
E}={\kappa\,e\over\sqrt{a\,(1-e^2)}}\,\sin v$ and all elements are
again functions of time: $a=a(t)$, $e=e(t)$, etc. Also the anomalies
-- eccentric $E$, true $v$ and mean $M$ -- are related to each other
in the same way as in the two-body problem.

\section{Gaussian perturbation equations}

{\small {\bf Summary:}\
The radial, tangential and transverse components of the disturbing force.
The Gaussian perturbation equations: the differential equations for the osculating
elements. Other variants of the Gaussian perturbation equations.
}\bigskip

Let us derive the equations for osculating elements for a general
disturbing force $\ve{F}$. First, we introduce a new Cartesian coordinate
system. The origin of the new system $(S,T,W)$ 
is the same as usual, but the orientation is different and depends
at each moment of time 
on the position $\ve{r}$ and velocity $\dot{\ve{r}}$ of the considered body. 
Axis $S$ is directed radially, that is parallel to $\ve{r}$.
Axis $T$ lies in the momentary orbital plane (the plane containing both
$\ve{r}$ and $\dot{\ve{r}}$), perpendicular to $S$ (and, therefore, to $\ve{r}$) 
and the angle between $T$ and $\dot{\ve{r}}$ does not exceed $90^\circ$.
Axis $W$ is perpendicular to both $S$ and $T$ (that is, perpendicular to both
$\ve{r}$ and $\dot{\ve{r}}$) and completes $S$ and $T$ to a right-hand 
coordinate system $(S,T,W)$. Coordinates $(S,T,W)$ rotate as the body moves 
along its orbit. 

The components of disturbing force $\ve{F}$ in axes $(S,T,W)$
are also denoted by $(S,T,W)$ and can be computed as
\begin{eqnarray}
\label{S-force}
S&=&{\ve{r}\over r}\cdot\ve{F},
\\
\label{T-force}
T&=&{\left(\ve{r}\times\dot{\ve{r}}\right)\times\ve{r}\over \left|\ve{r}\times\dot{\ve{r}}\right|\,r}\cdot\ve{F},
\\
\label{W-force}
W&=&{\ve{r}\times\dot{\ve{r}}\over \left|\ve{r}\times\dot{\ve{r}}\right|}\cdot\ve{F},
\end{eqnarray}
\noindent
where '$\cdot$' and '$\times$' denote the scalar and cross products of two vectors.
Obviously, the relation of $(S,T,W)$ and our usual coordinates $(x,y,z)$ reads

\begin{equation}
\label{xyz-STW}
\pmatrix{a_x\cr a_y\cr a_z}=
\mat{A}^T_z(\Omega)\,\mat{A}^T_x(i)\,\mat{A}^T_z(u)\,
\pmatrix{a_A\cr a_T\cr a_W},
\end{equation}

\noindent
where angle $u=v+\omega$ is called \concept{argument of latitude}. The matrix
$\mat{A}^T_z(\Omega)\,\mat{A}^T_x(i)\,\mat{A}^T_z(u)$
is given by (\ref{P-matrix}) with $u$
substituted for $\omega$. Here for an arbitrary vector $\ve{a}$
its components in coordinates $(x,y,z)$ are denoted as $(a_x,a_y,a_z)$
and the corresponding components in coordinates $(S,T,W)$ are
$(a_S,a_T,a_W)$.

\subsection{Derivation of differential equations for osculating elements}

Now, let us derive the required equations one by one. First, let
us consider the integral of areas 
$\ve{c}=\ve{r}\times\dot{\ve{r}}$. This leads to
$c^2=(\ve{r}\times\dot{\ve{r}})\,\cdot\,(\ve{r}\times\dot{\ve{r}})$,
where $c=|\ve{c}|$.
A time derivative of $c^2$ then reads
$${dc^2\over dt}=2\,(\ve{r}\times\dot{\ve{r}})\,\cdot\,(\ve{r}\times\ddot{\ve{r}})
=
2\,(\ve{r}\times\dot{\ve{r}})\,\cdot\,(\ve{r}\times\ve{F})
=
2\left(\left(\ve{r}\times\dot{\ve{r}}\right)\times\ve{r}\right)\cdot\ve{F}
=
2\,r^2\,\dot{\ve{r}}\cdot\ve{F}-2\,r\,\dot r\,\ve{r}\cdot\ve{F}$$
\noindent
($\dot{\ve{r}}\cdot\ve{r}=\dot r\, r$ is used here).
Since $\ve{F}=(S,T,W)$ and $\ve{r}=(r,0,0)$ in STW-coordinates one gets
$\ve{r}\cdot\ve{F}=r\,S$. Let us now consider $\dot{\ve{r}}\cdot\ve{F}$.
We need vectors $\dot{\ve{r}}$ and $\ve{F}$ in STW-coordinates.
Since the instantaneous plane of the orbit is
defined by the instantaneous position and velocity vectors of the body,
the $W$ component of the velocity $\dot{\ve{r}}$ vanish by definition.
Obviously, the $S$ component is $\dot r$ and the T component is
$r\dot v$. For the latter from the integral of
areas in polar coordinates $r^2\,\dot v=\kappa\,\sqrt{p}$, 
one has $r\dot v={\kappa\,\sqrt{p}\over r}$. Therefore,
in STW coordinates
$\dot{\ve{r}}=(\dot r,{\kappa\,\sqrt{p}\over r},0)$. 
Therefore,
$\dot{\ve{r}}\cdot\ve{F}=\dot r\,S+{\kappa\,\sqrt{p}\over r}\,T$.
Substituting $\ve{r}\cdot\ve{F}$ and $\dot{\ve{r}}\cdot\ve{F}$ into the
equation for ${dc^2\over dt}$ derived above and taking into account
that the semi-latus rectum $p=c^2/\kappa^2$ we get
\begin{equation}
\label{dot-p}
\dot p=2\,p\,r\,\left({T\over\kappa\,\sqrt{p}}\right).
\end{equation}

Now, let us consider the time derivative of the integral of area
itself.  One has $\ve{c}=\ve{r}\times\dot{\ve{r}}$ and, therefore,
$\dot{\ve{c}}={d\over dt}\,\left(\ve{r}\times\dot{\ve{r}}\right)
=\ve{r}\times\ddot{\ve{r}}=\ve{r}\times\ve{F}$.  It is clear that
$\ve{c}$ is parallel to axis W of the STW system (since $\ve{c}$ is
defined as $\ve{c}=\ve{r}\times\dot{\ve{r}}$ it is perpendicular to
both $\ve{r}$ and $\dot{\ve{r}}$).  For this reasons and considering
that $c=\kappa\,\sqrt{p}$, the components of $\ve{c}$ in STW axes read
$\ve{c}=(0,0,\kappa\,\sqrt{p})$. Using transformation
(\ref{xyz-STW}) to convert the $STW$-components into $xyz$-ones, one gets
\begin{equation}
\label{c-xyz}
\pmatrix{c_x\cr c_y\cr c_z}=
\left(
\begin{array}{r}
\sin i\,\sin\Omega\,\kappa\,\sqrt{p}\\
-\sin i\,\cos\Omega\,\kappa\,\sqrt{p}\\
\cos i\,\kappa\,\sqrt{p}
\end{array}
\right)\,.
\end{equation}
\noindent
In the STW system one has
$\ve{r}=(r,0,0)$ and $\ve{F}=(S,T,W)$. Therefore, in STW components
one gets $\ve{r}\times\ve{F}=(0,-r\,W,r\,T)$. Again  
using transformation (\ref{xyz-STW}) it is easy to calculate that the z component
of $\ve{r}\times\ve{F}$ reads $\left(\ve{r}\times\ve{F}\right)_z=
-\sin i\,\cos u\,r\,W+\cos i\,r\,T$. Considering the z component of
$\dot{\ve{c}}=\ve{r}\times\ve{F}$ one gets
$$
{d\over dt}\,c_z=-\sin i\,\cos u\,r\,W+\cos i\,r\,T\,.
$$
\noindent
On the other hand, from (\ref{c-xyz}) one gets
$$
{d\over dt}\,c_z=
{d\over dt}\,\left(\cos i\,\kappa\,\sqrt{p}\right)=
-\kappa\,\sqrt{p}\,\sin i{d\over dt}\,i
+\cos i\,{d\over dt}\,\left(\kappa\,\sqrt{p}\right)
\,.
$$ 
\noindent
From (\ref{dot-p}) it is easy to see that
$$
{d\over dt}\,\left(\kappa\,\sqrt{p}\right)
=\kappa\,{1\over 2\,\sqrt{p}}\ \dot p=r\,T\,,
$$
\noindent
and we finally get the following equation for the derivative of inclination $i$:
\begin{equation}
\label{dot-i}
{d\over dt}\, i=r\,\cos u\,\left({W\over\kappa\,\sqrt{p}}\right)\,.
\end{equation}

Analogously, considering the time derivative of $c_x=
\sin i\sin\Omega\,\kappa\,\sqrt{p}$ and computing the x
component of $\ve{r}\times\ve{F}$ from (\ref{xyz-STW})
one gets
$$
{d\over dt}\,c_x={d\over dt}\,\left(\sin i\,\sin\Omega\,\kappa\,\sqrt{p}\right)
=(-\cos\Omega\,\sin u-\cos i\,\sin\Omega\,\cos u)\,(-r\,W)
+\sin i\,\sin\Omega\,r\,T\,.
$$
\noindent
Using here equations 
(\ref{dot-p}) and (\ref{dot-i})
for $\dot p$ and ${d\over dt}\,i$ one
gets the equation for the time derivative of $\Omega$:
\begin{equation}
\label{dot-Omega}
{d\over dt}\, \Omega  =  {r\,\sin(v+\omega)\over \sin i}\,
\left({W\over\kappa\sqrt{p}}\right)\,.
\end{equation}
\noindent
Clearly, the y component of $\dot{\ve{c}}=\left(\ve{r}\times\ve{F}\right)$ gives no
new information since $c_y$ depends on the same elements as $c_x$. 
We have therefore got all possible equations from the integral of areas
of the two-body problem. Now let us turn to the integral of energy
(\ref{2body-energy-cartesian-in-plane}). This integral can be rewritten in the form
\begin{equation}
{1\over 2}\,\dot{\ve{r}}\cdot\dot{\ve{r}}-{\kappa^2\over r}=-{\kappa^2\over 2a}.
\end{equation}
\noindent
The derivative of this equation 
$$
\dot{\ve{r}}\cdot\ddot{\ve{r}}+{\kappa^2\over r^2}\,\dot{r}={\kappa^2\over 2a^2}\,\dot{a}
$$
\noindent 
can be simplified using
$$
\dot{\ve{r}}\cdot\ddot{\ve{r}}=\dot{\ve{r}}\cdot\left(-{\kappa^2\over r^3}\,\ve{r}+\ve{F}\right)
=-{\kappa^2\over r^3}\,\dot{\ve{r}}\cdot\ve{r}+\dot{\ve{r}}\cdot\ve{F},
$$
$$
\dot{\ve{r}}\cdot\ddot{\ve{r}}+{\kappa^2\over r^2}\,\dot{r}=
-{\kappa^2\over r^3}\,\dot{\ve{r}}\cdot\ve{r}+\dot{\ve{r}}\cdot\ve{F}+
{\kappa^2\over r^2}\,\dot{r}=\dot{\ve{r}}\cdot\ve{F}
$$
\noindent
(here we used that $\dot{r}\,r=\dot{\ve{r}}\cdot\ve{r}$) and computing
$\dot{\ve{r}}\cdot\ve{F}$ from the STW components of $\dot{\ve{r}}$ and $\ve{F}$
already given above
$$
\dot{\ve{r}}\cdot\ve{F}=\dot{r}\,S+{\kappa\,\sqrt{p}\over r}\,T\,.
$$
\noindent
This allows one to get the equation for $\dot{a}$:
\begin{equation}
\label{dot-a}
{d\over dt}\, a  =  2\,a^2\,e\,\sin v\,\left({S\over\kappa\sqrt{p}}\right)
+2\,a^2\,{p\over r}\,\left({T\over\kappa\sqrt{p}}\right)\,.
\end{equation}
\noindent
Here we used that $\dot{r}={\kappa e\over\sqrt{p}}\,\sin v$.
Having equations (\ref{dot-p}) and (\ref{dot-a}) 
for $\dot{p}$ and $\dot{a}$ it is easy to derive the equation
for $\dot{e}$. Indeed, from $p=a(1-e^2)$ one gets $\dot p=\dot a(1-e^2) -2e\,a\,\dot e$.
Solving for $\dot{e}$ and substituting (\ref{dot-p}) and (\ref{dot-a}) one gets
\begin{equation}
\label{dot-e}
{d\over dt}\, e  =  p\,\sin v\,\left({S\over\kappa\sqrt{p}}\right)
+p\,(\cos v+\cos E)\,\left({T\over\kappa\sqrt{p}}\right)\,.
\end{equation}

Now, let us turn to the derivation of $\dot{\omega}$. The derivation consists
of several steps. First, from
$$
r={p\over 1+e\cos v}
$$
\noindent
one gets
$$
1+e\,\cos v={p\over r}.
$$
\noindent
Computing time derivative of the latter equations
$$
-e\,\sin v\,\dot{v}+\dot{e}\,\cos v= {\dot{p}\over r}-{p\over r^2}\,\dot{r}
$$
\noindent
and using equations for $\dot{r}$, $\dot{p}$ and $\dot{e}$ derived
above one gets the derivative of true anomaly $v$ as
\begin{equation}
\label{dot-v}
\dot{v}={\kappa\,\sqrt{p}\over r^2}+{p\,\cos v\over e}\,\left({S\over \kappa\,\sqrt{p}}\right)
-{p+r\over e}\,\sin v\,\left({T\over \kappa\,\sqrt{p}}\right)\,.
\end{equation}
\noindent
As the second step in the derivation of $\dot{\omega}$, 
let us consider the formula for the z component of vector $\ve{r}$: 
$z=r\,\sin i\,\sin u$, where $u=v+\omega$. This formula can be derived e.g., from
the equation immediately before (\ref{cos-sin-v+omega}). On the one hand, $\dot{z}$ 
is the component of the velocity vector $\dot{\ve{r}}$ and can be calculated
considering all osculating elements as constants (from the integral of area in polar
coordinates $r^2\,\dot{v}=\kappa\,\sqrt{p}$ one has $\dot{v}={\kappa\,\sqrt{p}\over r^2}$):
$$
\dot{z}=\dot{r}\,\sin i\,\sin u+r\,\dot{v}\,\sin i\,\cos u
=
\dot{r}\,\sin i\,\sin u+r\,{\kappa\sqrt{p}\over r^2}\sin i\,\cos u
$$ 
\noindent 
However, the same $\dot{z}$ can be computed not assuming
that the osculating elements are constants (that is, explicitly 
considering the time derivatives
of the osculating elements). This should give the same
results according to the idea of osculating elements described in
Section \ref{Section-osculating-elements-general}. Therefore, one has
$$
\dot{z}=\dot{r}\,\sin i\,\sin u+r\,\cos i\,\sin u\,{di\over dt}+
r\,\sin i\,\cos u\,{du\over dt}\,.
$$ 
\noindent
Equating the last two expressions for $\dot{z}$ one gets
\begin{equation}
\label{dot-u}
\dot{u}={\kappa\sqrt{p}\over r^2}-\cot i\,\tan u\,{di\over dt}.
\end{equation}
\noindent
Finally, since $u=v+\omega$ one has $\dot{\omega}=\dot{u}-\dot{v}$ and
using equations (\ref{dot-u}) and (\ref{dot-v}) for
$\dot{u}$ and $\dot{v}$ and (\ref{dot-i}) for ${di\over dt}$ one gets
\begin{equation}
\label{dot-omega}
{d\over dt}\, \omega  = 
-{p\,\cos v\over e}\,\left({S\over\kappa\sqrt{p}}\right)
+{r+p\over e}\,\sin v\,\left({T\over\kappa\sqrt{p}}\right)
-r\,\sin(v+\omega)\,\cot i\,\left({W\over\kappa\sqrt{p}}\right)\,.
\end{equation}

The only equation still to be derived is that for the mean anomaly of an epoch.
Let us first consider the definition of the mean anomaly
\begin{equation}
\label{M}
M=M_0+n(t-t_0)\,,
\end{equation}
\noindent
where $t_0$ is a given fixed epoch for which $M=M_0$. 
In the framework of the two-body problem
$n=\kappa\,a^{-3/2}$ is constant. In case of osculating elements $n$ is time-dependent
(since $a$ is time-dependent) and the derivative of $M$ reads
\begin{equation}
\label{M-dot-osculating}
{d\over dt}\,M={d\over dt}\,M_0+n+{dn\over dt}\,(t-t_0).
\end{equation}
\noindent
This formula contains time $t$ explicitly. This is not convenient for many applications.
This can be avoided if instead of (\ref{M}) one defines the mean anomaly
as
\begin{equation}
\label{osculating-M}
M=\overline{M}_0+\int_{t_0}^tn\,dt\,.
\end{equation}
\noindent
In the framework of the two-body problem (\ref{osculating-M}) is fully equivalent
to (\ref{M}). However, for osculating elements derivative of $M$ from (\ref{osculating-M})
reads
\begin{equation}
\label{M-osculating-dot}
{d\over dt}\,M={d\over dt}\,\overline{M}_0+n\,.
\end{equation}
\noindent
Clearly, ${d\over dt}\,\overline{M}_0={d\over
dt}\,M_0+\dot{n}\,(t-t_0)$, where $\dot{n}={dn\over dt}=-{3\over
2}\,\kappa\,a^{-5/2}\,\dot{a}$.  Definition (\ref{osculating-M}) is
used below.

In order to derive the equation for ${d\over dt}\,\overline{M}_0$
let us consider two equations
\begin{eqnarray}
&&E-e\,\sin E=M\,,
\nonumber\\
&&r=a\,(1-e\,\cos E)\,.
\nonumber
\end{eqnarray}
\noindent
Differentiating these equations one gets
\begin{eqnarray}
&&\dot E\,(1-e\,\cos E)-\dot{e}\,\sin E=\dot M,
\nonumber\\
&&\dot{r}=\dot{a}\,(1-e\,\cos E)-a\,\dot{e}\,\cos E+a\,e\,\sin E\,\dot{E}.
\nonumber
\end{eqnarray}
\noindent
Considering that $\dot{r}={\kappa\over\sqrt{p}}\,e\,\sin v$ one gets
\begin{equation}
\label{dot-M}
\dot{M}=n+{\sqrt{1-e^2}\over e}\,\left(\dot{e}\,\cot v-\dot{a}\,{r\over a^2\,\sin v}\right)\,.
\end{equation}
\noindent
Comparing to (\ref{M-osculating-dot}) and substituting 
(\ref{dot-e}) and (\ref{dot-a}) for $\dot{e}$ and $\dot{a}$ one finally gets
\begin{equation}
\label{dot-overlineM0}
{d\over dt}\, \overline{M}_0 = {\sqrt{1-e^2}\over e}\,\left(
\left(p\cos v-2\,e\,r\right)\,\left({S\over\kappa\sqrt{p}}\right)
-(r+p)\sin v\,\left({T\over\kappa\sqrt{p}}\right)\right).
\end{equation}

\subsection{Discussion of the derived equations}

Gathering all equations for the derivatives of osculating elements
derived above one gets the full set of equations:
\begin{eqnarray}
\label{Euler-equation-a}
{d\over dt}\, a & = & 2\,a^2\,e\,\sin v\,\left({S\over\kappa\sqrt{p}}\right)
+2\,a^2\,{p\over r}\,\left({T\over\kappa\sqrt{p}}\right),
\\[5pt]
\label{Euler-equation-e}
{d\over dt}\, e & = & p\,\sin v\,\left({S\over\kappa\sqrt{p}}\right)
+p\,(\cos v+\cos E)\,\left({T\over\kappa\sqrt{p}}\right),
\\[5pt]
\label{Euler-equation-i}
{d\over dt}\, i & = & r\,\cos(v+\omega)\,\left({W\over\kappa\sqrt{p}}\right),
\\[5pt]
\label{Euler-equation-omega}
{d\over dt}\, \omega & = &
-{p\,\cos v\over e}\,\left({S\over\kappa\sqrt{p}}\right)
+{r+p\over e}\,\sin v\,\left({T\over\kappa\sqrt{p}}\right)
-r\,\sin(v+\omega)\,\cot i\,\left({W\over\kappa\sqrt{p}}\right),
\\[5pt]
\label{Euler-equation-Omega}
{d\over dt}\, \Omega & = & {r\,\sin(v+\omega)\over \sin i}\,
\left({W\over\kappa\sqrt{p}}\right),
\\[5pt]
\label{Euler-equation-overline-M0}
{d\over dt}\, \overline{M}_0 & = & {\sqrt{1-e^2}\over e}\,\left(
\left(p\cos v-2\,e\,r\right)\,\left({S\over\kappa\sqrt{p}}\right)
-(r+p)\sin v\,\left({T\over\kappa\sqrt{p}}\right)\right).
\end{eqnarray}
\noindent 
These equations were derived by Johann Carl Friedrich Gau\ss{} (1777
-- 1855) and called \concept{Gaussian perturbation equations}.  The
equations were also independently derived by Leonhard Euler some time
before Gauss and are sometimes called \concept{Euler equations}. We
will use these equations below when discussing the influence of
atmospheric drag on the motion of Earth's satellites.

Numerous alternative forms of the Gaussian perturbation equations are
given in Beutler (2005). In particular, one can use any other components
of the disturbing force $\ve{F}$ instead of $(S,T,W)$.
Eqs. (\ref{Euler-equation-a})--(\ref{Euler-equation-overline-M0}) 
are valid for elliptic motion ($e<1$). Similar equations for hyperbolic motion with
$e>1$ can be derived, but are rarely used in practice. Let us
note that Eqs. (\ref{Euler-equation-a})--(\ref{Euler-equation-overline-M0}) 
have singularities 
for $e=0$ or very small $e$ (equations for $\omega$ and $\overline{M}_0$),
$i=0$ and $i=\pi$ (equations for $\omega$ and $\Omega$),
and $e$ close to 1 (e.g. since $p$ is small for this case).
Indeed, the right-hand sides of the equations are not well defined in these cases.
These singularities are related to the fact that some of the Keplerian elements
are not well defined for $e=0$, $i=0$ or $i=\pi$. If the application requires
these cases to be included one introduces 
\begin{eqnarray}
\label{hk}
h&=&e\,\sin\omega,
\nonumber\\
k&=&e\,\cos\omega,
\end{eqnarray}
\noindent
instead of $e$ and $\omega$ and
\begin{eqnarray}
\label{pq}
p&=&\tan i\,\sin\Omega,
\nonumber\\
q&=&\tan i\,\cos\Omega.
\end{eqnarray}
\noindent
instead of $i$ and $\Omega$. The equations for the derivatives 
$h$, $k$, $p$ and $q$ can be easily derived from
(\ref{Euler-equation-a})--(\ref{Euler-equation-overline-M0}).
The equations for $\dot h$, $\dot k$, $\dot p$ and $\dot q$
and do not contain singularities.

\bigskip
{\small
{\bf Exercise.}
Derive equations for the derivatives of osculating $h$, $k$, $p$ and $q$ using corresponding
equations for $\dot\omega$, $\dot e$, $\dot i$ and $\dot\Omega$ in 
(\ref{Euler-equation-a})--(\ref{Euler-equation-overline-M0}).
}

\section{Lagrange equations}
\label{section-Lagrange-equations}

{\small {\bf Summary:}\
The potential disturbing force.
Lagrange equations for the osculating elements.
Properties of the Lagrange equations.
}\bigskip

The Gaussian perturbation equations are valid for any disturbing force
$\ve{F}$. For the special case when the perturbing force  
has a potential $R$ one has
$$
F^i={\partial\over \partial r^i}\,R\,.
$$
\noindent 
In this case one can modify the Gaussian perturbation equations so that the
disturbing potential $R$ appear in the equations instead of the
components of perturbing force $\ve{F}$. This case of potential
perturbations is very important for practical applications. For example, above we
have seen that the N-body problem can be considered as perturbed
two-body problem with disturbing force having potential
(\ref{perturbation-potential-N-body}).  

In general both $\ve{F}$ and $R$ can depend on time $t$, position
$\ve{r}$ and velocity $\dot{\ve{r}}$ of the body under study:
$R=R(t,\ve{r},\dot{\ve{r}})$.  Here we consider the simpler situation
when both $\ve{F}$ and $R$ do not depend of velocity
$\dot{\ve{r}}$. Therefore, we assume that $R=R(t,\ve{r})$.  This
covers the most important applications of potential perturbations.

First, we consider potential $R$ as function of osculating elements
$R=R(t,e_1,e_2,e_3,e_4,e_5,e_6)$.  This parametrization can be
directly derived by substituting (\ref{f-Kepler-osculating}) into
$R(t, \ve{r})$. Our goals is to replace the components $S$, $T$ and $W$
of the disturbing force in 
(\ref{Euler-equation-a})--(\ref{Euler-equation-overline-M0}) by partial derivatives 
of $R$. First, we should compute ${\partial R\over \partial e_i}$ as
functions of $S$, $T$ and $W$. One has
$$
{\partial R\over \partial e_i} =\sum_{i=1}^3 {\partial R\over \partial r^j}\,
{\partial r^j\over \partial e_i}
=\sum_{i=1}^3 F^j {\partial r^j\over \partial e_i}.
$$
\noindent 
Here $r^j$ is the $j$-th component of vector $\ve{r}$ and $F^j$ is the
$j$-th component of $\ve{F}$. Therefore one should only know
${\partial r^j\over \partial e_i}$ to compute ${\partial R\over
\partial e_i}$. 

Let us consider the example of semi-major axis $a$.  According to the
formulas of the two-body problem summarized in Section
\ref{Section-2-body-summary}, radius-vector $\ve{r}$ is linearly
proportional to $a$. Therefore, one has
${\partial r^j\over \partial a}={r^j\over a}$ and, finally,
\begin{equation}
\label{partial-R-partial-a}
{\partial R\over \partial a} 
=\sum_{i=1}^3 F^j {\partial r^j\over \partial a}
= {1\over a}\,\sum_{i=1}^3 F^j r^j={1\over a}\,\ve{F}\cdot\ve{r}=
{r\over a}\,S.
\end{equation}
\noindent 
Analogously, one gets 
\begin{eqnarray}
\label{partial-R-partial-e} 
{\partial R\over\partial e}&=&a\left(-S\,\cos v+T\,\left(1+{r\over p}\right)\sin v\right)\,,
\\ 
{\partial R\over\partial i}&=&r\,W\,\sin\left(v+\omega\right)\,, 
\\
{\partial R\over\partial \omega}&=&r\,T\,, 
\\ 
{\partial R\over\partial \Omega}&=&r\,T\,\cos i-r\,W\,\sin i\,\cos\left(v+\omega\right)\,, 
\\
\label{partial-R-partial-overlineM0} 
{\partial R\over\partial\overline{M}_0}&=&
{a\over \sqrt{1-e^2}}\, \left(e\,S\,\sin v+{p\over r}\,T\right)\,.  
\end{eqnarray} 
\noindent 
Now, one should invert
(\ref{partial-R-partial-a})--(\ref{partial-R-partial-overlineM0}) to
get three components $S$, $T$ and $W$ as functions of six partial
derivatives ${\partial R\over\partial e_i}$. Clearly, this inversion
is not unique. In principle, any possible inverse of 
(\ref{partial-R-partial-a})--(\ref{partial-R-partial-overlineM0})
can be inserted to
the Gaussian perturbation equations 
(\ref{Euler-equation-a})--(\ref{Euler-equation-overline-M0}) and lead
to a correct set of equations for osculating elements. However, the
resulting equations can be made especially simple if one requires that
the coefficients of ${\partial R\over\partial e_i}$ in the final
equations do not depend on time explicitly (e.g., do not contain
radius-vector $r$ and true anomaly $v$).  With this requirement the
inversion of
(\ref{partial-R-partial-a})--(\ref{partial-R-partial-overlineM0}) is
also unique. Finally, the equations read
\begin{eqnarray}
\label{Lagrange-a}
{d\over dt}\, a & = &
{2\over n\,a}\,{\partial R\over \partial \overline{M}_0},
\\[5pt]
\label{Lagrange-e}
{d\over dt}\, e & = &
{1-e^2\over e\,n\,a^2}\,{\partial R\over \partial \overline{M}_0}
-{\sqrt{1-e^2}\over e\,n\,a^2}\,{\partial R\over \partial \omega},
\\[5pt]
\label{Lagrange-i}
{d\over dt}\, i & = &
{\cot i\over n\,a^2\,\sqrt{1-e^2}}\,{\partial R\over \partial \omega}
-{{\rm cosec}\, i \over n\,a^2\,\sqrt{1-e^2}}\,{\partial R\over \partial \Omega},
\\[5pt]
\label{Lagrange-omega}
{d\over dt}\, \omega & = &
{\sqrt{1-e^2}\over e\,n\,a^2}\,{\partial R\over \partial e}
-{\cot i\over n\,a^2\,\sqrt{1-e^2}}\,{\partial R\over \partial i},
\\[5pt]
\label{Lagrange-Omega}
{d\over dt}\, \Omega & = &
{{\rm cosec}\, i \over n\,a^2\,\sqrt{1-e^2}}\,{\partial R\over \partial i},
\\[5pt]
\label{Lagrange-overlineM0}
{d\over dt}\, \overline{M}_0 & = &
-{2\over n\,a}\,{\partial R\over \partial a}
-{1-e^2\over e\,n\,a^2}\,{\partial R\over \partial e}\,,
\end{eqnarray}
\noindent 
where the potential $R$ is considered as function of
osculating elements and possibly time $t$: 
$R=R(t,a,e,i,\omega,\Omega,\overline{M}_0)$.
These equation were first derived by Joseph Louis Lagrange
(1736 -- 1813)
and are called \concept{Lagrange equations}.
Clearly, the Lagrange equations are fully
equivalent to the Gaussian perturbation equations.

The structure of the Lagrange equations is very interesting.
As noted above time appear in (\ref{Lagrange-a})--(\ref{Lagrange-overlineM0})
(if at all) only in $R$. The coefficients the partial derivatives depend only
on osculating elements that are constants in the two-body problem. 
Furthermore, the elements can be divided into two groups:
$\alpha_k=(a,e,i)$ and $\beta_k=(\omega,\Omega,\overline{M}_0)$, $k=1,2,3$.
Then the Lagrange equations can be symbolically written as
\begin{eqnarray}
\dot\alpha_k&=&\sum_{l=1}^3 A_{kl}\,{\partial R\over\partial\beta_l}\,,
\nonumber\\
\dot\beta_k&=&-\sum_{l=1}^3 A_{lk}\,{\partial R\over\partial\alpha_l}\,.
\nonumber
\end{eqnarray}
\noindent
This means that 
\begin{itemize}
\item[(a)] the derivatives of the elements of one group
depend only on the partial derivatives of $R$ with respect to the
elements of the other group and 
\item[(b)] if the coefficient of $\displaystyle{\partial
R\over\partial\beta_l}$ in $\dot\alpha_k$ is $A_{kl}$ then the same
coefficient (with minus) appears at $\displaystyle{\partial R\over\partial\alpha_k}$
in $\dot\beta_l$. 
\end{itemize}
\noindent
The last property should be illustrated. For example, the coefficient
of $\displaystyle{\partial R\over\partial\overline{M}_0}$ in ${d\over
dt}\,e$ is $\displaystyle{1-e^2\over e\,n\,a^2}$ (see
Eq. (\ref{Lagrange-e})). The same coefficient
$\displaystyle{1-e^2\over e\,n\,a^2}$ (with minus) appears in front of
$\displaystyle{\partial R\over\partial e}$ in ${d\over
dt}\,\overline{M}_0$ (see Eq. (\ref{Lagrange-overlineM0})). The same
symmetry holds for all coefficients.  Finally, among nine possible
$A_{kl}$ four vanish and one has only five different coefficients in
(\ref{Lagrange-a})--(\ref{Lagrange-overlineM0}). This structure of the
Lagrange equations is used to introduce the so-called canonical
elements which will not be considered here.

Osculating elements are very convenient for analytical assessments of
the effects of particular perturbations. They are also widely used for
practical representations of orbits of asteroids and artificial
satellites. Such a representation is especially efficient when the
osculating elements can be represented using simple functions of time
requiring a limited number of numerical parameters (i.e., when a
relatively ``lower-accuracy'' representation over relatively ``short''
period of time is required, the exact meaning of ``lower-accuracy''
and ``short'' depending on the problem). For example, the predicted
ephemerides (orbits) of GPS satellites are represented in form of
osculating elements using simple model for the osculating elements
(linear drift plus some selected periodic terms). The numerical
parameters are broadcasted in the GPS signals from each
satellite. Osculating elements are also used in many cases by the
Minor Planet Center of the IAU to represent orbits of asteroids and
comets.

\chapter{Three-body problem}

\section{The Lagrange solutions}

{\small {\bf Summary:}\
The case when the three-body motion can be described by the
equations of motion of the two-body problem: the five Lagrange
solutions. Examples of the Lagrange motion in the Solar system.
}\bigskip

\section{The restricted three-body problem}

{\small {\bf Summary:}\
The equations of motion. Corotating coordinates. The
equations of motion in the corotating coordinates. The Jacobi integral.
The Hill's surfaces of zero velocity.
}\bigskip

\section{Motion near the Lagrange equilibrium points}

{\small {\bf Summary:}\
The Lagrange points as equilibrium points. Libration.
The equations of motion for the motion near the Lagrange points.
The stability of the motion near the equilibrium points
}\bigskip

\chapter{Gravitational Potential of an Extended Body}

\section{Definition and expansion of the potential}

{\small {\bf Summary:}\
Gravitational potential of an extended body as an integral. The Laplace
equation. Legendre polynomials. Associated Legendre polynomials.
Expansion of the potential in terms of the associated Legendre polynomials.
}\bigskip

\subsection{Definition of the potential of an extended body}

\begin{figure}
\begin{center}
\resizebox{!}{3.0cm}{\includegraphics{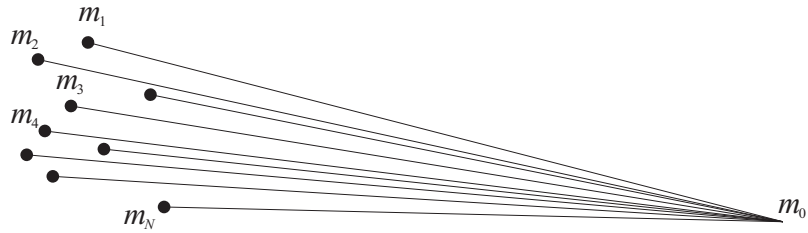}}
\end{center}
\caption[$N$ body system]{\small A system of $N$ massive bodies
acting on particle $m_0$. \label{Figure-N-body-a}}
\end{figure}

Let us again consider the equations of motion of a body with mass $m_0$
under the influence of $N$ bodies with masses $m_i$, $i=1,\dots,N$ as
we did in Section \ref{section-the-disturbing-function} (see
Fig. \ref{Figure-N-body-a}).  The first equation in
(\ref{eqm-preparation}) can be written as
\begin{equation}
\label{eqm-N-body-test-particle}
\ddot{\ve{\rho}}_0=-\sum_{i=1}^N {G\,m_i\,\ve{\rho}_{i0}\over \rho_{i0}^3}
=\grad_0\ \sum_{i=1}^N{G\,m_i\over \rho_{i0}},
\end{equation}
\noindent
where $\grad_0$ defined by (\ref{def-grad}) is the vector of partial derivatives
with respect to the components of the position $\ve{\rho}_0$ of body~$0$. Of course,
this equation does not depend on mass $m_0$. Therefore, $m_0$ can also be considered as zero.
In that case we can think of the influence of a system of $N$ massive bodies on a test
(massless) particle situated at $\ve{\rho}_0$. In this way, for any arbitrary position
$\ve{x}$ at each moment of time $t$ the gravitational potential of the system of $N$ bodies
reads
\begin{equation}
\label{potential-N-body}
U(t,\ve{x})=\sum_{i=1}^N{G\,m_i\over \rho_{ix}}, \quad \rho_{ix}=|\ve{\rho}_i(t)-\ve{x}|,
\end{equation}
where $\ve{\rho}_i(t)$ is the position of body $i$ as function of time
$t$. This potential, through Eq. (\ref{eqm-N-body-test-particle}),
gives the equations of motion of a test particle in the gravitational
field of $N$ massive bodies.

\begin{figure}
\begin{center}
\resizebox{!}{4.5cm}{\includegraphics{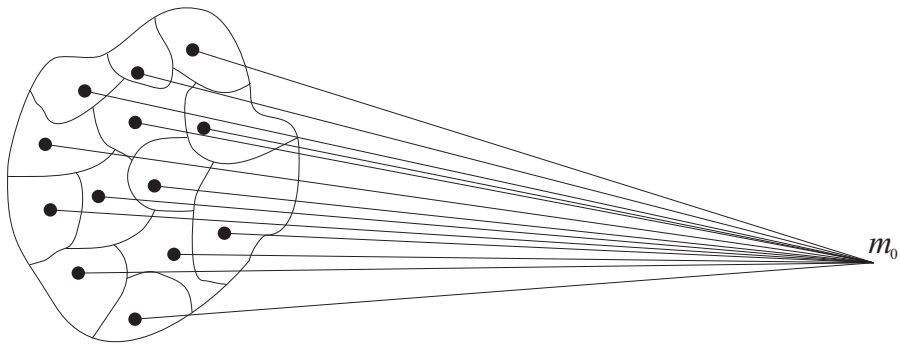}}
\end{center}
\caption[Extended body as an $N$ body system]{\small An extended body is split into a number of 
$N$ parts or cells. Each cell is then approximated by a point-like body 
situated at the center of mass of the cell and having mass $m_i$ equal to the mass of the cell.
In this way the extended body is approximated by a system of $N$ point-like bodies. 
\label{Figure-N-body-b}}
\end{figure}

Let us now consider an extended body with some continuous mass
distribution.  The task is to calculate its gravitational potential at
a point lying outside of the body. We can split the whole body into
$N$ parts or ``cells'' with some arbitrary $N\ge1$ (see
Fig. \ref{Figure-N-body-b}).  Now, if as an approximation, we replace
each cell by a point-like body situated at the center of mass of the
cell and having mass $m_i$ equal to the mass of the cell, we get a
system of $N$ point-like bodies instead of the extended body. The
gravitational potential of such a system is again given by
(\ref{potential-N-body}). Clearly, the larger is the number of cells
$N$, the better is the approximation. In order to get the potential of
an extended body we can simply consider the limit $N\to\infty$. It is
clear that such a limit means mathematically that we proceed from a
finite sum $\sum_{i=1}^N$ to an integral over the volume of the body
$\int_V$.  Masses $m_i$ of the cells should be replaced by mass
elements $dm=\sigma d^3x$, where $\sigma$ is the mass density of the
body (in general as function of time and position within the body) and
$d^3x$ is the volume element. In this way one gets
\begin{equation}
\label{potential-extended-body}
U(t,\ve{x})=\int_V {G\,dm\over \rho} 
=\int_V {G\,\sigma(t,\ve{x}^\prime)\,d^3x^\prime\over |\ve{x}-\ve{x}^\prime|},  
\end{equation}
where the integration goes over all points $\ve{x}^\prime$ inside the
body (that is, where $\sigma>0$).  This equation is valid for any
point $\ve{x}$ at which the potential is evaluated irrespective if
$\ve{x}$ is situated inside or outside the body. However, in the
following we consider only points $\ve{x}$ lying outside of the
body. Mathematically this can be written as follows. For any body
there exists such a radius $R$ so that for each points $\ve{x}$ such
that $|\ve{x}|>R$ the density $\sigma$ of the body vanishes:
\begin{equation}
\exists\,R:\,\forall\,\ve{x}:\,|\ve{x}|>R,\ \sigma(t,\ve{x})=0\,.
\end{equation}
\noindent
Radius $R$ can be called maximal radius of the body in the selected
reference system. In the following we consider $U(t,\ve{x})$ only for
such $\ve{x}$ that $|\ve{x}|>R$. It is easy to see that in this case
function $U(t,\ve{x})$ defined by (\ref{potential-extended-body})
satisfies the \concept{Laplace equation}
\begin{equation}
\label{Laplace-equation}
\Delta\,U(t,\ve{x})=0,
\end{equation}
\noindent
where $\Delta$ is the Laplace operator defined for any function $f$ as
\begin{equation}
\Delta\,f={\partial^2\over \partial x^2}\,f
+{\partial^2\over \partial y^2}\,f
+{\partial^2\over \partial z^2}\,f\,.
\end{equation}
\noindent
One can check directly that $U(t,\ve{x})$ defined by
(\ref{potential-extended-body}) satisfies
(\ref{Laplace-equation}). Indeed, considering that vectors $\ve{x}$
and $\ve{x}^\prime$ have components $(x,y,z)$ and $(x^\prime,y^\prime,z^\prime)$,
respectively, one gets
\begin{equation}
{\partial U\over\partial x}=-\int_V\,G\,\sigma(t,\ve{x}^\prime)\,{x-x^\prime\over |\ve{x}-\ve{x}^\prime|}\,d^2x^\prime
\end{equation}
\noindent
and
\begin{equation} 
{\partial^2 U\over\partial x^2}=\int_V\,G\,\sigma(t,\ve{x}^\prime)\,
{3\,(x-x^\prime)^2-|\ve{x}-\ve{x}^\prime|^2\over |\ve{x}-\ve{x}^\prime|^5}\,d^2x^\prime\,,
\end{equation}
\noindent
and analogous for ${\partial^2 U\over\partial y^2}$ and ${\partial^2
  U\over\partial z^2}$.  Summing up ${\partial^2 U\over\partial x^2}$,
${\partial^2 U\over\partial y^2}$ and ${\partial^2 U\over\partial
  z^2}$ one sees that (\ref{Laplace-equation}) is satisfied.
Functions satisfying Laplace equation are called \concept{harmonic
  functions}. Therefore, one can say that the gravitational potential
of an extended body is harmonic function outside of the body.

\begin{figure}
\begin{center}
\resizebox{!}{8.0cm}{\includegraphics{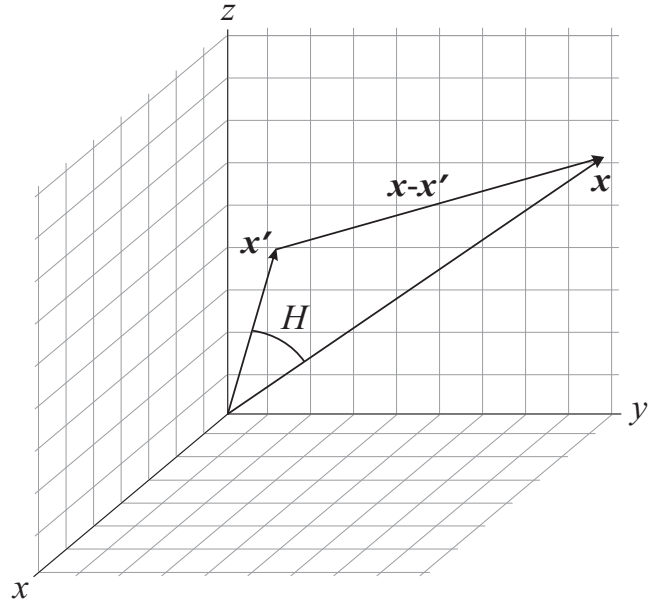}}
\end{center}
\caption[Definition of the angle H]{\small Point $\ve{x}^\prime$ is situated within the body, while
point $\ve{x}$ outside of it. The distance between two points is given by
$|\ve{x}-\ve{x}^\prime|$ and the angles between vectors $\ve{x}$ and $\ve{x}^\prime$ is denoted by
$H$ ($0\le H\le\pi$).
\label{Figure-H}}
\end{figure}

Now let us turn to the calculation of $U(t,\ve{x})$ for a given body.
If density $\sigma$ is given, for any given $t$ and $\ve{x}$ the
potential $U$ can in principle be calculated by numerical integration
of (\ref{potential-extended-body}). The involved integral over the
volume is a three-dimensional integral and its calculation is rather
complicated.  For each $\ve{x}$ the integral should be computed anew
and this is extremely inconvenient. It turns out, however, that this
integral can be represented in such a way that the whole dependence on
$\ve{x}$ is explicit. Let us write
\begin{equation}
\label{inverse-distance-1}
{1\over |\ve{x}-\ve{x}^\prime|}={1\over\sqrt{r^2+r^{\prime2}-2\,r\,r^{\prime}\,\cos H}},
\end{equation}
\noindent
where $r=|\ve{x}|$, $r^\prime=|\ve{x}^\prime|$ and $0\le H\le\pi$ is the angle between 
vectors $\ve{x}$ and $\ve{x}^\prime$ (see Fig. \ref{Figure-H}). Denoting 
$\displaystyle{r^\prime\over r}=z$ and $\cos H=x$ ($-1\le x\le 1$) one can write 
\begin{equation}
\label{inverse-distance-2}
{1\over |\ve{x}-\ve{x}^\prime|}={1\over\sqrt{r^2+r^{\prime2}-2\,r\,r^{\prime}\,\cos H}}
={1\over r}\,{1\over\sqrt{1+z^2-2\,x\,z}}.
\end{equation}
Note that notations $x$ and $z$ have now nothing to do with the components of vector $\ve{x}$.
Since we only consider points with $r=|\ve{x}|>R$ and since for any point of the body 
$r^\prime=|\ve{x}^\prime|\le R$, one has $z<1$.

\subsection{Legendre polynomials}

Mathematically, the second factor on the right-hand side of (\ref{inverse-distance-2}) can be expanded
as
\begin{equation}
\label{Legendre-polynomials-generating}
{1\over\sqrt{1+z^2-2\,x\,z}}=\sum_{n=0}^\infty P_n(x)\,z^n\,,
\end{equation}
\noindent
where $P_n(x)$ are the so-called \concept{Legendre polynomials}. The left-hand
side of (\ref{Legendre-polynomials-generating}) is called generating function for
the Legendre polynomials. Direct calculations show that
\begin{equation}
\label{Legendre-polynomials-generating-low-order}
{1\over\sqrt{1+z^2-2\,x\,z}}=1+x\,z+{3\,x^2-1\over 2}\,z^2+O(z^3)\,.
\end{equation}
\noindent
This allows one derive explicitly lower-order polynomials $P_n$:
\begin{eqnarray}
\label{Legendre-polynomials-low-order}
P_0(x)&=&1\,,
\nonumber\\
P_1(x)&=&x\,,
\nonumber\\
P_2(x)&=&{3\over 2}\,x^2-{1\over 2}\,.
\end{eqnarray}
\noindent
In principle, computing higher-order terms in (\ref{Legendre-polynomials-generating-low-order})
directly one can derive higher-order Legendre polynomials. However, the amount of computational
work grows very quickly and it is much better to proceed differently. Let us derive recurrent
relations for $P_n(x)$ starting from their definition (\ref{Legendre-polynomials-generating}).
To this end let us compute the derivative of (\ref{Legendre-polynomials-generating}) with respect to $z$:
\begin{equation}
\label{Legendre-polynomials-generating-derivative}
{d\over dz}\,{1\over\sqrt{1+z^2-2\,x\,z}}={d\over dz}\,\sum_{n=0}^\infty P_n(x)\,z^n\,.
\end{equation}
\noindent
This gives
\begin{equation}
\label{Legendre-polynomials-generating-derivative-1}
{x-z\over\left(\sqrt{1+z^2-2\,x\,z}\right)^3}=
\sum_{n=0}^\infty n\,P_n(x)\,z^{n-1}
\end{equation}
\noindent
or multiplying both sides by $1+z^2-2\,x\,z$
\begin{equation}
\label{Legendre-polynomials-generating-derivative-2}
{x-z\over \sqrt{1+z^2-2\,x\,z}}=(1+z^2-2\,x\,z)\,
\sum_{n=0}^\infty n\,P_n(x)\,z^{n-1}\,.
\end{equation}
\noindent
Using again (\ref{Legendre-polynomials-generating}) in the left-hand side one finally gets
\begin{equation}
\label{Legendre-polynomials-generating-derivative-3}
(x-z)\,\sum_{n=0}^\infty P_n(x)\,z^n=(1+z^2-2\,x\,z)\,
\sum_{n=0}^\infty n\,P_n(x)\,z^{n-1}\,.
\end{equation}
\noindent
Now, since this equations must be satisfied identically (that is, for any $z$),
the idea is to equate the coefficients at equal powers of $z$ on the left-hand and right-hand sides
of (\ref{Legendre-polynomials-generating-derivative-3}). Expanding 
(\ref{Legendre-polynomials-generating-derivative-3}) one gets
\begin{eqnarray}
\label{Legendre-polynomials-generating-derivative-4}
&&\phantom{=}
\sum_{n=0}^\infty x\,P_n(x)\,z^n
-\sum_{n=0}^\infty P_n(x)\,z^{n+1}
\nonumber \\
&&=
\sum_{n=0}^\infty n\,P_n(x)\,z^{n-1}
+\sum_{n=0}^\infty n\,P_n(x)\,z^{n+1}
-\sum_{n=0}^\infty 2\,n\,x\,P_n(x)\,z^{n}\,.
\end{eqnarray}
\noindent
Using that
\begin{eqnarray}
\label{sums-renumbering}
\sum_{n=0}^\infty P_n(x)\,z^{n+1}&=&\sum_{n=1}^\infty P_{n-1}(x)\,z^n\,,
\\
\sum_{n=0}^\infty n\,P_n(x)\,z^{n-1}&=&\sum_{n=0}^\infty (n+1)\,P_{n+1}(x)\,z^n\,,
\\
\sum_{n=0}^\infty n\,P_n(x)\,z^{n+1}&=&\sum_{n=1}^\infty (n-1)\,P_{n-1}(x)\,z^n
\end{eqnarray}
\nonumber
one gets
\begin{eqnarray}
\label{Legendre-polynomials-generating-derivative-5}
&&\phantom{=}
\sum_{n=0}^\infty x\,P_n(x)\,z^n
-\sum_{n=1}^\infty P_{n-1}(x)\,z^n
\nonumber \\
&&=
\sum_{n=0}^\infty (n+1)\,P_{n+1}(x)\,z^n
+\sum_{n=1}^\infty (n-1)\,P_{n-1}(x)\,z^n
-\sum_{n=0}^\infty 2\,n\,x\,P_n(x)\,z^{n}\,.
\end{eqnarray}
\noindent
Now, equating the coefficients at equal powers of $z$ one gets
\begin{eqnarray}
\label{Legendre-recurrencies-0}
&&x\,P_0(x)=P_1(x),\ {\rm for}\ n=0\,, 
\\
\label{Legendre-recurrencies->0}
&&x\,P_n(x)-P_{n-1}(x)=(n+1)P_{n+1}(x)+(n-1)P_{n-1}(x)-2nx P_n(x),\ {\rm for}\ n>0\,. 
\end{eqnarray}
\noindent
From (\ref{Legendre-polynomials-low-order}) one can see that the first
equation is satisfied by $P_0(x)$ and $P_1(x)$ and, therefore, gives
no additional information. The second equation can be written as
\begin{equation}
\label{Legendre-recurrent}
(n+1)\,P_{n+1}(x)=(2\,n+1)\,x\,P_n(x)-n\,P_{n-1}(x),\quad n\ge 1\,. 
\end{equation}
\noindent
This equation allows one to compute $P_n(x)$ for a given $x$ if 
$P_{n-1}(x)$ and $P_{n-2}(x)$ are given. This means that, starting from
$P_0(x)=1$ and $P_1(x)=x$ as given by (\ref{Legendre-polynomials-low-order}),
$P_n(x)$ can be computed for any $x$ and $n\ge2$. Below we will
need also the following properties of Legendre polynomials:
\begin{equation}
\label{Legendre-odd-even}
P_n(-x)=(-1)^n\,P_n(x)\,,
\end{equation}
\begin{equation}
\label{Legendre-orthogonality}
\int_{-1}^{1}P_n(x)P_m(x)dx=\biggl[\matrix{0,\quad n\neq m\cr {2\over 2n+1},\quad n=m}\,.
\end{equation}
\noindent
Eq. (\ref{Legendre-odd-even})
means that the Legendre polynomials $P_n$ are even functions of $x$
for even $n$ and odd functions for odd $n$. For $n=0$ and $n=1$ this
relation can be seen directly. For $n>1$ it is easy to prove 
(\ref{Legendre-odd-even}) from the recurrent formula (\ref{Legendre-recurrent}).
\bigskip

{\small
{\bf Exercise.}
Prove (\ref{Legendre-odd-even}) from (\ref{Legendre-recurrent}) for $n>1$ using
that (\ref{Legendre-odd-even}) is correct for $n=0$ and $n=1$.
}
\bigskip

\begin{figure}
\begin{center}
\resizebox{!}{8.0cm}{\includegraphics{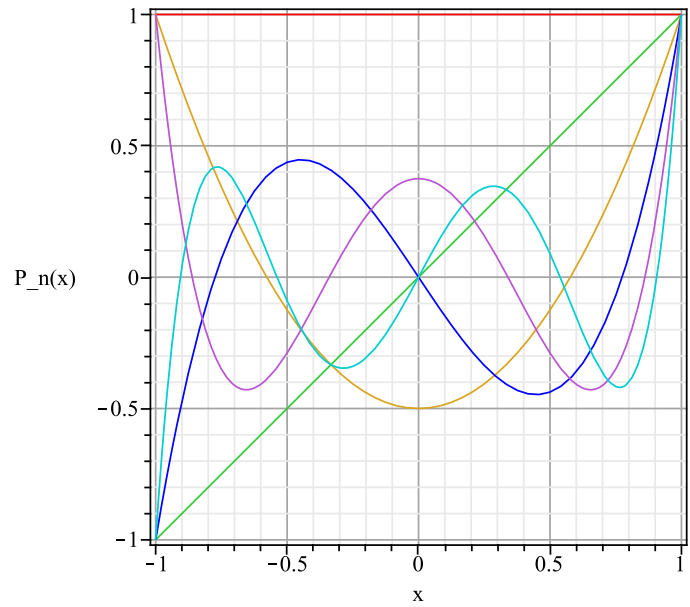}}
\end{center}
\caption[Legendre polynomials]{\small Legendre polynomials $P_n(x)$ are shown here
for $-1\le x\le 1$ and for $n=0, 1, \dots, 5$ (red, green, brown, dark blue, magenta 
and light blue, respectively).
\label{Figure-Pn}}
\end{figure}

The proof of (\ref{Legendre-orthogonality}) and
further properties of Legendre polynomials 
can be found, e.g., in Chapter 8 of Abramowitz \& Stegun (1965).
Fig. \ref{Figure-Pn} shows several first Legendre polynomials. 
One can prove that $-1\le P_n(x)\le 1$ for any $n$. 

\begin{figure}
\begin{center}
\resizebox{!}{8.0cm}{\includegraphics{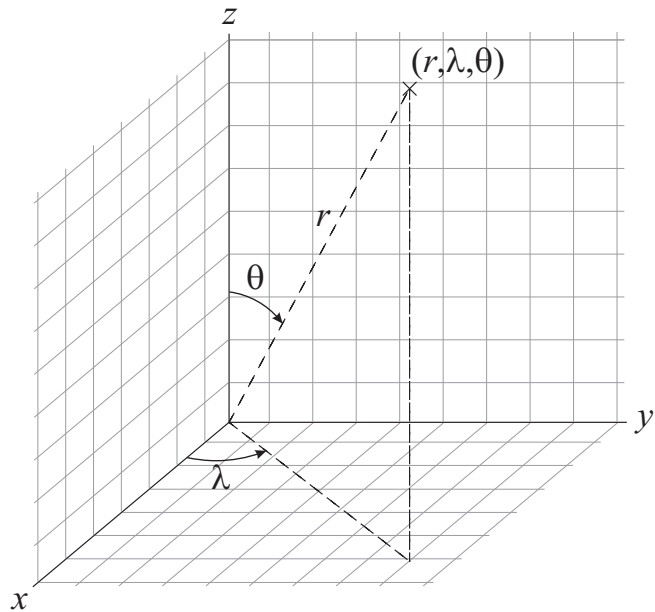}}
\end{center}
\caption[Spherical coordinates]{\small Definition of the spherical coordinates:
longitude $\lambda$ ($0\le\lambda\le2\pi$) and co-latitude $\theta$ ($0\le\theta\le\pi$).
\label{Figure-spherical}}
\end{figure}

\subsection{Expansion of the potential}

Using the expansion (\ref{Legendre-polynomials-generating}) 
in (\ref{inverse-distance-2}) the potential (\ref{potential-extended-body})
can be written as 
\begin{equation}
\label{U-expansion-preliminary}
U=\sum_{n=0}^\infty{1\over r^{n+1}}\,\int_V {r^{\prime}}^n\,P_n(\cos H)\, G\,\sigma(t,\ve{x}^\prime)\,d^3x^\prime\,.
\end{equation}
\noindent
Here we have reached our goal -- to get an expression for $U(t,\ve{x})$
in which the dependence on $\ve{x}$ is explicit -- only
partially. Indeed, the dependence on $r=|\ve{x}|$ is explicitly
written since all quantities under the integral are independent of
$r$. On the other hand $\cos H$ under the integral does depend on the
orientation of $\ve{x}$, that is, on the unit vector $\ve{x}/r$. Such
unit vector can be characterized for example by the corresponding
spherical coordinates.  Introducing longitude $\lambda$
($0\le\lambda\le2\pi$) and co-latitude $\theta$ ($0\le\theta\le\pi$)
(see Fig. \ref{Figure-spherical}) -- one has
\begin{equation}
\cos H=\cos\theta\,\cos\theta^\prime+\sin\theta\,\sin\theta^\prime\,\cos(\lambda-\lambda^\prime),
\end{equation}
\noindent
where $\lambda$ and $\theta$ correspond to $\ve{x}$ while $\lambda^\prime$ and $\theta^\prime$ correspond to 
$\ve{x}^\prime$. This formula can be easily derived using
\begin{equation}
\ve{x}=\pmatrix{r\,\cos\lambda\,\sin\theta\cr r\,\sin\lambda\,\sin\theta\cr r\,\cos\theta}\,,
\end{equation}
\noindent
similar formula for $\ve{x}^\prime$, and noting that $\cos
H=\displaystyle{\ve{x}\cdot\ve{x}^\prime\over r\,r^\prime}$, where
'$\cdot$' denotes scalar product of two vectors. Furthermore, one has
\begin{equation}
\label{Legendre-cos-H}
P_n(\cos H)=\sum_{k=0}^n(2-\delta_{k0})\,{(n-k)!\over (n+k)!}\,P_{nk}(\cos\theta)\,P_{nk}(\cos\theta^\prime)\,
\cos k(\lambda-\lambda^\prime),
\end{equation}
\noindent
where 
\begin{equation}
\label{delta}
\delta_{k0}=\biggl[\matrix{1,\quad k=0\cr 0,\quad k>0}
\end{equation}
\noindent
and $P_{nk}(x)$ are the \concept{associated Legendre polynomials} defined as
\begin{equation}
%change: delete (-1)^k
P_{nk}(x)={\left(1-x^2\right)}^{k/2}\,{d^k\over dx^k}\,P_n(x),\quad 0\le k\le n.
\end{equation}
\noindent
Clearly, $P_{n0}(x)=P_n(x)$. Mathematical proof of
(\ref{Legendre-cos-H}) and further properties of the associated
Legendre polynomials can be found, e.g., in Chapter 8 of Abramowitz \&
Stegun (1965). Note that in the literature one can find different sign
conventions related to the associated Legendre polynomials, in
particular regarding an additional factor $(-1)^k$.

Substituting (\ref{Legendre-cos-H}) into (\ref{U-expansion-preliminary}) one finally
gets the expansion of the gravitational potential of an extended body in the form
\begin{equation}
\label{U-expansion-final}
U=\sum_{n=0}^\infty\sum_{k=0}^n\,G\,{P_{nk}(\cos\theta)\over r^{n+1}}\,\left(C_{nk}\cos k\lambda+S_{nk}\sin k\lambda\right).
\end{equation}
\noindent
Coefficients $C_{nk}$ and $S_{nk}$ are real numbers that fully characterize the gravitational
potential of the body and defined as
\begin{eqnarray}
\label{C-nk-def}
C_{nk}=(2-\delta_{k0})\,{(n-k)!\over (n+k)!}\,\int_V {r^\prime}^n\,P_{nk}(\cos\theta^\prime)\,\cos k\lambda^\prime\,
\sigma(t,\ve{x}^\prime)\,d^3x^\prime\,,
\\
\label{S-nk-def}
S_{nk}=(2-\delta_{k0})\,{(n-k)!\over (n+k)!}\,\int_V {r^\prime}^n\,P_{nk}(\cos\theta^\prime)\,\sin k\lambda^\prime\,
\sigma(t,\ve{x}^\prime)\,d^3x^\prime\,.
\end{eqnarray}
\noindent
It is clear that $C_{nk}$ and $S_{nk}$ do not depend on $\ve{x}$, but
only on the mass distribution inside the body. Coefficient $S_{n0}=0$ for any $n$ because of
the factor $\sin k\lambda^\prime$ under the integral in its definition (the latter factor vanishes for $k=0$). 
One has, therefore, $2n+1$ coefficients $C_{nk}$ and $S_{nk}$ characterizing the gravitational field of
an extended body for each $n$. Note that $C_{nk}$ and $S_{nk}$ are not dimensionless. Their dimensionality 
is ${\rm kg}\cdot{\rm m}^n$. 

The expansion (\ref{U-expansion-final}) for $U(t,\ve{x})$ gives
explicitly the dependence of $U$ on the point $\ve{x}$ at which $U$
should be evaluated. This is done using spherical coordinates
$(r,\lambda,\theta)$ of $\ve{x}$. We have thus achieved our goal.

\section{First terms of the expansion}

{\small {\bf Summary:}\
The mass of the body. The relations between the potential and the choice of the origin and the 
orientation of the coordinate system.
}\bigskip

Up to now the reference system where we described the body and its
gravitational potential $U$ was arbitrary. Here we show that the
lower-order terms in (\ref{U-expansion-final}) are closely related to
the choice of the origin and the orientation of the reference
system. In the following we consider that $\displaystyle{U=\sum_{n=0}^\infty U_n}$,
where $U_n$ are all terms in (\ref{U-expansion-final}) or
(\ref{U-expansion-preliminary}) corresponding to a given $n$, that is
\begin{eqnarray}
\label{U-n-def}
U_n&=&{1\over r^{n+1}}\,G\,\int_V {r^{\prime}}^n\,P_n(\cos H)\, \sigma(t,\ve{x}^\prime)\,d^3x^\prime
\nonumber\\
&=&\sum_{k=0}^n\,G\,{P_{nk}(\cos\theta)\over r^{n+1}}\,\left(C_{nk}\cos k\lambda+S_{nk}\sin k\lambda\right)
\,.
\end{eqnarray}
\noindent
Note that $U_n$ is the part of potential $U$ falling off as $\displaystyle{1\over r^{n+1}}$ for $r$ 
going to infinity.

\subsection{The term for $n=0$}
\label{section-n=0}

For $n=0$ the first line of (\ref{U-n-def}) gives (note that $P_0(\cos H)=1$)
\begin{equation}
\label{U-0}
U_0={1\over r}\,\int_V G\,\sigma(t,\ve{x}^\prime)\,d^3x^\prime={G\,M\over r}\,,
\end{equation}
\noindent
where $M=\int_V \sigma(t,\ve{x}^\prime)\,d^3x^\prime$ is the total mass of the body.
From the second line of (\ref{U-n-def}) we see that for $n=0$ there is only one
coefficient $C_{00}$. Definition (\ref{C-nk-def}) again shows that 
\begin{equation}
\label{C-00}
C_{00}=M\,.  
\end{equation}
\noindent
Thus, we get an important result: the main $r^{-1}$ term of the gravitational potential
of an arbitrary extended body is the same as if the body would be a mass point with mass 
$M$ equal to the total mass of the extended body. This term cannot be affected by
any change of the reference system.

\subsection{The terms for $n=1$}

For $n=1$ the first line of (\ref{U-n-def}) gives (note that $P_1(\cos H)=\cos H$ and 
$r^\prime\,\cos H=\displaystyle{\ve{x}\cdot\ve{x}^\prime\over r}$)
\begin{equation}
\label{U-1}
U_1={1\over r^2}\,G\,\int_V {\ve{x}\cdot\ve{x}^\prime\over r}\,\sigma(t,\ve{x}^\prime)\,d^3x^\prime=
{G\,M\over r^3}\,\ve{x}\cdot \ve{x}_c,
\end{equation}
where $\ve{x}_c=\displaystyle{{1\over M}\,\int_V
\sigma(t,\ve{x}^\prime)\,\ve{x}^\prime\,d^3x^\prime}$ are the
coordinates of the center of mass of the body in the chosen reference
system. On the other hand, considering the second line of 
(\ref{U-n-def}) one sees that $U_1$ is characterized by $C_{10}$, $C_{11}$, and
$S_{11}$ (recall that $S_{n0}=0$ for any $n$). These three coefficients
are equivalent to the three components of $\ve{x}_c$. Indeed, it is easy to see that
\begin{eqnarray}
\label{x-c-C1k-S1k}
C_{10}&=&M\,z_c\,,
\nonumber \\
%change: remove "-"
C_{11}&=&M\,x_c\,,
\nonumber \\
%change: remove "-"
S_{11}&=&M\,y_c\,,
\end{eqnarray}
\noindent
where $(x_c,y_c,z_c)$ are the components of vector $\ve{x}_c$.
It is clear that choosing the reference system in such a way
that its origin coincides with the center of mass of the body under
study, one gets $\ve{x}_c=0$. In this case, we have $U_1=0$
and, therefore, $C_{10}=C_{11}=S_{11}=0$. Usually, this possibility is indeed
used and one puts $C_{10}$, $C_{11}$, $S_{11}$ to zero.

\subsection{The terms for $n=2$}

For $n=2$ the first line of (\ref{U-n-def}) gives (note that $P_2(\cos H)={3\over2}\cos^2H-{1\over 2}$)
\begin{eqnarray}
\label{U-2}
U_2&=&{1\over r^3}\,G\,\int_V {r^\prime}^2\left({3\over2}\cos^2H-{1\over 2}\right)\,\sigma(t,\ve{x}^\prime)\,d^3x^\prime
\nonumber\\
&=&{1\over r^3}\,G\,\int_V {3\,{\left(\ve{x}\cdot\ve{x}^\prime\right)}^2
-r^2\,{r^\prime}^2\over 2\,r^2}\,\sigma(t,\ve{x}^\prime)\,d^3x^\prime
\nonumber\\
&=&-{3\over 2}\,G\,\sum_{i=1}^3\sum_{j=1}^3 \hat I^{ij}\,{x^i\,x^j\over r^5}\,,
\end{eqnarray}
\noindent
where $\hat I^{ij}$ is the trace-free part of the tensor of inertia
$I^{ij}$ of the body.  Namely, $\hat I^{ij}=I^{ij}-{1\over
  3}\,\delta^{ij}\,I^{kk}$, where $I^{kk}=I^{11}+I^{22}+I^{33}$ is the
trace of $I^{ij}$ and $\delta^{ij}$ are the components of identity
matrix ($\delta^{ij}=1$ for $i=j$ and $\delta^{ij}=0$ for $i\neq
j$). The tensor of inertia $I^{ij}$ has its usual definition
\begin{eqnarray}
\label{I-ij}
I^{ij}=\int_V \left(\delta^{ij}\,|\ve{x}|^2-x^i\,x^j\right)\,\sigma(t,\ve{x})\,d^3x\,.
\end{eqnarray}

For $n=2$ there are five coefficients $C_{nk}$ and $S_{nk}$: 
$C_{20}$, $C_{21}$, $C_{22}$, $S_{21}$, and $S_{22}$. On the other hand 
the symmetric trace-free matrix $\hat I^{ij}$ can be written as
\begin{equation}
\label{hat-I-ij-matrix}
\hat I^{ij}=-M\,R^2\,\pmatrix{a & b & c\cr b & d & e\cr c& e& -a-d}
\end{equation}
and also has five independent components.
Dimensionless numbers $a$, $b$, $c$, $d$ and $e$ fully characterize
the gravitational potential $U_2$ and are directly related to the five
coefficients $C_{20}$, $C_{21}$, $C_{22}$, $S_{21}$, and $S_{22}$. One can
demonstrate that
\begin{eqnarray}
\label{Cnk-Snk-hatI}
C_{20}&=& -{3\over 2}\,(a+d)\,M\,R^2\,,
\nonumber\\
%change: remove "-"
C_{21}&=& c\,M\,R^2\,,
\nonumber\\
C_{22}&=& {1\over 4}\,(a-d)\,M\,R^2\,,
\nonumber\\
%change: remove "-"
S_{21}&=& e\,M\,R^2\,,
\nonumber\\
S_{22}&=& {1\over 2}\,b\,M\,R^2
\end{eqnarray}
\noindent
or inverting
\begin{eqnarray}
\label{Cnk-Snk-hatI-inverse}
M\,R^2\,a&=& -{1\over 3}\,C_{20}+2\,C_{22}\,,
\nonumber\\
M\,R^2\,b&=& 2\,S_{22}\,,
\nonumber\\
%change: remove "-"
M\,R^2\,c&=& C_{21}\,,
\nonumber\\
M\,R^2\,d&=& -{1\over 3}\,C_{20}-2\,C_{22}\,,
\nonumber\\
%change: remove "-"
M\,R^2\,e&=& S_{21}\,.
\end{eqnarray}
\noindent
Since we consider rectangular Cartesian right-handed system there are
only two sorts of freedom to define it. First, the choice of the origin
used for $n=1$ above. Second, the choice of spatial orientation of the
axes.  Clearly, the components of $\hat I^{ij}$ depend on the
orientation of the coordinate system. Generally, a rotation in
three-dimensional space is defined by three parameters. Therefore,
three coefficients among $C_{20}$, $C_{21}$, $C_{22}$, $S_{21}$, and
$S_{22}$ can be made zero by choosing some special orientation of the
coordinate system.  For example, the matrix $\hat I^{ij}$ can be
diagonalized by a suitable rotation to get $C_{21}$, $S_{21}$ and
$S_{22}$ to zero (see (\ref{Cnk-Snk-hatI}) and consider
that for a diagonal matrix $b=c=e=0$ in (\ref{hat-I-ij-matrix})). 
More detailed analysis shows that the orientation
can be chosen in such a way that $C_{21}$, $S_{21}$ and either
$C_{22}$ or $S_{22}$ vanish.  This possibility is usually not used
since the orientation of the reference system is fixed from the
consideration of continuity and convenience.

Let us note that it makes no sense to consider separately terms $U_n$
for $n>2$ since the freedom in the definition of the coordinate system
is already exhausted.

\section{Symmetric bodies}
\label{section-symmetric-bodies}

{\small {\bf Summary:}\
Axial symmetry. Axial symmetry and the symmetry between the north and
the south. Spherical symmetry. Symmetry with respect to three coordinate planes.
}\bigskip

In this Section we will simplify the general expansion 
(\ref{U-expansion-final}) for the gravitational potential $U$ for the
case of extended bodies having several sorts of symmetries. This consideration
allows one to understand which coefficients $C_{nk}$ and $S_{nk}$ describe 
which properties of the mass distribution within the body.

\subsection{Axial symmetry}
\label{section-axial-symmetry}

Let us first consider a body that is symmetric with respect to some axis and
let us choose that symmetry axis as z-axis of our coordinate system. Then the symmetry
means that the density $\sigma$ does not depend on the spherical coordinate $\lambda$
(see Fig. \ref{Figure-spherical}): $\sigma\neq\sigma(\lambda)$. Therefore, $\sigma=\sigma(t,r,\theta)$
and the integrals (\ref{C-nk-def})--(\ref{S-nk-def}) can be written as
\begin{eqnarray}
\label{C-nk-axial}
C_{nk}&=&(2-\delta_{k0})\,{(n-k)!\over (n+k)!}\,
\nonumber\\
&&
\times \left(\int_0^R dr^\prime {r^\prime}^{n+2} \int_0^\pi d\theta^\prime 
\sin\theta^\prime\,P_{nk}(\cos\theta^\prime)\,\sigma(t,r^\prime,\theta^\prime)\right)
\times\left(\int_0^{2\pi} d\lambda^\prime\,\cos k\lambda^\prime\right)\,,
\\
\label{S-nk-axial}
S_{nk}&=&(2-\delta_{k0})\,{(n-k)!\over (n+k)!}\,
\nonumber\\
&&
\times \left(\int_0^R dr^\prime {r^\prime}^{n+2} \int_0^\pi d\theta^\prime 
\sin\theta^\prime\,P_{nk}(\cos\theta^\prime)\,\sigma(t,r^\prime,\theta^\prime)\right)
\times\left(\int_0^{2\pi} d\lambda^\prime\,\sin k\lambda^\prime\right)\,.
\end{eqnarray}
Here we used the expression for the volume element in spherical coordinates: 
$d^3x^\prime={r^\prime}^2\,\sin\theta^\prime\,dr^\prime\,d\theta^\prime\,d\lambda^\prime$. 
The last integral in (\ref{C-nk-axial})--(\ref{S-nk-axial}) over
$d\lambda^\prime$ is zero for any $k$ in $S_{nk}$ and for $k>0$ in $C_{nk}$. Therefore, for an axially
symmetric body $C_{nk}=0$ for $k>0$ and $S_{nk}=0$ for any $k$. This holds for any $n$. It means that
only coefficients $C_{n0}$ are not zero and fully characterize the gravitational field of
the body. The expansion of $U$ in this case takes the form
\begin{equation}
\label{U-expansion-final-axial}
U=\sum_{n=0}^\infty\,G\,C_{n0}\,{P_n(\cos\theta)\over r^{n+1}}\,.
\end{equation}
\noindent
Therefore, we conclude that $C_{nk}$ for any $n$ and $k>0$ and $S_{nk}$ for any $n$ and $k$
characterize the deviation of the body from axial symmetry.

\subsection{Axial symmetry and the symmetry about $xy$-plane}

Let us now consider the case when, in addition to the axial symmetry
(about $z$-axis), the body is symmetric about the
$xy$-plane (one can speak also of ``symmetry between the north and
the south'').  In this case density has the property that $\sigma(t,r,\theta)=\sigma(t,r,\pi-\theta)$.
Using (\ref{C-nk-axial}) for $k=0$ one gets
\begin{eqnarray}
\label{C-nk-axial-xy}
C_{n0}&=&2\pi\,\int_0^R dr^\prime {r^\prime}^{n+2} \int_0^\pi d\theta^\prime 
\sin\theta^\prime\,P_n(\cos\theta^\prime)\,\sigma(t,r^\prime,\theta^\prime)\,,
\end{eqnarray}
\noindent
The integral over $d\theta^\prime$ can be split into two parts:
\begin{eqnarray}
\label{integral-split}
\int_0^{\pi/2} d\theta^\prime 
\sin\theta^\prime\,P_n(\cos\theta^\prime)\,\sigma(t,r^\prime,\theta^\prime)
+
\int_{\pi/2}^\pi d\theta^\prime 
\sin\theta^\prime\,P_n(\cos\theta^\prime)\,\sigma(t,r^\prime,\theta^\prime)\,.
\end{eqnarray}
\noindent
Now, using the symmetry of the body one can show that these two
integrals are related to each other.
Indeed, let us introduce $\widetilde\theta^\prime=\pi-\theta^\prime$ and 
replace $\theta^\prime$ through $\widetilde\theta^\prime$ in the integral
from $\pi/2$ to $\pi$:
\begin{eqnarray}
\label{integral-symmetry}
&&\int_{\pi/2}^\pi d\theta^\prime 
\sin\theta^\prime\,P_n(\cos\theta^\prime)\,\sigma(t,r^\prime,\theta^\prime)
\nonumber\\
&&
\qquad
=
-\int_{\pi/2}^0d\widetilde\theta^\prime 
\sin\widetilde\theta^\prime\,(-1)^n\,P_n(\cos\widetilde\theta^\prime)\,\sigma(t,r^\prime,\widetilde\theta^\prime)
\nonumber\\
&&
\qquad
=
(-1)^n\,\int_0^{\pi/2}d\widetilde\theta^\prime 
\sin\widetilde\theta^\prime\,P_n(\cos\widetilde\theta^\prime)\,\sigma(t,r^\prime,\widetilde\theta^\prime)
\nonumber\\
&&
\qquad
=
(-1)^n\,\int_0^{\pi/2}d\theta^\prime 
\sin\theta^\prime\,P_n(\cos\theta^\prime)\,\sigma(t,r^\prime,\theta^\prime).
\nonumber
\end{eqnarray}
\noindent
Here we used that $d\theta^\prime=-d\widetilde\theta^\prime$,
$\sigma(t,r^\prime,\theta^\prime)=\sigma(t,r^\prime,\pi-\widetilde\theta^\prime)=\sigma(t,r^\prime,\widetilde\theta^\prime)$,
$\sin\theta^\prime=\sin(\pi-\widetilde\theta^\prime)=\sin\widetilde\theta^\prime$,
$\cos\theta^\prime=\cos(\pi-\widetilde\theta^\prime)=-\cos\widetilde\theta^\prime$,
and that, according to (\ref{Legendre-odd-even}),
$P_n(-\cos\widetilde\theta^\prime)=(-1)^n\,P_n(\cos\widetilde\theta^\prime)$. Finally,
for the boundaries of integration we get that $\theta^\prime=\pi/2$
and $\theta^\prime=\pi$ correspond to $\widetilde\theta^\prime=\pi/2$
and $\widetilde\theta^\prime=0$, respectively.  In the last equality
we simply replaced $\widetilde\theta^\prime$ by $\theta^\prime$ as a
change of notation.

It is, therefore, clear that with this symmetry $C_{n0}=0$ for odd $n$.
The expansion of $U$ thus reads
\begin{equation}
\label{U-expansion-final-axial-xy}
U=\sum_{n=0}^\infty\,G\,C_{2n,0}\,{P_{2n}(\cos\theta)\over r^{2n+1}}\,.
\end{equation}
\noindent
We conclude that $C_{2n+1,0}$ characterize the deviation of the (axially symmetric) body from 
the symmetry about the $xy$-plane.

\subsection{Spherical symmetry}

We assume now an even stronger symmetry. Namely the spherical one. In this case
the density only depends on the radial coordinate and dies not depend
on $\lambda$ and $\theta$: $\sigma=\sigma(t,r)$. In this case 
from (\ref{C-nk-axial-xy}) one has
\begin{eqnarray}
\label{C-nk-axial-spherical}
C_{n0}&=&2\pi\,\left(\int_0^R dr^\prime {r^\prime}^{n+2} \sigma(t,r^\prime)\right)\times\left(\int_0^\pi d\theta^\prime 
\sin\theta^\prime\,P_n(\cos\theta^\prime)\right)\,,
\end{eqnarray}
\noindent
The integral over $d\theta^\prime$ can be written as
\begin{eqnarray}
\label{integral-theta-spherical}
\int_0^\pi d\theta^\prime  \sin\theta^\prime\,P_n(\cos\theta^\prime)
=-\int_0^\pi P_n(\cos\theta^\prime)\,d\cos\theta^\prime
=-\int_{-1}^{1} P_n(s)\,ds,
\end{eqnarray}
\noindent
where $s=\cos\theta^\prime$. According to (\ref{Legendre-orthogonality})
where we can put $m=0$ and $P_m(x)=1$ one gets
\begin{eqnarray}
\label{integral-P-n}
\int_{-1}^{1} P_n(s)\,ds=\left[
\matrix{2,\quad n=0\cr 0,\quad n\ge1}\right.\,.
\end{eqnarray}
\noindent
Therefore, only $C_{00}=M$ (see Section \ref{section-n=0}) does not vanish for 
a spherical symmetric body and the gravitational field of a spherically
symmetric body reads
\begin{equation}
\label{U-expansion-final-spherical}
U={G\,C_{00}\over r}={G\,M\over r}\,.
\end{equation}
\noindent
Note that we have now proved this formula for any spherically
symmetric distribution of the density $\sigma=\sigma(t,r)$.  This is a
substantial generalization with respect to the case of point-like
bodies. We conclude that $C_{nk}$ for $n>0$ and $S_{nk}$ for any $n$
and $k$ characterize the deviation of the body from the spherical
symmetry.

\subsection{Symmetry with respect to three coordinate planes}

One more interesting case is a body symmetric with respect to all three
coordinate planes: $xy$-plane, $xz$-plane and $yz$-plane. For example,
a triaxial ellipsoid possesses such a symmetry. One can demonstrate
that in this case the expansion of $U$ reads
\begin{equation}
\label{U-expansion-final-xyz}
U=G\,\sum_{n=0}^\infty\,{1\over r^{2n+1}}\sum_{k=0}^n\,C_{2n,2k}\,P_{2n,2k}(\cos\theta)\cos2k\lambda\,.
\end{equation}
\noindent
We conclude that $C_{nk}$ with odd $n$ and/or $k$ and all $S_{nk}$
characterize the deviation of the body from the symmetry about all
three coordinate planes. The proof of 
(\ref{U-expansion-final-xyz}) can be found, e.g., in Chandrasekhar (1987).

\section{Spherical functions and the classification of the coefficients}

{\small {\bf Summary:}\
Definition of spherical functions. The expansion of the
gravitational potential in terms of spherical functions.
The principal part of the potential: spherically-symmetric
gravitational field. Zonal coefficients. Sectorial and tesseral
coefficients.
}\bigskip

Expansion (\ref{U-expansion-final}) is a special case of general expansion of
an arbitrary complex function of spherical coordinates in terms of spherical functions. Namely,
the factors $P_{nk}(\cos\theta)$, $\cos k\lambda$ and $\sin k\lambda$ can be combined
into a single complex function:
\begin{equation}
\label{Ykn}
%change: add (-1)^k
Y_{nk}(\lambda,\theta)=(-1)^k\,\sqrt{{2n+1\over 4\pi}\,{(n-k)!\over(n+k)!}}\,P_{nk}(\cos\theta)\,
e^{\mi k\lambda}\,.
\end{equation}
\noindent
Note that $e^{\mi k\lambda}=\cos k\lambda+\mi\,\sin
k\lambda$, $\mi$ being the imaginary unit $\mi=\sqrt{-1}$. 
Functions $Y_{nk}$ depend only on $\lambda$ and $\theta$
and are, therefore, defined on a sphere (the radius of which plays no
role). The numerical factor under the square root in (\ref{Ykn}) is
only for a specific normalization that the integral $\displaystyle{\int_0^\pi
d\theta\,\sin\theta\,\int_0^{2\pi}d\lambda\,Y_{nk}(\lambda,\theta)\,Y^*_{n^\prime
  k^\prime}(\lambda,\theta)}$ is equal to 1 if $n=n^\prime$ and
$k=k^\prime$ and vanishes otherwise (``$*$'' meaning complex conjugate). 
This factor does not play any role in 
the following and will not be further discussed.  Functions $Y_{nk}$
constitute full functional basis on a sphere. It means that any
complex function $g(\lambda,\theta)$ defined on a sphere can be
represented as
\begin{equation}
\label{Ykn-expansion}
g(\lambda,\theta)=\sum_{l=0}^\infty\sum_{m=-l}^l\,A_{lm}\,Y_{lm}(\lambda,\theta),
\end{equation}
\noindent
where $A_{lm}$ are some complex numbers. If function
$g(\lambda,\theta)$ is real, coefficients $A_{lm}$ have some symmetry
properties so that the whole sum on the right-hand side of
(\ref{Ykn-expansion}) remains real. Let us also note that since $\cos
k\left(\lambda-{\pi\over 2k}\right)=\sin k\lambda$ the real and imaginary parts of $Y_{nk}$ are
related to each other in a simple way: $\Im\,
Y_{nk}(\lambda,\theta)=\Re\,Y_{nk}\left(\lambda-{\pi\over 2k},\theta\right)$.

Spherical functions are convenient to discuss the character of the
coefficients $C_{nk}$ and $S_{nk}$.  The expansion
(\ref{U-expansion-final}) is often written in the form
\begin{eqnarray}
\label{U-expansion-final-Re}
U&=&{G\,M\over r}\,\left\{\,\left(1-\sum_{n=2}^\infty J_n\,{\left(R\over r\right)}^n\,P_n(\cos\theta)\right)\right.
\nonumber\\
&&\qquad\qquad
\left.
+\sum_{n=2}^\infty\sum_{k=1}^n\,{\left(R\over r\right)}^n\,P_{nk}(\cos\theta)\,\left(\overline C_{nk}\cos k\lambda+
\overline S_{nk}\sin k\lambda\right)
\right\},
\end{eqnarray}
\noindent
where $R$, as before, is the radius of a sphere encompassing the body
and $M=C_{00}$ is the total mass of the body.  Here one uses the
coordinate system the origin of which coincides with the center of
mass of the body, so that all terms in (\ref{U-expansion-final}) with
$n=1$ vanish and do not appear in (\ref{U-expansion-final-Re}). 
Coefficients $J_n$, $\overline C_{nk}$ and $\overline
S_{nk}$ are dimensionless real numbers that are trivially related to
$C_{nk}$ and $S_{nk}$:
\begin{eqnarray}
\label{Jn}
J_n&=&-{1\over M\,R^n}\,C_{n0}\,,
\\
\label{overline-Cnk}
\overline C_{nk}&=&{1\over M\,R^n}\,C_{nk}\,,
\\
\label{overline-Snk}
\overline S_{nk}&=&{1\over M\,R^n}\,S_{nk}\,.
\end{eqnarray}

\begin{figure}
\begin{center}
\begin{tabular}{cc}
\resizebox{!}{6.0cm}{\includegraphics{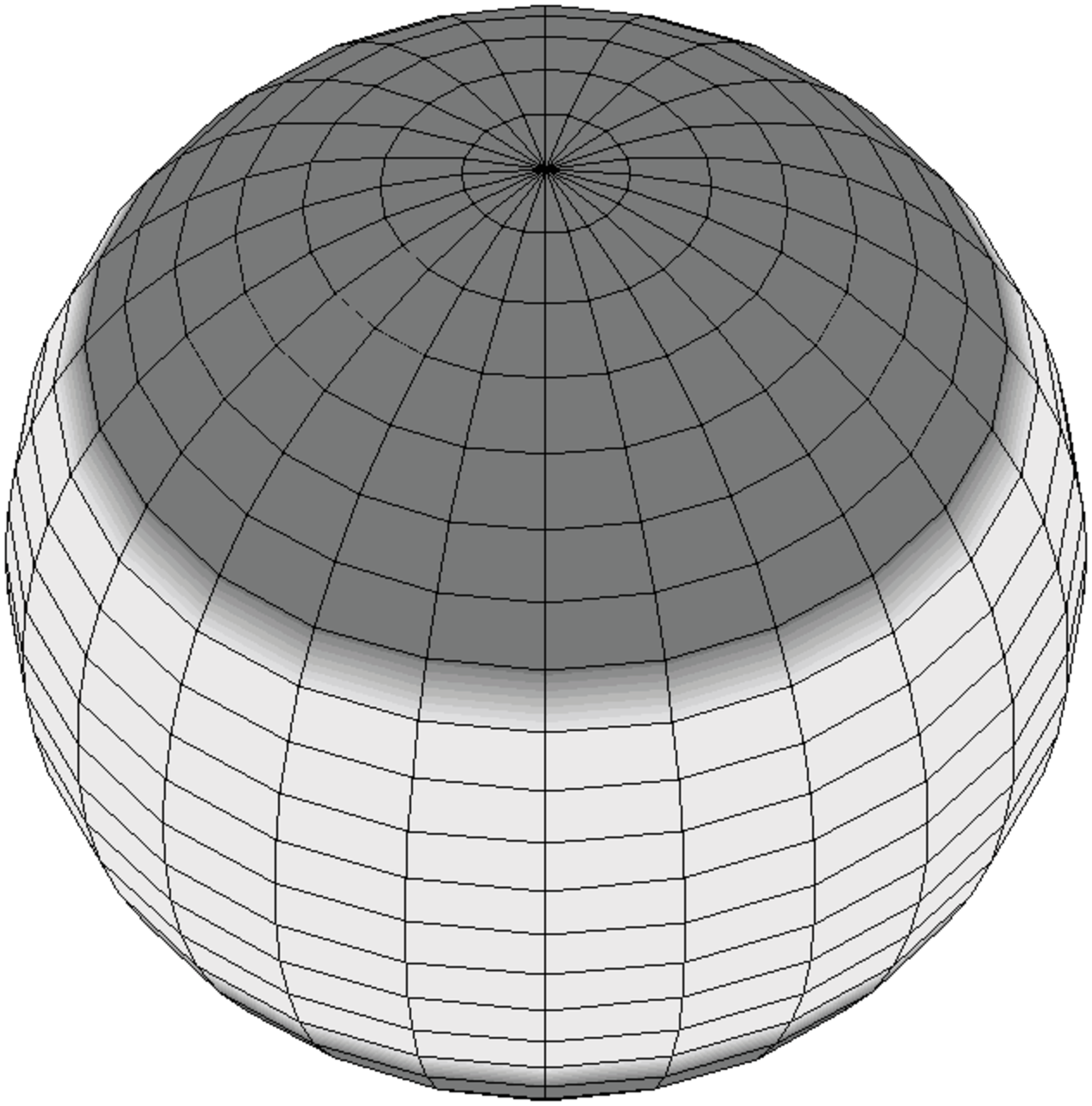}}&\resizebox{!}{6.0cm}{\includegraphics{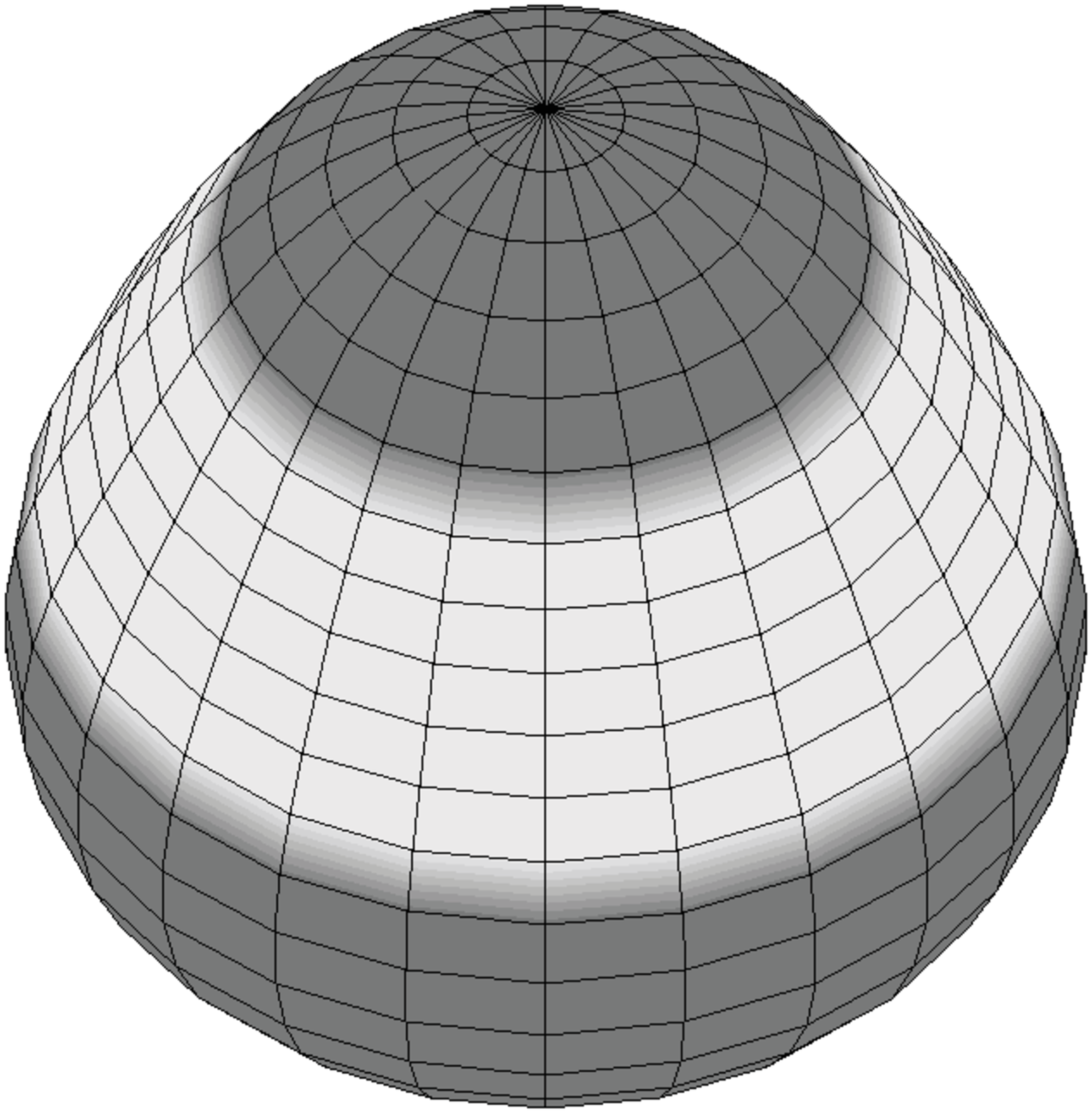}} \\
$n=2$\ $k=0$ & $n=3$\ $k=0$ \\[15pt] 
\resizebox{!}{6.0cm}{\includegraphics{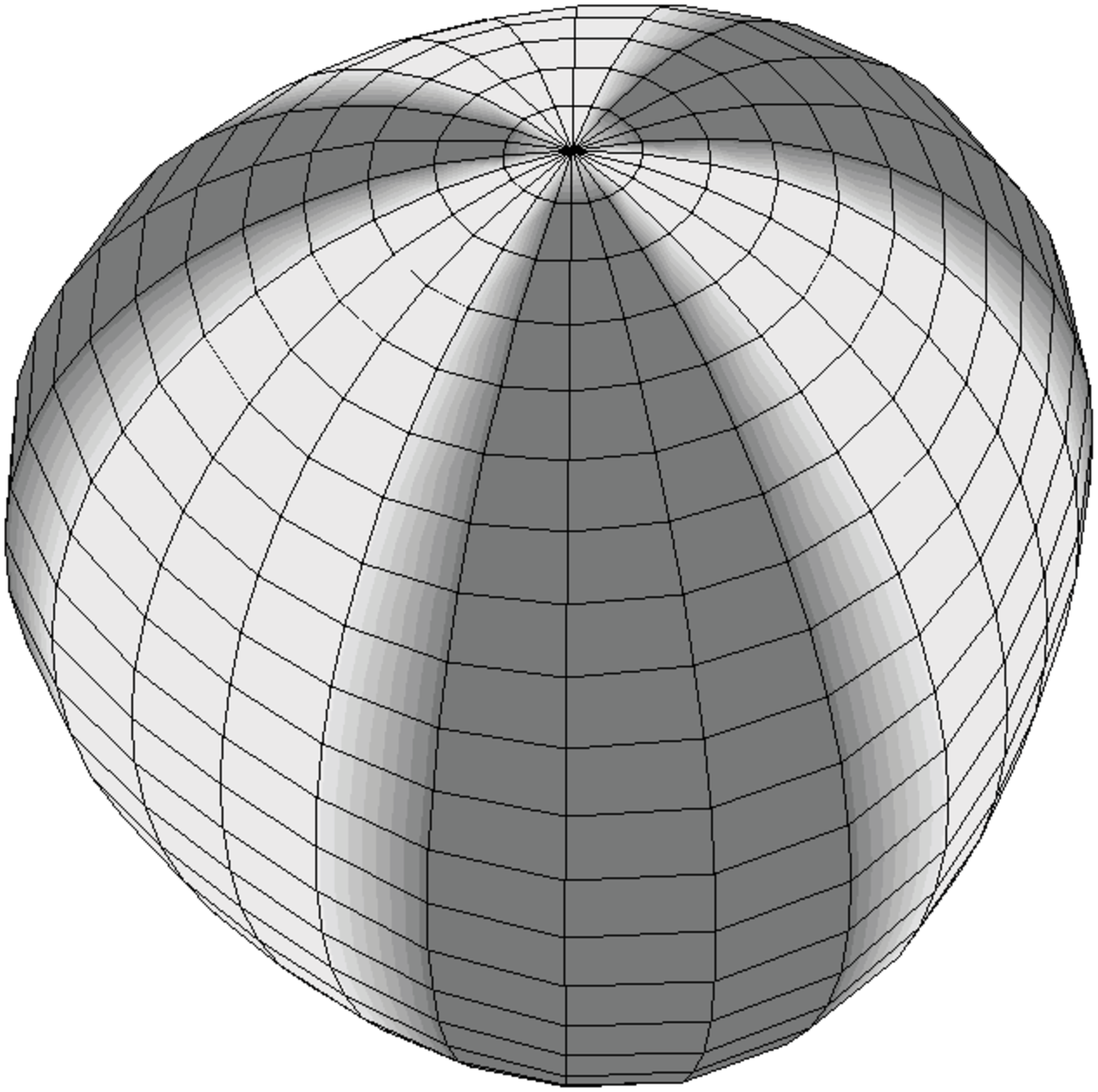}}&\resizebox{!}{6.0cm}{\includegraphics{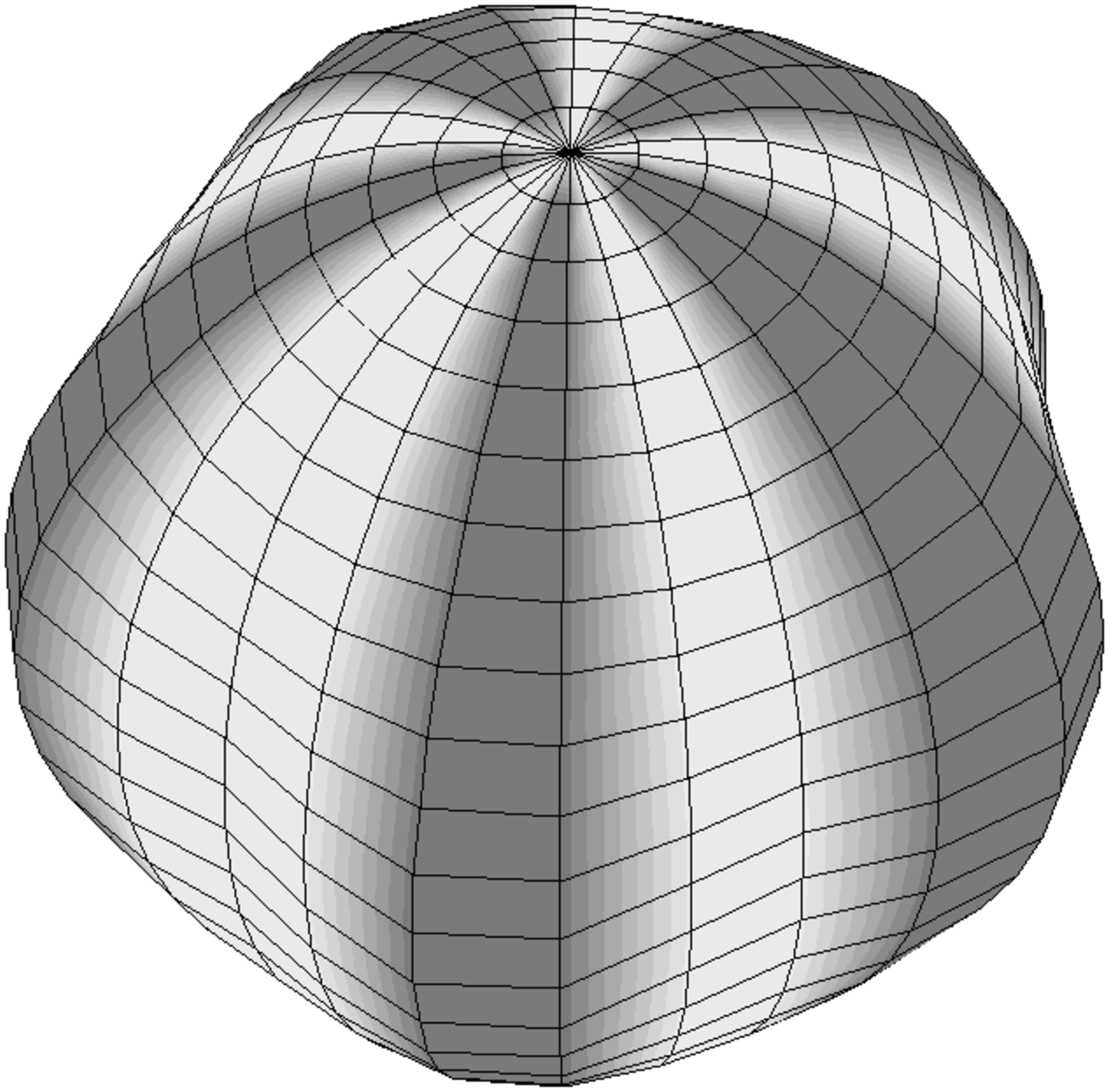}} \\
$n=3$\ $k=3$ & $n=5$\ $k=5$ \\[15pt] 
\resizebox{!}{6.0cm}{\includegraphics{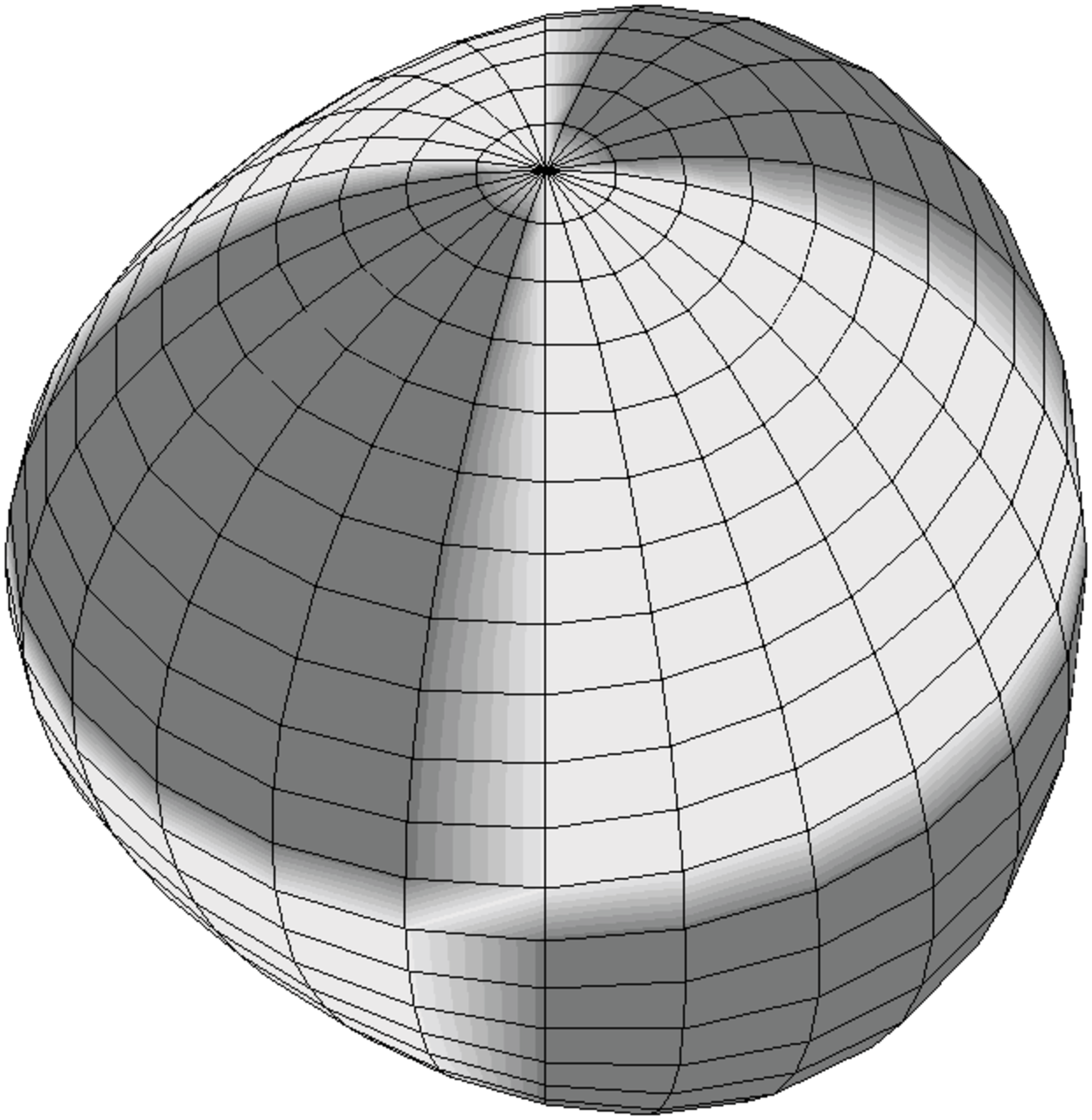}}&\resizebox{!}{6.0cm}{\includegraphics{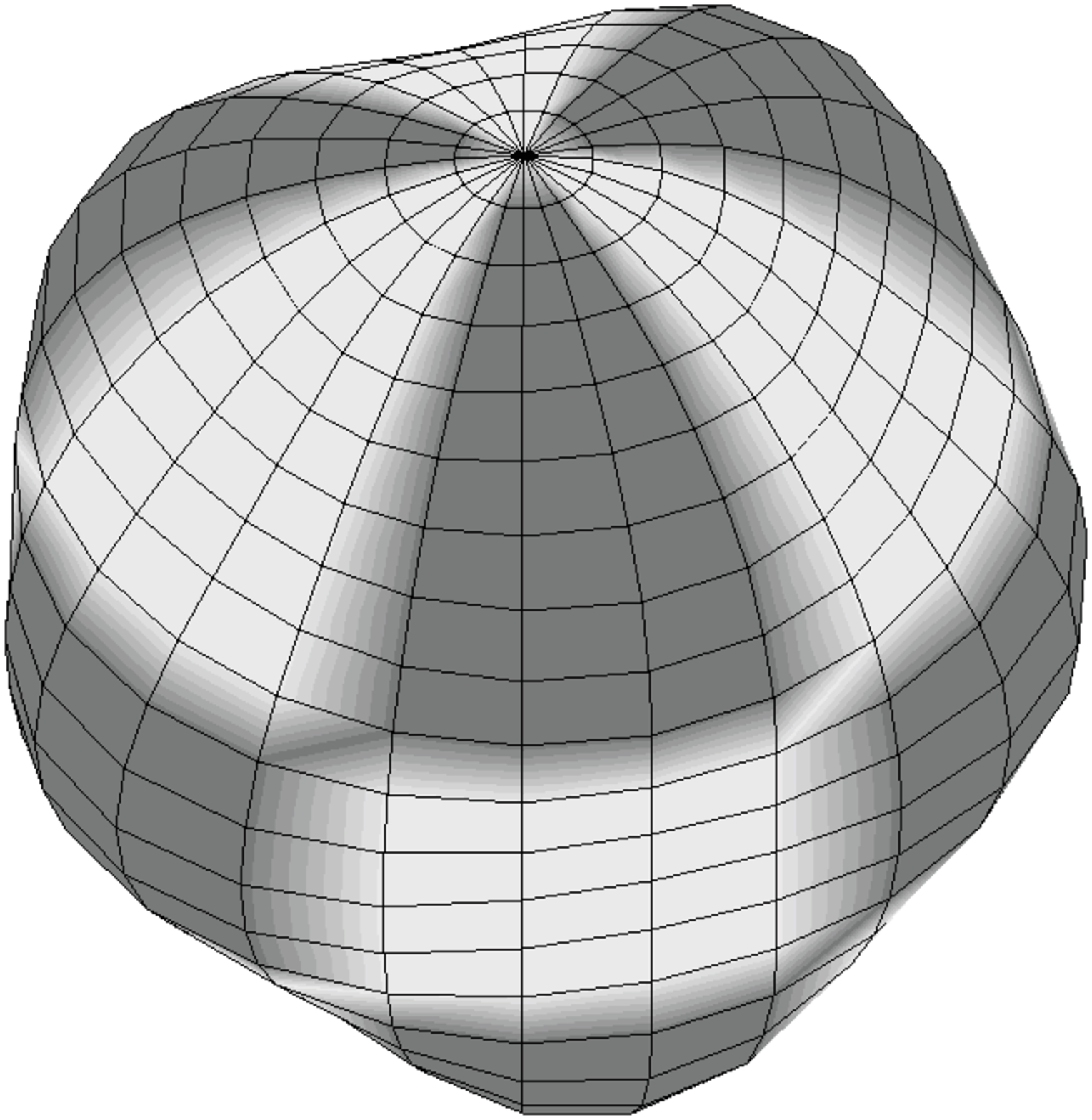}} \\
$n=3$\ $k=2$ & $n=5$\ $k=3$ 
\end{tabular}
\end{center}
\caption[Spherical functions on a sphere]{\small Real parts of spherical functions
$Y_{nk}$ for several sets of $n$ and $k$ plotted on a sphere. The character of the functions is shown using
the color code: dark gray corresponds to the areas where $Y_{nk}$ is positive, 
light gray is used where $Y_{nk}$ is negative.
\label{figure-spherical-functions}}
\end{figure}

Coefficients $C_{nk}$ and $S_{nk}$ (or $J_n$, $\overline C_{nk}$
and $\overline S_{nk}$) can be divided into four parts:
\begin{itemize}
\item[1.] The main part of the potential $\displaystyle{GM\over r}$ corresponds to
  $C_{00}=M$ and to $Y_{00}={1\over\sqrt{4\pi}}={\rm const}$.
\item[2.] The \concept{zonal harmonics} $J_n$ or $C_{n0}$ correspond
  to $Y_{n0}=\sqrt{2n+1\over4\pi}\,P_n(\cos\theta)$ that do not
depend on $\lambda$ (see Fig. \ref{figure-spherical-functions}). For
odd $n$ the coefficients $J_n$ characterize the oblateness of the
body. For even $n$ they characterize the asymmetry of the body with
respect to the $xy$-plane (that is, asymmetry between the north and
the south). This has been discussed in Section
\ref{section-symmetric-bodies}.
\item[3.] The \concept{sectorial harmonics} $\overline C_{nn}$ and
  $\overline S_{nn}$ correspond to $Y_{nn}$ and describe the effects
  that depend only on longitude $\lambda$ (see
  Fig. \ref{figure-spherical-functions}).
\item[4.] The \concept{tesseral harmonics} $\overline C_{nk}$ and
  $\overline S_{nk}$ for $n\ge 2$ and $1\le k<n$ correspond to
  $Y_{nk}$ with the same indices and describe the effects that depend
  both on longitude $\lambda$ and co-latitude $\theta$. The word
  'tessera' means 'four' or 'rectangle'. The surface of a sphere with
  $Y_{nk}$ plotted on it appears to be divided into various ``rectangles'' (see
  Fig. \ref{figure-spherical-functions}).
\end{itemize}
\noindent
In general, the coefficients with larger $n$ and $k$ describes finer details
of the potential $U$.

\chapter{Satellite Motion}

\section{Typical perturbations in satellite motion}

{\small {\bf Summary:}\
Non-sphericity of the gravitational field of the Earth. Gravitational forces of the Moon and
the Sun. Planetary perturbations. Atmospheric drag. Light pressure.
Magnetic field of the Earth. Neutral and charged particles.
Relativistic perturbations.
}\bigskip

It is obvious that nowadays the task to model the motion of Earth satellites
plays an important role and even influences our every-day life.
It is sufficient to mention communication satellites as well as the GPS, GLONASS and Galileo satellites. 
Also for scientific work Earth satellites play important role. It is well known,
for example, that the analysis of the motion of specially designed satellites (LAGEOS, CHAMP, GRACE, GOCE, etc.)
is the best way to measure the gravitational field of the Earth.

If the Earth were spherically symmetric, had no atmosphere and were
the only massive body in the universe, the orbit of a satellite in the
framework of Newtonian physics would be one of the solutions of the
two-body problem discussed in Chapter
\ref{chapter-two-body-problem}. None of these conditions is met in
reality. An orbit that would be an ellipse in the ideal case, is still
close to an ellipse in real world. However, several sorts of
perturbations lead to time-dependent
osculating elements of the orbit.  Which perturbations have to be taken
into account to describe the motion of a satellite depends both on the
parameters of the orbit and on the required accuracy. The main
perturbations that must be taken into account for most
satellites are:
\begin{itemize}
\item[-] the effects of the deviation of the Earth's gravity from spherical symmetry
as described by coefficients $J_n$, $\overline C_{nk}$ and $\overline S_{nk}$ in (\ref{U-expansion-final-Re});
\item[-] air drag caused by the motion of the satellite through the rarefied upper atmosphere;
\item[-] $N$-body perturbations due to the Moon and the Sun. 
\end{itemize}
\noindent
The first perturbation in this list is the largest one for asteroids
flying at the altitudes between approximately 300 and 30000 km above
the surface of the Earth. Air drag is the largest perturbation for
lower satellites (whose with the altitudes lower than about 300 km).
For higher flying satellites (altitude over about 30000 km) the $N$-body
perturbations from the Moon and the Sun are the most important ones.

For high accuracy modelling (mostly important for scientific satellites of
various kinds) more subtle effects should be taken into account:
\begin{itemize}
\item[-] the effects of the tidal deformations of the gravitational
  field (tidally induced temporal variations of $J_n$, $\overline
  C_{nk}$ and $\overline S_{nk}$); both solid tides and ocean tides
  should be taken into account here;
\item[-] $N$-body perturbations due to other bodies of the Solar
  system (mostly due to Jupiter and Venus)
\item[-] light pressure from the light from the Sun and reflected
  light from the Earth; the effects of umbra (shadow) and penumbra
  (the part of a shadow where the light source is only partially
  blocked) should be taken into account; the ``light'' should be
  considered not only in the ``visual band'', but also in other
  wavebands (especially, infrared);
\item[-] effects of the general theory of relativity;
\item[-] effects of the magnetic field of the Earth;
\item[-] effects of cosmic particles (both neutral and charged ones).
\end{itemize}
\noindent
Several perturbations mentioned above (e.g., light pressure and
magnetic field) require to model simultaneously also the dynamics of
the attitude (spatial orientation) of the satellite. In general,
high-accuracy modelling of satellite motion is a truly complicated
task. In this Chapter we will discuss only two sorts of perturbations
in their simplest form: the perturbations due to the oblateness of the
Earth and those due to the atmospheric drag.

\section{Motion in the quadrupole field}

{\small {\bf Summary:}\
The disturbing function for the oblate Earth. Solution in osculating
elements. Secular perturbations. Numerical example. Periodical perturbations.
}\bigskip

For the Earth (and all major planets of the solar system) the second
zonal harmonic $J_2$ is significantly larger than all other
coefficients in the expansion (\ref{U-expansion-final-Re}) of the
gravitational potential. Indeed, for the Earth $J_2\approx
1.083\times10^{-3}$ while other coefficients are of the order of
$10^{-6}$. For Earth satellites with the altitudes between about 300 and 30000 km
the effects of $J_2$ is the largest perturbation effect. Let us consider it
in more detail.

\subsection{Disturbing potential due to $J_2$}

Let us neglect 
all other terms in (\ref{U-expansion-final-Re}) and write the gravitational potential
of the Earth as
\begin{equation}
\label{Potential-J2}
U={GM\over r}\left(1-J_2\,{R^2\over r^2}\,P_2(\sin\varphi)\right),
\end{equation}
\noindent
where $\varphi=\displaystyle{\pi\over 2}-\theta$ is the geographic
latitude, $R$ is now the equatorial radius of the Earth ($R\approx
6378\,{\rm km}$), and $M$ is the mass of the Earth
($GM\approx 3.986004\times 10^{14}\,{\rm m}^3/{\rm s}^2$). The 
gravitational potential does not explicitly depend on
time (we assume $J_2={\rm const}$) and does not depend on the longitude
$\lambda$. The latter circumstance is related to the fact that if we
only consider $J_2$ the gravitational field is axially symmetric (see
Section \ref{section-axial-symmetry} above).  The gravitational
potential $U$ is considered now as a function of coordinates $r$ and
$\varphi$. The potential (\ref{Potential-J2}) can be written as
\begin{eqnarray}
\label{Potential-J2-perturbation}
U&=&{G\,M\over r}+{\cal R}\,,
\\
\label{cal-R}
{\cal R}&=&-{G\,M\over r}\,J_2\,{R^2\over r^2}\,P_2(\sin\varphi),
\end{eqnarray}
\noindent
where ${\cal R}$ can be considered as a perturbation of the two-body
motion with potential ${GM\over r}$. Then the equations of motion of
the satellite are given by (\ref{eqm-N-body-relative}) with $m_0=0$
(the influence of the satellite on the motion of the Earth is
neglected) and with gradient of ${\cal R}$ on the right-hand side.

The analysis of the motion can be done using the Lagrange equations
discussed in Section \ref{section-Lagrange-equations}. In order to use those equations we have
to express ${\cal R}$ as function of orbital elements. Position $\ve{r}=(x,y,z)$ 
of the satellite can be described using its geocentric distance $r$ and geographical longitude $\lambda$ 
and latitude $\varphi$ as
\begin{displaymath}
\pmatrix{x\cr y\cr z}=\pmatrix{r\,\cos\lambda\,\cos\varphi\cr r\,\sin\lambda\,\cos\varphi\cr r\,\sin\varphi}\,.
\end{displaymath}
\noindent
Using (\ref{XYZ2xyz}), (\ref{P-matrix}) and (\ref{X-E})--(\ref{Y-E}) one gets
\begin{equation}
\sin\varphi={z\over r}=\sin i\,\sin(v+\omega),
\end{equation}
\noindent
where $i$ is the inclination of the orbit, $\omega$ is the argument of geocenter and
$v$ is the true anomaly. Therefore, one gets
\begin{equation}
P_2(\sin\varphi)={3\over 2}\,\sin^2i\,\sin^2(v+\omega)-{1\over 2}
={3\,\sin^2i-2\over 4}-{3\over 4}\,\sin^2i\,\cos2(v+\omega)\,.
\end{equation}
\noindent
The disturbing potential can than be written as
\begin{eqnarray}
\label{cal-R-elements}
{\cal R}&=&{\mu\over a^3}\,{a^3\over r^3}\,\left({3\,\sin^2i-2\over 4}-{3\over 4}\,\sin^2i\,\cos2(v+\omega)\right)\,,
\\
\label{mu-def}
\mu&=&J_2\,G\,M\,R^2,
\end{eqnarray}
\noindent
where $a$ is the semi-major axis or the orbit. 

\subsection{Exact consequence of the axially symmetric perturbation}

Since
\begin{displaymath}
{a\over r}={1+e\,\cos v\over 1-e^2},
\end{displaymath}
\noindent
$e$ being the eccentricity, and the true anomaly is a function of the
mean anomaly $M$ and the eccentricity $e$ only, one can see that the
disturbing potential ${\cal R}$ depends on $a$, $e$, $i$, $\omega$ and
$M$, but does not depend on the longitude of the node $\Omega$.  This
is again the consequence of the fact that the assumed model for the
Earth gravity is axially symmetric.  Indeed, the only effect of a
change of $\Omega$ is a rotation of the orbit as a whole with respect
to the z-axis (see Fig.~\ref{Fig-Orbit-in-space}), but, in our model, the
gravitational force does not depend on such a rotation. One can demonstrate that
if ${\cal R}\neq{\cal R}(\Omega)$ the following combination of osculating
elements remains constant:
\begin{equation}
\label{combination-constant}
\sqrt{a\,(1-e^2)}\,\cos i={\rm const}\,.
\end{equation}
\bigskip

{\small
{\bf Exercise.}
Demonstrate that 
\begin{equation}
{d\over dt}\,\left(\sqrt{a\,(1-e^2)}\,\cos i\right)={1\over \kappa}\,{\partial R\over\partial\Omega}\,,
\end{equation}
where $\kappa^2=n^2\,a^3$, $n$ being the mean motion. 
{\it Hint:} use the Lagrange equations (\ref{Lagrange-a})--(\ref{Lagrange-i}). 
}
\bigskip

This means that as soon as the disturbing potential is axially symmetric, the combination 
$\sqrt{a\,(1-e^2)}\,\cos i$ remains constant.

\subsection{Secular part of the disturbing potential}

Now, let us confine our discussion to secular changes of the osculating elements, that is to 
the changes that polynomially depend on time. To this end let us consider the Fourier 
expansion of ${\cal R}$ in multiples of the mean anomaly $M$:
\begin{equation}
\label{cal-R-Fourier}
{\cal R}={\cal R}_0+\sum_{k=1}^\infty\left({\cal R}_k\cos kM+\widetilde{\cal R}_k\sin kM\right)\,,
\end{equation}
\noindent
where ${\cal R}_k={\cal R}_k(a,e,i,\omega)$ for $k\ge 0$.
We are only interested in the effect of ${\cal R}_0$.

Let us calculate the Fourier expansion of (\ref{cal-R-elements}). 
Two functions should be expanded: $\displaystyle{a^3\over r^3}$ and
$\displaystyle{a^3\over r^3}\,\cos2(v+\omega)$.  Eq. (\ref{Hansen})
allows one to write
\begin{eqnarray}
\label{hansen-ar3}
{a^3\over r^3}&=&\sum_{k=-\infty}^\infty X^{-3,0}_{k}(e)\,\cos kM
=X^{-3,0}_0(e)+\sum_{k=1}^\infty 2\,X^{-3,0}_{k}(e)\,\cos kM.
\end{eqnarray}
\noindent
The first equality uses the fact that the Hansen coefficients $X^{n,m}_{k}$ are real and, therefore, 
the real parts and imaginary parts of (\ref{Hansen}) also hold:
\begin{eqnarray}
\label{hansen-cos}
\left({r\over a}\right)^n\,\cos m\,v&=&\sum_{k=-\infty}^\infty X_k^{n,m}(e)\,\cos k\,M\,,
\\
\label{hansen-sin}
\left({r\over a}\right)^n\,\sin m\,v&=&\sum_{k=-\infty}^\infty X_k^{n,m}(e)\,\sin k\,M\,.
\end{eqnarray}
\noindent
For the second equality in (\ref{hansen-ar3}) 
we use the well-known general symmetry property of the Hansen coefficients
\begin{equation}
\label{Hansen-symmetry}
X^{n,m}_{k}=X^{n,-m}_{-k}.
\end{equation}
\noindent
Indeed, the substitution $m \to -m$, $k \to -k$ and $\mi \to -\mi$ in (\ref{Hansen})
changes nothing and demonstrate (\ref{Hansen-symmetry}). For $m=0$ as in (\ref{hansen-ar3})
one has $X^{n,0}_{-k}=X^{n,0}_{k}$ so that each term with $k<0$ can be written together with the 
corresponding term with $k>0$. This leads to (\ref{hansen-ar3}).

On the other hand, using (\ref{Hansen}) for $n=-3$ and $m=2$ 
($\mi$ being the imaginary unit, $\mi^2=-1$)
\begin{eqnarray}
\label{hansen-ar3-exp-i-2v}
\left({a\over r}\right)^3\,e^{\mi\,2\,v}&=&\sum_{k=-\infty}^\infty
X_k^{-3,2}(e)\,e^{\mi\,k\,M}\,,
\end{eqnarray}
\noindent
one can write
\begin{eqnarray}
\label{hansen-ar3-exp-i-2v-2omega}
\left({a\over r}\right)^3\,e^{\mi\,2\,v}\,e^{\mi\,2\,\omega}\,&=&\sum_{k=-\infty}^\infty
X_k^{-3,2}(e)\,e^{\mi\,k\,M}\,e^{\mi\,2\,\omega}\,,
\end{eqnarray}
\noindent
or
\begin{eqnarray}
\label{hansen-ar3-exp-i-2v-combined}
\left({a\over r}\right)^3\,e^{\mi\,2\,(v+\omega)}\,&=&\sum_{k=-\infty}^\infty
X_k^{-3,2}(e)\,e^{\mi\,(kM+2\omega)}
\end{eqnarray}
\noindent
and, finally,
\begin{eqnarray}
\label{hansen-ar3-cos2-v+omega}
\left({a\over r}\right)^3\,\cos2(v+\omega)\,&=&\sum_{k=-\infty}^\infty
X_k^{-3,2}(e)\,\cos(kM+2\omega)
\nonumber\\
&=&
X_0^{-3,2}(e)\,\cos2\omega+\sum_{k\neq0} X_k^{-3,2}(e)\,\cos(kM+2\omega)\,.
\end{eqnarray}
\noindent
This means that
\begin{equation}
\label{cal-R0-Hansen}
{\cal R}_0={\mu\over a^3}\,\left({3\,\sin^2i-2\over 4}\,X^{-3,0}_0-{3\over 4}\,\sin^2i\,X^{-3,2}_0\,\cos2\omega\right)\,.
\end{equation}
\noindent
The Hansen coefficients $X_k^{n,m}(e)$ can be computed in a variety of ways (see, e.g. Giacaglia, 1976). 
One possible (but, generally speaking, inefficient) way is to compute the Hansen coefficients
from their definition as a Fourier coefficient:
\begin{equation}
\label{Hansen-direct}
X^{n,m}_k={1\over 2\pi}\,\int_0^{2\pi}{\left({1-e^2\over 1+e\cos v}\right)}^n\,\cos(m\,v-k\,M)\,dM\,.
\end{equation}
\noindent
For $k=0$ it is sufficient to use (\ref{Kepler-in-true-anomaly}) to replace $dM$ by $dv$ as
\begin{equation}\label{dM-dv}
dM={\left(1-e^2\right)^{3/2}\over (1+e\,\cos v)^2}\,dv
\end{equation}
\noindent
to get
\begin{equation}
\label{Hansen-direct-k=0}
X^{n,m}_0={1\over 2\pi}\,\left(1-e^2\right)^{n+3/2}\,\int_0^{2\pi}{\cos m\,v\over{\left({1+e\cos v}\right)}^{n+2}}\,dv\,.
\end{equation}
\noindent
Computing this integral for $n=-3$, and $m=0$ and $m=2$ is trivial and one gets
\begin{eqnarray}
\label{X-3,0,0}
X^{-3,0}_0&=&(1-e^2)^{-3/2}\,,
\\
\label{X-3,2,0}
X^{-3,2}_0&=&0\,.
\end{eqnarray}
\noindent
With these expressions for the required Hansen coefficients one, finally, gets
\begin{equation}
\label{cal-R0-final}
{\cal R}_0={\mu\over a^3}\,{3\,\sin^2i-2\over 4}\,(1-e^2)^{-3/2}\,.
\end{equation}

\subsection{Secular perturbations of osculating elements}

Since ${\cal R}_0$ depends only on $a$, $e$ and $i$, one has
\begin{displaymath}
{\partial{\cal R}_0\over\partial\overline M_0}={\partial{\cal R}_0\over\partial\omega}={\partial{\cal R}_0\over\partial\Omega}=0\,.
\end{displaymath}
\noindent
The Lagrange equations (\ref{Lagrange-a})--(\ref{Lagrange-i}) give in this case 
\begin{eqnarray}
\label{a0=const}
a^{(0)}={\rm const}\,,
\\
\label{e0=const}
e^{(0)}={\rm const}\,,
\\
\label{i0=const}
i^{(0)}={\rm const}\,.
\end{eqnarray}
\noindent
This means that the semi-major axis $a$, eccentricity $e$ and
inclination $i$ remain constant. Index '$(0)$' in
(\ref{a0=const})--(\ref{i0=const}) and in equations for $\omega$,
$\Omega$ and $\overline M_0$ below stresses that the equations are valid
only in the approximation ${\cal R}={\cal R}_0$. 

The partial derivatives of (\ref{cal-R0-final}) read
\begin{eqnarray}
\label{partial-R0-a}
{\partial {\cal R}_0\over \partial a}&=&-\mu\,{3\,(2-3\,\sin^2i)\over 4\,a^4}\,{\left(1-e^2\right)}^{-3/2}\,,
\\
\label{partial-R0-e}
{\partial {\cal R}_0\over \partial e}&=&\ \mu\,{3\,e\,(2-3\,\sin^2i)\over 4\,a^3}\,{\left(1-e^2\right)}^{-5/2}\,,
\\
\label{partial-R0-i}
{\partial {\cal R}_0\over \partial i}&=&-\mu\,{3\,\sin i\,\cos i\over 2\,a^3}\,{\left(1-e^2\right)}^{-3/2}\,.
\end{eqnarray}
\noindent
Substituting these partial derivatives in the Lagrange equations
(\ref{Lagrange-omega})--(\ref{Lagrange-overlineM0}) and considering
that semi-major axis $a$, eccentricity $e$ and inclination $i$ remain
constant according to (\ref{a0=const})--(\ref{i0=const}) one get the
following simple solution for other three osculating elements:
\begin{eqnarray}
\label{omega0-linear}
\omega^{(0)}(t)&=&n_\omega\,(t-t_0)+\omega^{(0)}(t_0)\,,
\\
\label{Omega0-linear}
\Omega^{(0)}(t)&=&n_\Omega\,(t-t_0)+\Omega^{(0)}(t_0)\,,
\\
\label{overline-M00-linear}
\overline M_0^{(0)}(t)&=&n_{\overline M_0}\,(t-t_0)+\overline M_0^{(0)}(t_0)\,,
\end{eqnarray}
\noindent
where the drift rates read
\begin{eqnarray}
\label{n-omega}
n_\omega&=&{3\mu\,(4-5\,\sin^2i)\over 4\,\kappa\,a^{7/2}{(1-e^2)}^2}\,,
\\
\label{n-Omega}
n_\Omega&=&-{3\mu\,\cos i\over 2\,\kappa\,a^{7/2}{(1-e^2)}^2}\,,
\\
\label{n-overline-M00}
n_{\overline M_0}&=&{3\mu\,(2-3\,\sin^2i)\over 4\,\kappa\,a^{7/2}{(1-e^2)}^{3/2}}\,.
\end{eqnarray}
\noindent
Thus, in the general case the averaged elliptical orbit of a satellite in the field of oblate Earth
characterized by $J_2$ is an ellipse that (a) linearly precesses around the $z$-axis (linear change of $\Omega$), 
(b) linearly precesses in the orbital plane (linear change of $\omega$), and (c) has the period 
different by a constant from $P={2\pi\over n}$, $n=\kappa\,a^{-3/2}$ (linear change in $\overline M_0$). 

\subsection{Analysis of the secular perturbations}

Let us give a numerical value for the drifts for a particular
satellite (we recall that $\kappa=\sqrt{GM}$ and $\mu$ is defined by
(\ref{mu-def})):
\begin{equation}
\label{drifts-numerical-example}
\left\{
\begin{array}{ccl}
a&=&12000\ {\rm km}\\
i&=&20^\circ\\
e&=&0.1
\end{array}
\right.
\Longrightarrow
\left\{
\begin{array}{ccl}
n_\omega&=&\phantom{-}1.9^\circ/{\rm day}\\
n_\Omega&=&-1.4^\circ/{\rm day}\\
n_{\overline M_0}&=&\phantom{-}0.9^\circ/{\rm day}
\end{array}
\right.\,.
\end{equation}
\noindent
This demonstrates that the effects are significant even for relatively
high satellites ($a=12000\ {\rm km}$, $e=0.1$ corresponds to the
altitude of about 4500 km in the perigee).

\begin{figure}
\begin{center}
\begin{tabular}{ccc}
\resizebox{!}{5.0cm}{\includegraphics{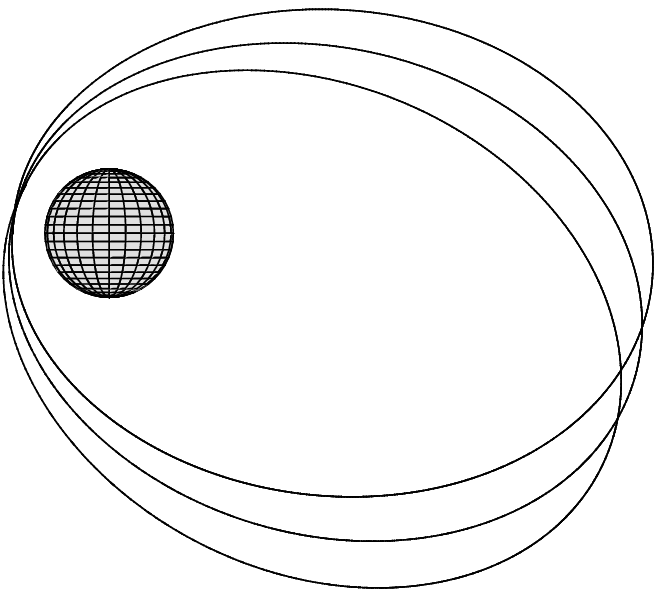}}
&
\resizebox{!}{5.0cm}{\includegraphics{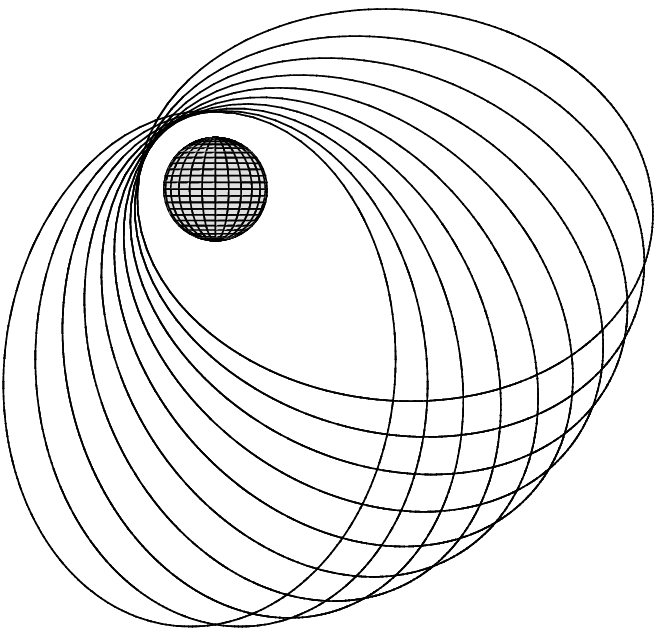}}
&
\resizebox{!}{5.0cm}{\includegraphics{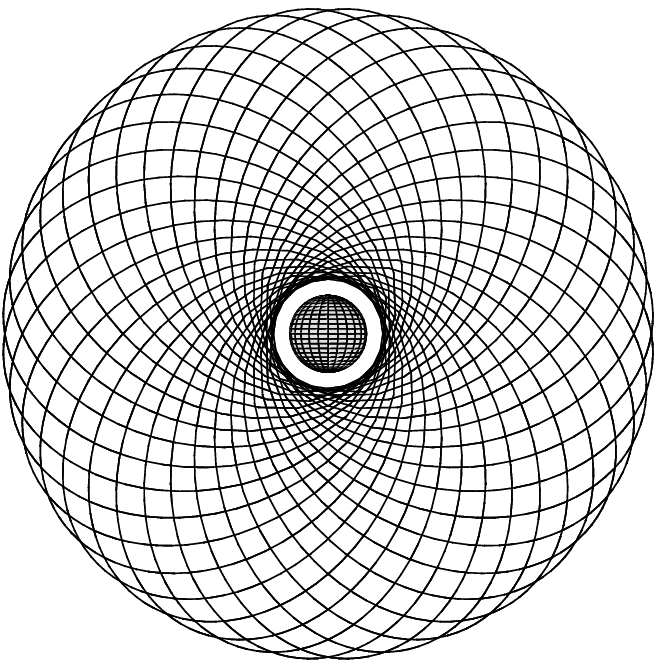}}
\end{tabular}
\end{center}
\caption[Precession of the pericenter]{\small 
An orbit with precessing argument of pericenter $\omega$ is shown 
for 3, 10 and 36 periods of motion.
The change of $\omega$ is taken to be $10^\circ$ per period. 
\label{figure-omega-precess}}
\end{figure}

\begin{figure}
\begin{center}
\begin{tabular}{ccc}
\resizebox{!}{5.6cm}{\includegraphics{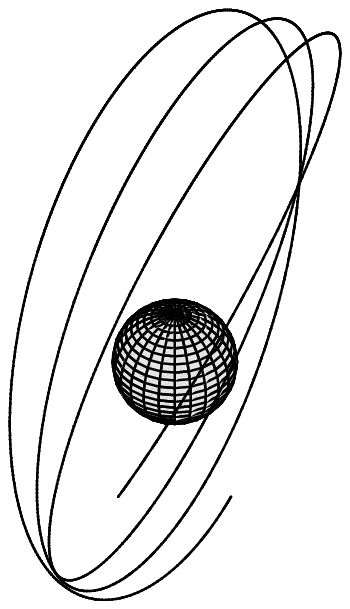}}
&
\resizebox{!}{5.6cm}{\includegraphics{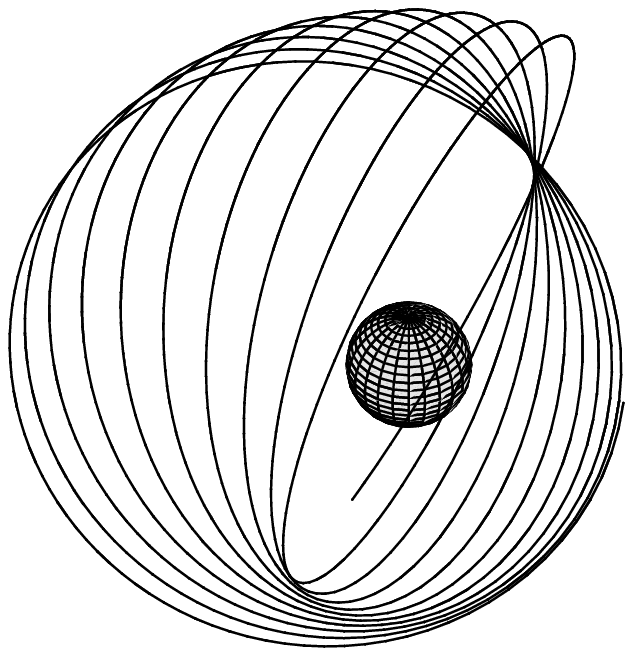}}
&
\resizebox{!}{5.6cm}{\includegraphics{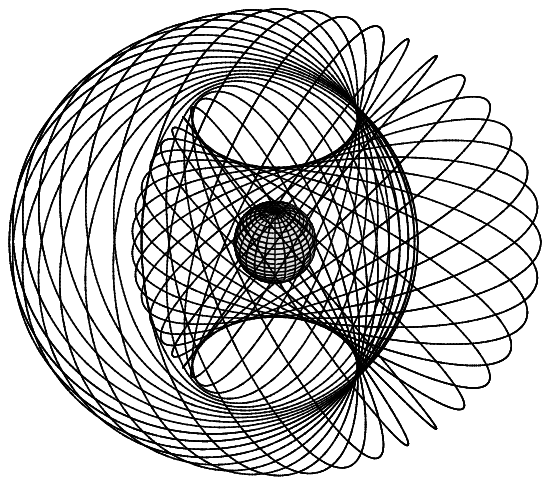}}
\end{tabular}
\end{center}
\caption[Precession of the longitude of the node]{\small 
An orbit with precessing longitude of the node $\Omega$ is shown
for 3, 10 and 36 periods of motion.
The change of $\Omega$ is taken to be $10^\circ$ per period. 
\label{figure-Omega-precess}}
\end{figure}

From (\ref{n-omega})--(\ref{n-overline-M00}) one can see that choosing some
specific inclinations $i$ each of the drifts (\ref{n-omega})--(\ref{n-overline-M00})
can be made zero. Indeed, 
\begin{itemize}

\item[-] For $i=90^\circ$ one has no rotation of the orbital plane
  $n_\Omega=0$.  The orbits with $i=90^\circ$ go straightly over the
  poles of the Earth and are called \concept{polar orbits}. These
  orbits are used e.g., for special scientific satellites that have to
  observe some objects fixed in space. For example, the mission GP-B
  (Gravity Probe B) aimed at high-accuracy testing of general
  relativity, the future astrometric mission J-MAPS (Joint
  Milli-Arcsecond Pathfinder Survey) of the US Naval Observatory and
  the mission GRACE to monitor the gravity field of the Earth all use
  such polar orbits. The argument of perigee still changes for such
  orbits.  Fig. \ref{figure-omega-precess} illustrates the form of the
  orbit in the orbital plane.
 
\item[-] For
  $i=\arcsin\displaystyle{2\over\sqrt{5}}\approx63^\circ26^\prime5.82^{\prime\prime}$
  and
  $i=\pi-\arcsin\displaystyle{2\over\sqrt{5}}\approx116^\circ33^\prime54.18^{\prime\prime}$
  the perigee $\omega$ does not precess since $n_\omega=0$. The
  orbits with such inclinations are used e.g. for special
  communication satellites especially useful for polar regions.

  Indeed, usual communication or broadcasting satellites are placed on
  the so-called \concept{geostationary orbit}.  Geostationary orbit is
  a circular orbit ($e=0$) with inclination $i=0$ (orbital plane of
  the satellite coincides with the equatorial plane of the Earth) and
  semi-major axis $a=42164\ {\rm km}$ chosen in such a way that the
  orbital period of the satellite exactly coincides with the
  rotational period of the Earth. From the point of view of an
  observer on the surface of the Earth a geostationary satellite is
  ``seen'' at a fixed position on the sky. This allows the observer to
  orient his communication antennas only once. The altitude of a
  geostationary satellite over the horizon is $90^\circ-|\varphi|$,
  where $\varphi$ is the geographical latitude of the observer. If the
  observer is situated in polar regions with $|\varphi|>70^\circ$ the
  geostationary satellites are ``seen'' too low over the horizon. This
  would require significant power increase to guarantee reliable
  communications.
 
  Already since the middle of 1960s a series of Soviet/Russian Molniya
  (Russian: ``Lightning'') communications satellites for polar regions
  used a different principle. Molniya satellites have orbits with
  inclination $i\approx63^\circ26^\prime$, high eccentricity
  $e\approx0.722$ and periods of 12 hours. Such an orbit is often called \concept{Molniya orbit}.
  Because of high eccentricity the satellite spend most of time far away from the Earth
  moving relatively slowly with respect to the Earth surface. As a result one such 
  satellite provides communication and broadcasting services for about 8 hours per day.
  Three such satellites are sufficient to make the service available at each moment of time.
  
  Fig. \ref{figure-Omega-precess} illustrates the form of the orbit
  with precessing $\Omega$.

\item[-] Finally, for $i=\arcsin\displaystyle{\sqrt{2\over3}}\approx54^\circ44^\prime8.20^{\prime\prime}$
and $i=\pi-\arcsin\displaystyle{2\over\sqrt{5}}\approx125^\circ15^\prime51.80^{\prime\prime}$
the orbital period of the satellite is given by the unperturbed two-body relations 
$P={2\pi\over n}$, $n=\kappa\,a^{-3/2}$ since $n_{\overline M_0}=0$. No practical applications
of such orbits are known.
\end{itemize}

\subsection{Additional remarks}

The problem of motion in the gravitational field of $J_2$ can be
solved exactly without averaging. In addition to secular effects that
were considered above one can consider all periodic terms in
(\ref{cal-R-Fourier}). In a more elegant way this can be done by
considering the whole perturbation given by (\ref{cal-R-elements}) and
using simultaneously both true and mean anomalies in the resulting
formulas. This allows one to derive the formulas for the first-order
variations of osculating elements in closed form (see Roy, 2005,
pp. 317--318).  The first-order variations mean here that terms that 
are at least quadratic in $J_2$ are neglected.

The effects of other coefficients $J_n$, $n\ge3$ as well as
$\overline C_{nk}$ and $\overline S_{nk}$ in (\ref{U-expansion-final-Re})
can be analyzed in a similar way. One can show, for example,
that the coefficients $J_n$ with odd $n$ lead only to periodic effects.
For Lageos all coefficients with $n$ up to $n=50$ should be taken into account. 
The motion of missions like 
CHAMP, GRACE, GOCE, etc. is sensitive to the coefficients with much higher values of $n$.
Thus, the data of CHAMP is sensitive to the coefficients with $n$ up to $n=140$,
that of CRACE up to $n=180$ and GOCE up to $n=250$.

\section{Atmospheric drag}

{\small {\bf Summary:}\
The model of the perturbing force. Models for the atmospheric pressure.
Gaussian perturbation equations for the atmospheric drag. Averaging of the equations.
The simplified Gaussian perturbation equations for small eccentricities. Solution
and its properties.
}\bigskip

Let us now turn to the perturbations due to the atmospheric
drag. Although the air in the upper atmosphere has very low density,
it influences significantly the motion of satellites (recall that the
velocity of satellites amounts to several km/s). For low
satellites with altitude 300 km and less the atmospheric drag is the
largest perturbation. Air drag can be understood as a result of
friction between the air and the body of the satellite. As with all
other friction forces the mechanical energy of the system does not
remain constant, but is partially transformed into other sorts of
energy (thermal one, etc.). It means that the air drag force does not
have a potential and should be considered using the Gaussian
perturbation equations
(\ref{Euler-equation-a})--(\ref{Euler-equation-overline-M0}).  In
general modelling of atmospheric drag is a very complex
problem. Here we consider the simplest case. More detailed discussion
can be found e.g. in King-Hele (1987).

\subsection{Model for the drag force}

\begin{figure}
\begin{center}
\begin{tabular}{ccc}
\resizebox{!}{2.7cm}{\includegraphics{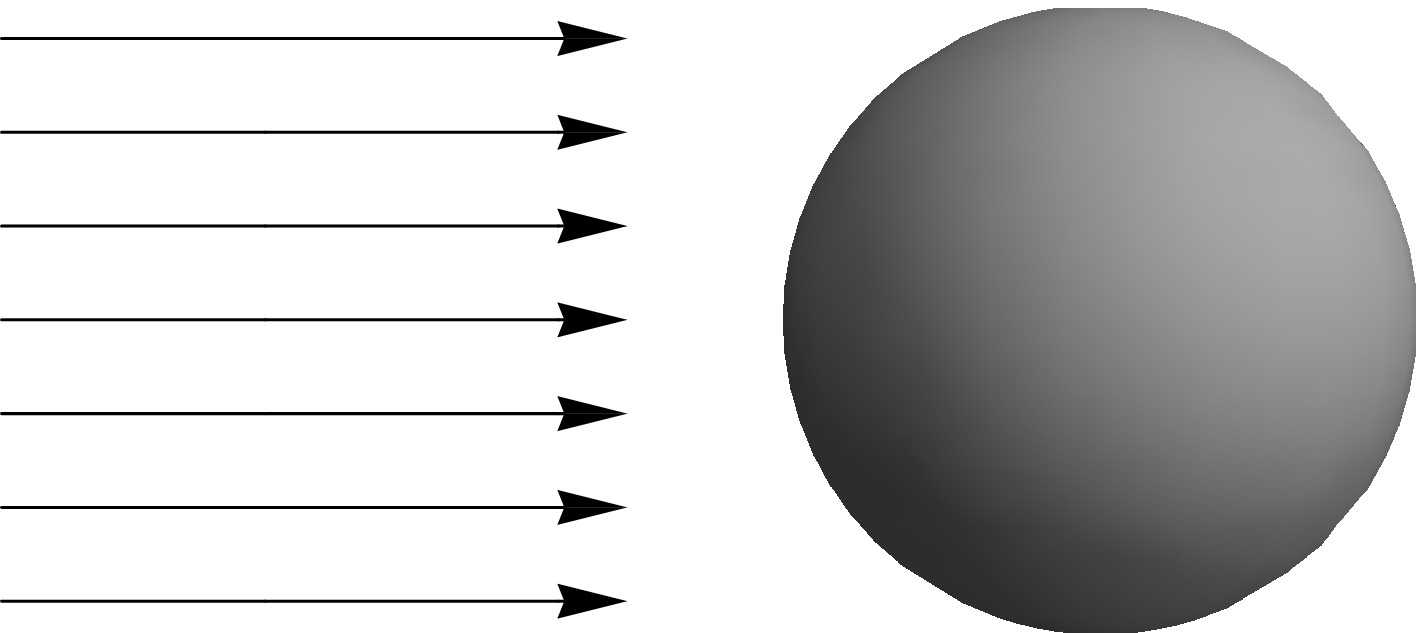}}
&
\resizebox{!}{2.7cm}{\includegraphics{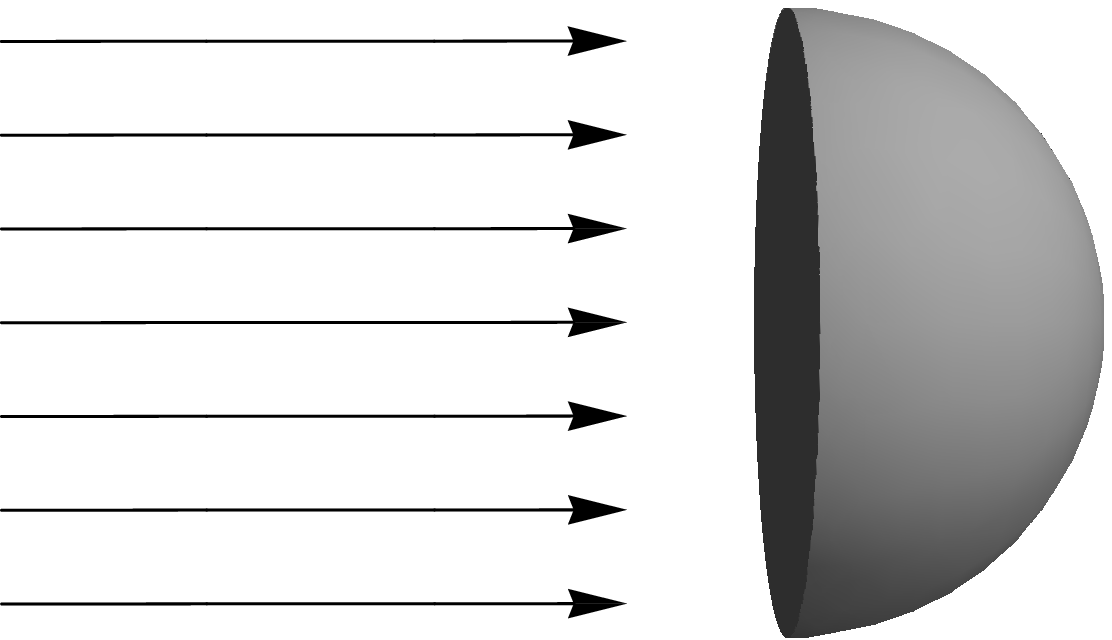}}
&
\resizebox{!}{2.7cm}{\includegraphics{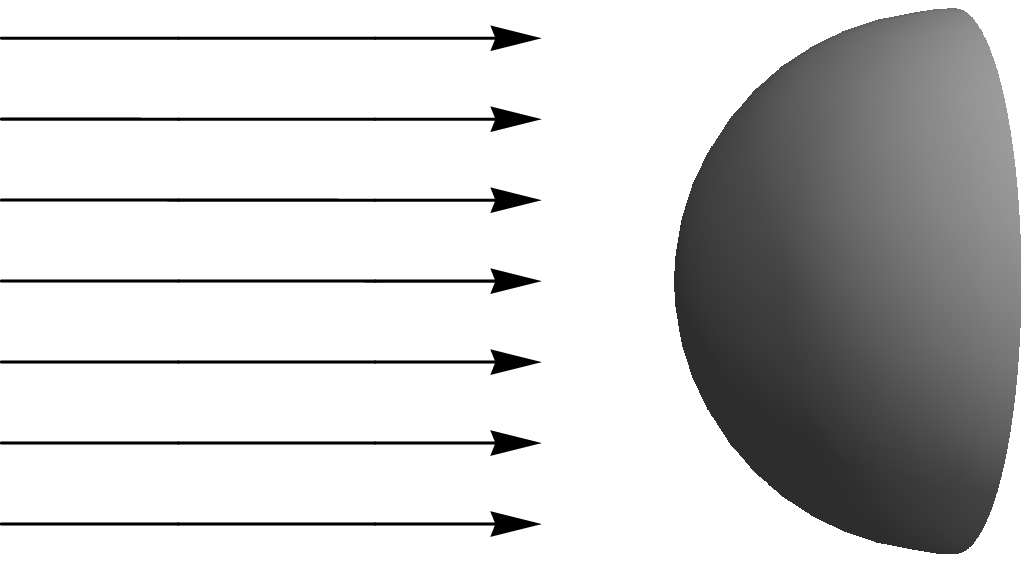}}
\\[10pt]
$C_d=0.4$ & $C_d=1.3$ & $C_d=0.3$ 
\end{tabular}
\end{center}
\caption[Drag coefficients for some bodies]{\small 
Drag coefficients for some bodies: a sphere and a half-sphere moving along its symmetry axis in two
directions.
\label{figure-drag-coefficient}}
\end{figure}

We consider here the problem of motion of a body through a medium (fluid or gas).
The relative velocity of this motion is denoted $\ve{v}$. 
It is well known that for the case of small velocities $v=|\ve{v}|$ the flow of the media  
is laminar and the friction force is linearly proportional to $v$: $F_S\propto v$.
This is called Stokes' friction model (George Gabriel Stokes 1819--1903).
For larger velocities $v$ the flow becomes turbulent and the drag force is described
by the Newtonian friction model:
\begin{equation}
\label{Newtonian-drag-force}
F_N={1\over 2}\,\rho\,C_d\,S\,v^2,
\end{equation}
\noindent
where $\rho$ is the air density, $S$ is the area of the cross
section of the body perpendicular to the direction of motion, $C_d$ is the
numerical drag coefficient that depends on the geometry of the body
(see Fig. \ref{figure-drag-coefficient} for a few examples).

In reality the situation is more difficult. The effect of atmosphere results in several effects:
(1)~air drag force with a magnitude given by (\ref{Newtonian-drag-force}) and directed along $-\ve{v}$, 
(2)~lift force directed perpendicular
to $\ve{v}$ (similar to the lift force for airplanes), (3)~angular moment leading to a rotation
of the body. We completely ignore the lift forces since they are important for higher densities
of air when the motion in lower atmosphere is considered. As for the angular moment, we simply average
the force (\ref{Newtonian-drag-force}) over several rotations of the satellite. Finally we get the
disturbing force due to atmospheric drag in the form:
\begin{equation}
\label{Newtonian-drag-force-final}
\ve{F}=-C\,\rho\,v^2\,{\ve{v}\over v},
\end{equation}
\noindent
where $C={1\over 2}\,\overline C_d\,\overline S$, and an overline means averaging over the rotation of the satellite.
We consider parameter $C$ to be known. Since velocity $\ve{v}$ can be computed from the osculating elements
of the orbit, the only unknown in (\ref{Newtonian-drag-force-final}) is the air density $\rho$.

\subsection{Model for the air density}

\begin{figure}
\begin{center}
\resizebox{!}{8.0cm}{\includegraphics{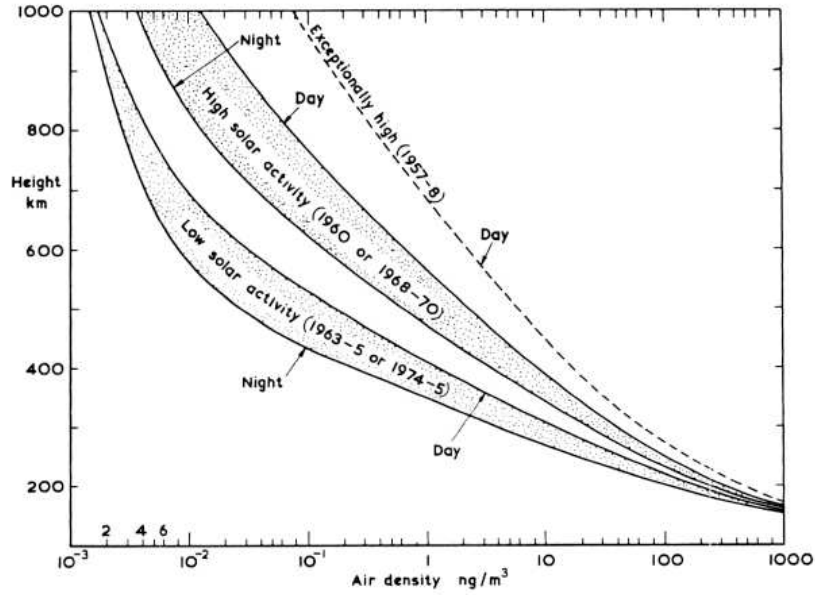}}
\end{center}
\caption[Air density as function of altitude]{\small 
The density of air is shown as function of the altitude between 150 km and 1000 km for
different conditions (day and night, low, high and exceptionally high solar activity).
The plot is based on COSPAR International Reference Atmosphere 1972 (CIRA 1972). 
\label{figure-atmosphere}}
\end{figure}

The air density can be taken from available models for the Earth's
atmosphere.  These models are based both of theoretical modelling and
on the results of measurements of various atmospheric parameters. The
models are thus semi-empirical. Fig. \ref{figure-atmosphere} shows the
air density as a function of the height for different conditions.  We
see that the air density $\rho$ at a given altitude $h$ depends on the
time (it is higher during daytime and lower at night): at a given
altitude the air density can be different by a factor 2--10 depending
on the time of the day.  It also depends on the level of solar
activity (the higher is the solar activity the higher is the density):
$\rho$ can be different by a factor 2--100 depending on the solar
activity.  There are also a number of smaller effects: (1)~the
atmosphere rotates, (2)~$\rho$ depends not only on the height $h$,
but, to a smaller degree, also on the geographical longitude $\lambda$
and the latitude $\phi$.

Here we ignore all these complications and consider a simple model of
exponential decay of the density:
\begin{equation}
\label{density-h}
\rho=A\,\exp\left(-{h\over H}\right),
\end{equation}
\noindent
where $\exp x\equiv e^x$, $h$ is the height over the Earth's surface, 
$A=\rho(0)$ is the density at $h=0$, and $H$ is height scale
(that is, the height difference for which the density decreases by a
factor of $e=2.71\dots$). At the surface of the Earth $A\approx1.3\
{\rm kg}/{\rm m}^3$ and $H\approx 8\ {\rm km}$.  Such a model
describes the air density quite well if the considered region of heights is
sufficiently small. Density $\rho$ is shown on Fig.~\ref{figure-atmosphere} in
logarithmic scale. Therefore, the model (\ref{density-h}) looks on
Fig. \ref{figure-atmosphere} as a straight line. Clearly, model
(\ref{density-h}) can be used for a sequence of layers $h\in[h_i,h_{i+1}]$,
$i=1,\dots,K$ with some boundaries $h_i$ and with constants $A=A_i$ and
$H=H_i$ depending on the layer. Such a layered model means that the curves on
Fig. \ref{figure-atmosphere} are approximated by a piecewise linear
curve. This is always possible, and the thinner are the layers the
better is the approximation. Thus, the height scale $H$ can be
considered as function $h$. Both theoretical considerations and
Fig. \ref{figure-atmosphere} demonstrate that $dH/dh>0$ that is, the
density decreases slower than exponentially (slower than prescribed by
(\ref{density-h})). In the following we simplify the model further and
consider that the orbit lies within one layer of $h$ and
the coefficients $A$ and $H$ in (\ref{density-h}) are some given constants.
Then we should express $\rho$ as function of osculating elements of the orbit.
Considering that the radial distance of the satellite $r=R+h$ and,
on the other side, $r=a(1-e\,\cos E)$, cf. Eq.(\ref{r-E}), we can write
\begin{eqnarray}
\label{rho-satellite}
\rho&=&A\,\exp\left(-{h\over H}\right)
=A\,\exp\left(-{r-R\over H}\right)
=A\,\exp\left({R\over H}\right)\,\exp\left(-{r\over H}\right)
\nonumber\\
&=&A\,\exp\left({R\over H}\right)\,\exp\left(-{a-a\,e\,\cos E\over H}\right)
=A\,\exp\left(-{a-R\over H}\right)\,\exp\left({a\,e\,\cos E\over H}\right)\,.
\end{eqnarray}
\noindent
Introducing
\begin{equation}
\label{B-def}
B=A\,\exp\left(-{a-R\over H}\right)
\end{equation}
\noindent
one finally gets
\begin{eqnarray}
\label{rho-satellite-def}
\rho&=&B\,\exp\left({\mu\,\cos E}\right)\,,
\\
\label{satellite-mu-def}
\mu&=&{a\,e\over H}\,.
\end{eqnarray}
\noindent
Note that $B$ has a simple meaning: $B$ is the air density $\rho$ at
the height equal to $a-R$, i.e. at the mean altitude of the satellite.
From the discussion above it is clear that this model works better for
orbit with smaller eccentricities $e$. Below we will see that the air
drag reduces the eccentricity $e$ of the orbit. Therefore, the model
(\ref{rho-satellite-def}) for $\rho$ works better when one considers
later stages of the orbital evolution.

\subsection{Gaussian perturbation equations in the axes aligned with the velocity vector}

\begin{figure}
\begin{center}
\resizebox{!}{7.0cm}{\includegraphics{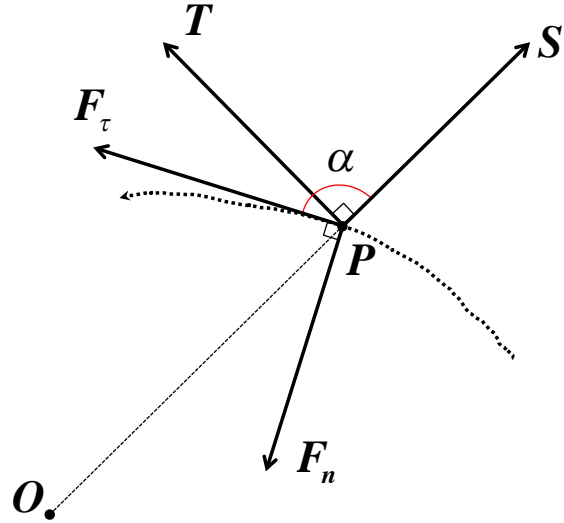}}
\end{center}
\caption[Relation between $(S,T)$ and $(F_\tau,F_n)$]{\small
The plot shows the relation between the components of the perturbing force
in two coordinate systems $(S,T)$ and $(F_\tau,F_n)$. Both systems are rectangular Cartesian
ones so that the components are mutually orthogonal. 
Axis $S$ is directed parallel to the positional vector $\ve{r}$ (from the origin $O$ to the current 
position $P$ of the body).
Axis $\tau$ is directed along the instantaneous velocity of the body. The trajectory
of the body is shown by a dotted curve and the arrow on that curve shows the direction of
motion. Axis $\tau$ is tangential to the trajectory at a given moment of time.
Finally, the angle between axes $S$ and $\tau$ is denoted by $\alpha$. Clearly, the relation
between $(S,T)$ and $(F_\tau,F_n)$ is a simple rotation given by (\ref{ST-tauN}). 
\label{figure-ST-tauN}}
\end{figure}

We have seen above that the direction of the disturbing force is
related to the direction of the orbital velocity $\ve{v}$. On the other
hand, the Gaussian perturbation equations
(\ref{Euler-equation-a})--(\ref{Euler-equation-overline-M0}) are
expressed through the components $S$, $T$, and $W$ of the disturbing
force $\ve{F}$ as defined by (\ref{S-force})--(\ref{W-force}). The components $(S,T,W)$ 
are immediately related to the vector $\ve{r}$ of orbital position, $S$ being
parallel to $\ve{r}$. Let us
introduce another coordinate system $(\tau,n,W)$ instead of
$(S,T,W)$. Namely, let the axis $\tau$ be directly along the orbital
velocity $\ve{v}$ and the axis $n$ perpendicular to $\tau$ in the
instantaneous orbital plane given by the vectors of orbital position
$\ve{r}$ and velocity $\dot{\ve{r}}=\ve{v}$.  Let $F_\tau$ and $F_n$
be components of $\ve{F}$ in the axes $\tau$ and $n$, respectively. One can write:
\begin{eqnarray}
\label{tau-force}
F_\tau&=&{\dot{\ve{r}}\over |\dot{\ve{r}}|}\cdot\ve{F},
\\
\label{n-force}
F_n&=&{\left(\ve{r}\times\dot{\ve{r}}\right)\times\dot{\ve{r}}\over \left|\ve{r}\times\dot{\ve{r}}\right|\,|\dot{\ve{r}}|}\cdot\ve{F}.
\end{eqnarray}
\noindent
Our goal is now to express the components $S$ and $T$ of $\ve{F}$ 
as given by (\ref{S-force})--(\ref{T-force}) in terms of $F_\tau$ and $F_n$.
The relation between the axes is shown on Fig. \ref{figure-ST-tauN}. It is clear from 
the Figure that the components $(S,T)$ can be derived from $(F_\tau,F_n)$ by a rotation in the orbital plane
by the angle $-\alpha$, $\alpha$ being the angle between vectors $\ve{r}$ and $\dot{\ve{r}}$.
It means
\begin{eqnarray}
\label{ST-tauN}
S&=&F_\tau\,\cos\alpha-F_n\,\sin\alpha\,,
\nonumber\\
T&=&F_\tau\,\sin\alpha+F_n\,\cos\alpha\,.
\end{eqnarray}
\noindent
where $\cos\alpha$ and $\sin\alpha$ are defined as
\begin{eqnarray}
\label{cos-alpha}
\cos\alpha&=&{\dot{\ve{r}}\cdot\ve{r}\over |\dot{\ve{r}}|\,|\ve{r}|}={e\,\sin E\over\sqrt{1-e^2\,\cos^2E}}\,,
\nonumber\\
\label{sin-alpha}
\sin\alpha&=&{\dot{\ve{r}}\times\ve{r}\over |\dot{\ve{r}}|\,|\ve{r}|}={\sqrt{1-e^2}\over\sqrt{1-e^2\,\cos^2E}}\,.
\end{eqnarray}
\noindent
Here we used (\ref{X-E})--(\ref{Y-E}) and (\ref{dotX-E})--(\ref{dotY-E}) for the
components of $\ve{r}$ and $\dot{\ve{r}}$, respectively. Substituting this transformation into the Gaussian
perturbation equations 
(\ref{Euler-equation-a})--(\ref{Euler-equation-overline-M0}) one get a version of the latter
with components $F_\tau$ and $F_n$:
\begin{eqnarray}
\label{Euler-equation-a-tauN}
{d\over dt}\, a & = &{2\over n}\,\sqrt{1+e\,\cos E\over 1-e\,\cos E}\,F_\tau\,,
\\[5pt]
\label{Euler-equation-e-tauN}
{d\over dt}\, e & = &{2(1-e^2)\,\cos E\over a\,n\,\sqrt{1-e^2\cos^2E}}\,F_\tau
-{\sqrt{1-e^2}\,\sin E\over a\,n}\,\sqrt{1-e\,\cos E\over 1+e\,\cos E}\,F_n\,,
\\[5pt]
\label{Euler-equation-omega-tauN}
{d\over dt}\, \omega & = &{2\over n\,a}\,{\sqrt{1-e^2}\over e}\,{\sin E\over \sqrt{1-e^2\,\cos^2E}}\,F_\tau
+{\cos E+e\over a\,n\,e}\,\sqrt{1-e\,\cos E\over 1+e\,\cos E}\,F_n
\nonumber\\
&&
-r\,\sin(v+\omega)\,\cot i\,\left({W\over\kappa\sqrt{p}}\right),
\\[5pt]
\label{Euler-equation-overline-M0-tauN}
{d\over dt}\, \overline{M}_0 & = & 
-{2\over a\,n\,e}\,{(1-e^3\cos E)\,\sin E\over \sqrt{1-e^2\cos^2E}}\,F_\tau
\nonumber\\
&&-{\sqrt{1-e^2}\,(\cos E-e)\over a\,n\,e}\,\sqrt{1-e\,\cos E\over 1+e\,\cos E}\,F_n\,.
\end{eqnarray}
\noindent 
The equations (\ref{Euler-equation-i}) and
(\ref{Euler-equation-Omega}) for $i$ and $\Omega$ remain unchanged.

\subsection{Osculating elements for the air drag}

The disturbing force from the air drag is given by
(\ref{Newtonian-drag-force-final}) and (\ref{rho-satellite-def})--(\ref{satellite-mu-def}). 
The components of the disturbing force $\ve{F}$ can be written as:
\begin{eqnarray}
\label{drag-F-tau}
F_\tau&=&-C\,\rho\,v^2\,,
\\
\label{drag-F-n}
F_n&=&0\,,
\\
\label{drag-F-W}
W&=&0\,.
\end{eqnarray}
\noindent 
Therefore, the air drag does not change $i$ and $\Omega$, i.e. the orbital plane
remains unchanged. In (\ref{Euler-equation-a-tauN})--(\ref{Euler-equation-overline-M0-tauN})
only the terms proportional to $F_\tau$ should be considered. From (\ref{dotX-E})--(\ref{dotY-E})
one gets
\begin{equation}
\label{v2-kepler}
v^2=n^2\,a^2\,{1+e\,\cos E\over 1-e\,\cos E}.
\end{equation}
\noindent
Let us also change the left-hand side of (\ref{Euler-equation-a-tauN})--(\ref{Euler-equation-overline-M0-tauN})
from the derivatives with respect to time $t$ to the corresponding derivatives with respect to the 
eccentric anomaly $E$. Using (\ref{E-dot}) for any element $\epsilon$ one has
\begin{eqnarray}
\label{change-derivatives}
{d\epsilon\over dE}&=&{d\epsilon\over dt}\,\left({dE\over dt}\right)^{-1}=
{1-e\cos E\over n}\,{d\epsilon\over dt}.
\end{eqnarray}
\noindent
In the following we are only interested in the form of the trajectory.
Therefore, the element $\overline M_0$ plays no role and will be
ignored below. Substituting (\ref{drag-F-tau}) and (\ref{v2-kepler})
into (\ref{Euler-equation-a-tauN})--(\ref{Euler-equation-omega-tauN})
and using (\ref{change-derivatives}) for all elements one gets:
\begin{eqnarray}
{da\over dE}&=&-2a^2\,C\,\rho\,
\sqrt{(1+e\,\cos E)^3\over 1-e\,\cos E},
\\
{de\over dE}&=&-2a\,C\,\rho\,(1-e^2)\,
\cos E\,\sqrt{1+e\,\cos E\over 1-e\,\cos E},
\\
{d\omega\over dE}&=&-2a\,C\,\rho\,e^{-1}\,\left(1-e^2\right)^{1/2}\,\sin E\,
\sqrt{1+e\,\cos E\over 1-e\,\cos E}.
\end{eqnarray}
\noindent
Now we can substitute the density model
(\ref{rho-satellite-def})--(\ref{satellite-mu-def}) and get
\begin{eqnarray}
\label{da-dE}
{da\over dE}&=&-2a^2\,C\,B\,\exp\left(\mu\,\cos E\right)\,
\sqrt{(1+e\,\cos E)^3\over 1-e\,\cos E},
\\
\label{de-dE}
{de\over dE}&=&-2a\,C\,B\,(1-e^2)\,\exp\left(\mu\,\cos E\right)\,
\cos E\,\sqrt{1+e\,\cos E\over 1-e\,\cos E},
\\
\label{domega-dE}
{d\omega\over dE}&=&-2a\,C\,B\,e^{-1}\,\left(1-e^2\right)^{1/2}\,\sin E\,
\exp\left(\mu\,\cos E\right)\,
\sqrt{1+e\,\cos E\over 1-e\,\cos E}.
\end{eqnarray}

\subsection{Averaged equations for the osculating elements}

Let us investigate now only secular perturbations of $a$, $e$ and
$\omega$. First, let us calculate the mean value of the derivatives of
these three elements over a period of motion as
\begin{eqnarray}
\left[{d\epsilon\over dE}\right]={1\over 2\,\pi}\,
\int_0^{2\pi} {d\epsilon\over dE}\, dE=
{1\over 2\,\pi}\,\int_{-\pi}^{\pi} {d\epsilon\over dE}\, dE,
\end{eqnarray}
\noindent
where again $\epsilon$ is any of the elements. Since $d\omega/dE$ given by
(\ref{domega-dE}) is an odd function of $E$ one has
\begin{equation}
\label{domega-dE-averaged}
\left[{d\omega\over dE}\right]=0\,.
\end{equation}
\noindent
This means that the osculating argument of perigee does not have secular variations, but only
periodic ones. For the other two elements one gets
\begin{eqnarray}
\label{da-dE-averaged}
\left[{da\over dE}\right]&=&-{a^2\,C\,B\over \pi}\,\int_0^{2\pi} \exp\left(\mu\,\cos E\right)\,
\sqrt{(1+e\,\cos E)^3\over 1-e\,\cos E}\,dE\,,
\\
\label{de-dE-averaged}
\left[{de\over dE}\right]&=&-{a\,C\,B\over \pi}\,(1-e^2)\,\int_0^{2\pi} \exp\left(\mu\,\cos E\right)\,
\cos E\,\sqrt{1+e\,\cos E\over 1-e\,\cos E}\,dE\,.
\end{eqnarray}

\subsection{Averaged osculating elements for small eccentricities}

\begin{figure}
\begin{center}
\resizebox{!}{8.0cm}{\includegraphics{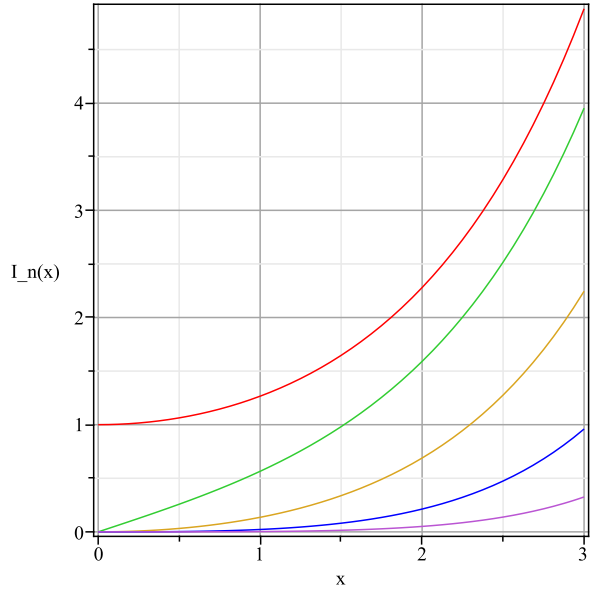}}
\end{center}
\caption[Modified Bessel functions of the first kind]{\small
Modified Bessel functions of the first kind
$I_n(x)$ are shown for $0\le x\le 3$ and $k=0$, $1$, $2$ and $3$
(red, green, yellow, blue, magenta). 
\label{figure-BesselI}}
\end{figure}

At this point we need one more sort of special functions. Namely, $I_m(x)$ defined as
\begin{equation}
\label{Bessel-I-m}
I_m(x)={1\over 2\pi}\int_0^{2\pi} \exp(\mu\,\cos E)\,\cos mE\,dE
\end{equation}
\noindent
are called \concept{modified Bessel functions of the first kind}
(sometimes \concept{hyperbolic Bessel functions of the first
 kind}). Many properties of these functions are known (see Abramowitz
\& Stegun (1965), Chapter 9). They are related to the Bessel functions
of the first kind given by (\ref{Bessel-J-n}) as
\begin{equation}
\label{Bessel-I-m-J-n}
I_m(x)=\mi^{-m} J_m(\mi x),
\end{equation}
\noindent
and can be calculated through the following power series
\begin{equation}
\label{Bessel-I-m-series}
I_m(x)=\sum_{l=0}^\infty {1\over l!(l+m)!}\,{\left({1\over 2}\,x\right)}^{m+2l}.
\end{equation}
\noindent
Functions $I_m(x)$ are shown on Fig. \ref{figure-BesselI} for a few values of $m$. Below
it will be important that 
\begin{itemize}
\item[1.] $I_m(0)\neq 0$ only for $m=0$ (one has $I_0(0)=1$);
\item[2.] $I_m(x)\ge0$ for any $m$ and $x\ge 0$;
\item[3.] $dI_m(x)/dx\ge0$ for any $m$ and $x\ge 0$.
\end{itemize}
\noindent
These properties can be seen directly from (\ref{Bessel-I-m-J-n}).  

Using $I_m(x)$ and considering the Taylor series
\begin{eqnarray}
\label{da-dE-taylor}
\sqrt{(1+e\,\cos E)^3\over 1-e\,\cos E}
&=&
1+2\cos E\,e+{3\over 4}\,(1+\cos 2E)\,e^2+{\cal O}(e^3)\,,
\\
\label{de-dE-taylor}
\cos E\,\sqrt{1+e\,\cos E\over 1-e\,\cos E}
&=&
\cos E+{1\over 2}(1+\cos 2E)\,e+{1\over 8}\left(3\cos E+\cos 3E\right)\,e^2+{\cal O}(e^3)
\end{eqnarray}
\noindent
one gets
\begin{eqnarray}
\label{da-dE-averaged-expanded}
\left[{da\over dE}\right]&=&-2a^2\,C\,B\,
\left(I_0(\mu)+2e\,I_1(\mu)+{3\over 4}\,e^2\,\left(I_0(\mu)+I_2(\mu)\right)+{\cal O}(e^3)\right)\,,
\\
\label{de-dE-averaged-expanded}
\left[{de\over dE}\right]&=&-2a\,C\,B\,(1-e^2)\,
\left(I_1(\mu)+{1\over 2}\,e\,\left(I_0(\mu)+I_2(\mu)\right)\right.
\nonumber\\
&&
\phantom{-2a\,C\,B\,(1-e^2)\,\biggl(\ }
\left.+{1\over 8}\,e^2\,\left(3I_1(\mu)+I_3(\mu)\right)
+{\cal O}(e^3)\right)\,.
\end{eqnarray}
\noindent
Here we expanded the equations in power of eccentricity $e$ and
neglected terms of the order of $e^3$ and higher.  It means that
(\ref{da-dE-averaged-expanded})-(\ref{de-dE-averaged-expanded}) are
valid only for sufficiently small eccentricities. In principle,
higher-order terms in $e$ can be calculated. However, for large $e$ it
is better to solve (e.g., numerically) directly
(\ref{da-dE-averaged})-(\ref{de-dE-averaged}) or, directly,
(\ref{da-dE})-(\ref{domega-dE}).

\subsection{Discussion of the solution for osculating elements}

\begin{figure}
\begin{center}
\begin{tabular}{cc}
\resizebox{!}{8.0cm}{\includegraphics{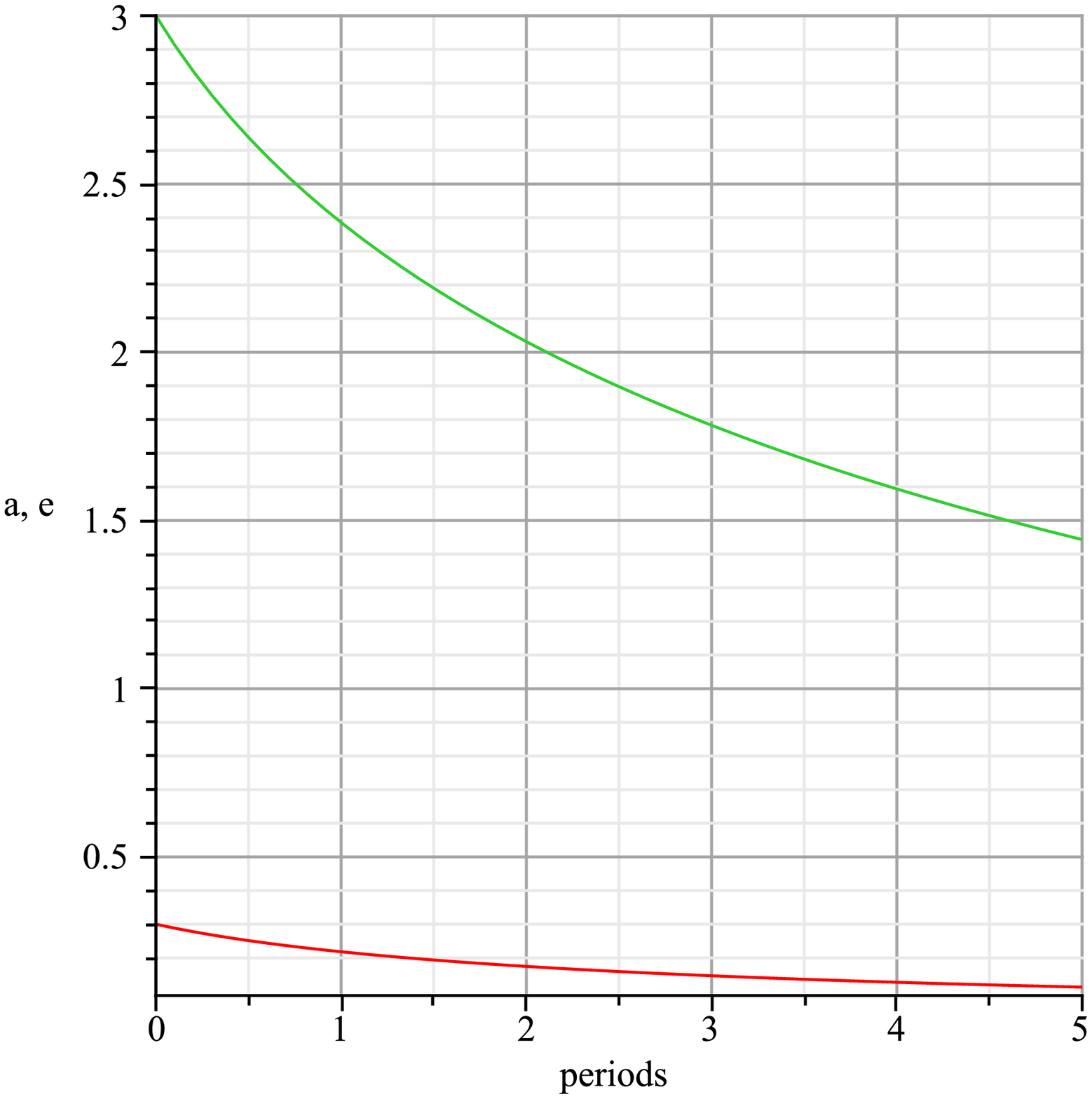}}
&
\resizebox{!}{8.0cm}{\includegraphics{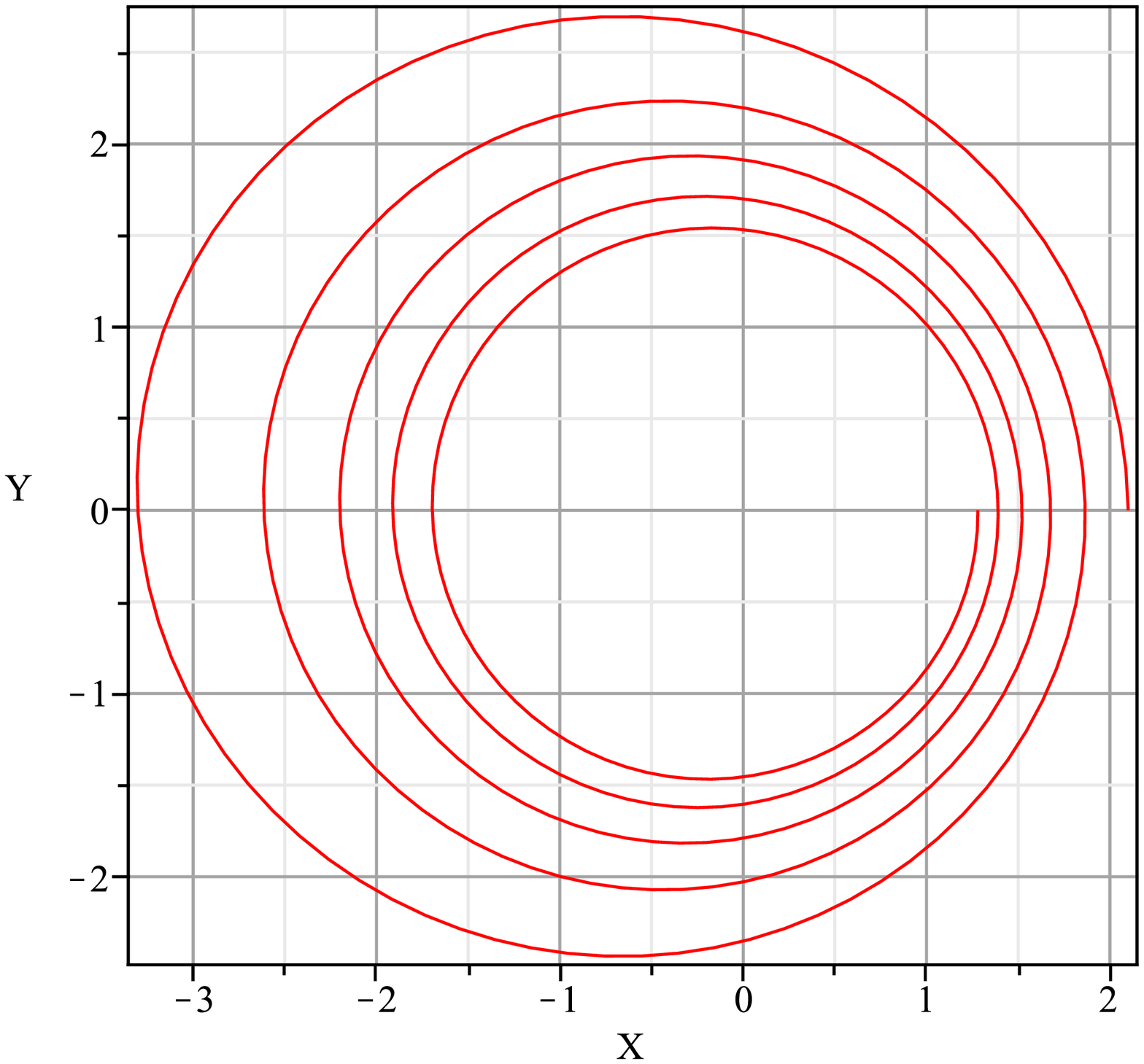}}
\end{tabular}
\end{center}
\caption[The effect of the atmospheric drag on a satellite orbit]{\small
The left pane shows the variation of the osculating semi-major axes $a$ (upper green curve) and eccentricity $e$ 
(lower red curve) over several orbital periods.
On the right pane the orbit of a satellite affected by the atmospheric drag is shown over the same
interval of time. 
One can see that both the semi-major axis and the eccentricity decrease with time.
The effect is exaggerated to make it better visible. 
Initial eccentricity is $e=0.3$.
\label{figure-satellite-atmosphere}}
\end{figure}

Equations
(\ref{da-dE-averaged-expanded})-(\ref{de-dE-averaged-expanded}) are
two differential equations for two unknown functions $a(E)$ and
$e(E)$. These equations are coupled. Note also that $\mu$ depends on
both $a$ and $e$ as given by (\ref{satellite-mu-def}). For a numerical example, the solution of
(\ref{da-dE-averaged-expanded})-(\ref{de-dE-averaged-expanded}) is
shown on Fig. \ref{figure-satellite-atmosphere}.  Let us make several
remarks.
\begin{itemize}

\item[1.] Since $I_m(\mu)\ge 0$ the right-hand sides of (\ref{da-dE-averaged-expanded})-(\ref{de-dE-averaged-expanded})
are non-positive. It means that the averaged values of $a$ and $e$ are non-increasing function of $E$ 
(and, therefore, of time $t$). That is, both $a$ and $e$ becomes smaller during the evolution of the orbit.

\item[2.] For circular orbits ($e=0$) the equations can be drastically simplified:
\begin{eqnarray}
\label{da-dE-averaged-expanded-e=0}
\left[{da\over dE}\right]&=&-2a^2\,C\,B\,,
\\
\label{de-dE-averaged-expanded-e=0}
\left[{de\over dE}\right]&=&0\,.
\end{eqnarray}
\noindent
It means that the orbit remains circular with $e=0$ and that 
the semi-major axis decreases hyperbolically:
\begin{eqnarray}
\label{da-averaged-e=0}
a(E)&=&{1\over {1\over a(E_0)}+2C\,B\,(E-E_0)}\,,
\end{eqnarray}
\noindent
$a(E_0)$, being an integration constant, is the value of $a$ for some initial moment $E=E_0$.

\item[3.] The smaller becomes the eccentricity $e$, the smaller is the absolute value of
the right-hand sides of (\ref{da-dE-averaged-expanded})-(\ref{de-dE-averaged-expanded}).
Here we used that ${d I_m(x)\over dx}\ge 0$ for any $n$ and $x\ge 0$. This means that
the rate of change of both $a$ and $e$ decreases with time.

\item[4.] Finally, let us note that because of the air drag, satellites become faster
and not slower as one could expect from a friction force. This is of course related 
with the gravitational character of motion: if semi-major axes $a$ decreases, the velocity
increases as $a^{-1/2}$. Indeed, Eq. (\ref{v2-kepler}) can be written as 
\begin{equation}
\label{v-kepler}
v=\sqrt{GM\over a}\,\sqrt{1+e\,\cos E\over 1-e\,\cos E}.
\end{equation}
\noindent
Although the air drag as any friction force decreases the total mechanical energy of the system
Earth-satellite, the potential energy $-GM/r$ of the system is transformed 
into the kinetic energy $v^2/2$ and the latter increases.
\end{itemize}

\chapter[Numerical integration]{Numerical integration of ordinary differential equations}

\section{Basic notions}

{\small {\bf Summary:}\
Euler step for the differential equations of the first order.
Discretization. Three kinds of errors: the local truncation error, the
global error and the roundoff error.
}\bigskip

\section{Methods of numerical integration}

{\small {\bf Summary:}\
The method of Taylor expansion. The Runge-Kutta method.
Stepsize control for the Runge-Kutta methods (Fehlberg method). The
Runge-Kutta-Nystr\"om method. Multistep methods. Explicit and implicit
methods. Predictor-corrector methods. Adams-Bashforth and Adams-Moulton
methods. Extrapolation methods.
}\bigskip

\section{Reliability of numerical integration}

{\small {\bf Summary:}\
Close encounters. Regularization. Accuracy control.
}\bigskip

\addcontentsline{toc}{chapter}{Index}

\printindex

\end{document}